\documentclass[11pt,a4paper]{article}

\usepackage[T1]{fontenc}
\usepackage[utf8]{inputenc}
\usepackage[sc]{mathpazo}          
\usepackage[final]{microtype}
\usepackage[a4paper,margin=27mm,top=30mm,bottom=32mm]{geometry}

\usepackage{amsmath,amssymb,amsthm,mathtools}
\usepackage{mathrsfs}
\usepackage{courier}

\usepackage{booktabs,array}
\usepackage{caption}
\usepackage{enumitem}
\setlist{itemsep=2pt,topsep=4pt}
\usepackage{tikz}
\usetikzlibrary{positioning,arrows.meta,calc,shapes.geometric}
\usepackage{xcolor}

\usepackage{fancyhdr}
\usepackage{titlesec}
\titleformat{\section}{\normalfont\large\scshape\centering}{\thesection.}{0.6em}{}
\titleformat{\subsection}{\normalfont\normalsize\bfseries}{\thesubsection.}{0.5em}{}
\titleformat{\subsubsection}[runin]{\normalfont\normalsize\itshape}{}{0pt}{}[.]
\titlespacing*{\section}{0pt}{3.0ex plus 1ex minus .2ex}{1.8ex plus .2ex}

\usepackage[colorlinks=true,linkcolor=blue!45!black,citecolor=green!35!black,
            urlcolor=blue!55!black,pdftitle={Topology and Information in Regularized Transport on Graphs},
            pdfauthor={Shintaro Yoshizawa}]{hyperref}
\usepackage[capitalise,nameinlink]{cleveref}

\theoremstyle{plain}
\newtheorem{theorem}{Theorem}[section]
\newtheorem{proposition}[theorem]{Proposition}
\newtheorem{lemma}[theorem]{Lemma}
\newtheorem{corollary}[theorem]{Corollary}
\newtheorem*{theoremBN}{Theorem (Baker--Norine)}
\theoremstyle{definition}
\newtheorem{definition}[theorem]{Definition}
\newtheorem{example}[theorem]{Example}

\newtheorem{exercise}{Exercise}[section]
\theoremstyle{remark}
\newtheorem{remark}[theorem]{Remark}
\crefname{exercise}{Exercise}{Exercises}
\crefname{question}{Question}{Questions}

\newcommand{\Z}{\mathbb{Z}}
\newcommand{\R}{\mathbb{R}}

\newcommand{\MobIdx}{\mathfrak{M}}
\newcommand{\Div}{\operatorname{Div}}

\newcommand{\Pic}{\operatorname{Pic}}
\newcommand{\Jac}{\operatorname{Jac}}
\newcommand{\diam}{\operatorname{diam}}
\newcommand{\TV}{\mathrm{TV}}
\newcommand{\lin}[1]{\lvert #1\rvert}
\newcommand{\one}{\mathbf{1}}
\DeclareMathOperator{\adj}{adj}

\tikzset{
  vertex/.style={circle,draw=black!70,fill=white,minimum size=7.5mm,inner sep=0pt,font=\small},
  vhub/.style={circle,draw=blue!60!black,fill=blue!10,minimum size=7.5mm,inner sep=0pt,font=\small},
  vleaf/.style={circle,draw=red!55!black,fill=red!7,minimum size=7.5mm,inner sep=0pt,font=\small},
  gedge/.style={semithick,draw=black!60},
  chord/.style={thick,draw=red!60!black,dashed},
  anno/.style={font=\footnotesize,text=black!70}
}

\usepackage{pgfplots}\pgfplotsset{compat=1.18}
\newcommand{\Dq}{\mathrm{D}_q}
\newcommand{\Wq}{\mathcal W_{q,\gamma}}

\tikzset{
  hubn/.style={circle,fill=blue!55!black,draw=white,line width=0.2pt,inner sep=0pt,minimum size=3.2pt},
  transn/.style={circle,fill=black!45,draw=white,line width=0.2pt,inner sep=0pt,minimum size=3.2pt},
  termn/.style={circle,fill=red!70!black,draw=white,line width=0.2pt,inner sep=0pt,minimum size=3.2pt},
  srcn/.style={rectangle,draw=black,fill=yellow!80!orange,line width=0.5pt,inner sep=0pt,minimum size=5pt},
  tgtn/.style={diamond,draw=black,fill=green!60!black,line width=0.5pt,inner sep=0pt,minimum size=6pt}
}
\DeclareMathOperator{\dist}{dist}
\usepackage[section]{placeins}
\begin{document}

\thispagestyle{plain}
\begin{center}
  \vspace*{1.0em}
  {\LARGE\scshape Topology and Information\\[0.3em]
   in Regularized Transport on Graphs}\\[1.7em]
  {\large\scshape Shintaro Yoshizawa}\footnote{E-mail:
  \href{mailto:shintaro.yoshizawa.net@gmail.com}{\texttt{shintaro.yoshizawa.net@gmail.com}},
  \href{mailto:shintaro_yoshizawa@mail.toyota.co.jp}{\texttt{shintaro\_yoshizawa@mail.toyota.co.jp}}}\\[0.4em]
  {\small Frontier Research Center, Toyota Motor Corporation}\\[1.6em]
  {\small\itshape Dedicated to Professor Shun-ichi Amari\\
  on the occasion of his 90th birthday}
\end{center}

\vspace{1.0em}
\begin{center}
\begin{minipage}{0.87\textwidth}
\small
\noindent\textsc{Abstract.}\;
We study how two geometries of mass on a finite graph interact: the
information geometry of the generalized entropy $F_q$ used for regularized
transport, and the arithmetic of divisors governed by the Riemann--Roch theorem
of Baker and Norine and by the Jacobian. Because the correction term of $F_q$
is affine, $F_q$-regularization at strength $\gamma$ is Tsallis regularization at
temperature $\gamma/q$. Over the reals, chip-firing connects all configurations
of equal degree, so the Jacobian enters transport only through integrality.
We introduce the transport defect, the worst-case cost of steering $d$ chips
towards a target by chip-firing, and prove
$1/d\le\delta_{\TV}(D)\le(d-r(D))/d=(g-1-r(K-D))/d\le\min(1,g/d)$ whenever
$g\ge1$; vertex multiplicities give a lower bound and an equality criterion. A
regularized topological gap $\Gamma_{q,\gamma}(D,E)$ tends to the chip-move
defect as $\gamma\to0$, lies in an explicit window, has a high-temperature
slope given by Tsallis coupling entropies, and admits a positive regime if and
only if $g\ge1$. Dual potentials certify exactly the convex relaxation of a
linear system; the remainder is an integrality gap. Zeros in optimal plans
occur only for $q>1$, by a mechanism different from the sign change of $\Gamma$.
Exact computations on $1784$ divisor classes find the rank bound $(d-r(D))/d$
attained in $1769$ of them, an observed frequency. Experiments on honeycomb and
Erd\H{o}s--Rényi networks with $10^2$ to $10^5$ nodes test the statements on
regularized plans, not $\Gamma$; over $10^3\le n\le10^5$ their transport costs
are consistent with growth like $n^{1/2}$ and $\log n$, respectively.


\end{minipage}
\end{center}

\newpage
{\small\tableofcontents}
\newpage

\section{Introduction}\label{sec:intro}

\subsection{Two geometries of mass on a graph}
Entropic regularization has made optimal transport a practical tool: adding a
strictly convex penalty to the Kantorovich problem yields unique, smooth and
efficiently computable plans~\cite{Cuturi2013,PeyreCuturi2019}. From the
viewpoint of information geometry~\cite{AmariNagaoka2000,Amari2016}, the
regularizer generates a dually flat structure on the cone of plans: the plan
itself is the mixture coordinate, the derivative of the regularizer is the
exponential coordinate, and the optimal plan is a Bregman projection. Among
the regularizers proposed beyond Shannon's, the generalized entropy
\[
  F_q(x)=\frac{x^q+q(1-x)-1}{q(q-1)}-(2q-1)(1-x)
\]
of Kurose, Yoshizawa and Amari~\cite{KYA2022} has recently been applied to
particle transport on networks~\cite{JSAI2026}, where the parameter $q$ was
interpreted as a curvature controller and the correction term
$-(2q-1)(1-x)$ as a deformation of the geometry.

A graph carries a second, quite different geometry of mass. A configuration of
chips is a \emph{divisor}; chip-firing generates \emph{linear equivalence};
and the theorem of Baker and Norine~\cite{BakerNorine2007},
\begin{equation}\label{eq:RR}
  r(D)-r(K-D)=\deg D+1-g,
\end{equation}
endows the graph with a Riemann--Roch theory in which the genus is the
cyclomatic number $g=\lvert E\rvert-\lvert V\rvert+1$ and the Picard group is a
finite \emph{Jacobian} of order equal to the number of spanning
trees~\cite{Biggs1999,Kirchhoff1847}. The right-hand side
$\MobIdx(D)=\deg D+1-g$ we call the \emph{mobility index}.

Both geometries describe how mass may be moved, and it is natural to ask how
they interact. This paper studies one concrete problem that connects them,
optimization over a linear equivalence class, and proves a number of
quantitative results about it. Several of these follow directly from known
theory and are marked as such (\cref{ssec:prior}); the main contributions are
summarized in \cref{ssec:core}.

\subsection{Main results}
Throughout, $G$ is a finite connected graph of genus $g$ with $n$ vertices,
and $D$ is an effective divisor of degree $d$ with normalized measure
$\mu_D=D/d$.

\begin{enumerate}[label=\textup{(\Roman*)},leftmargin=2.2em]
\item \emph{Gauge invariance} (\cref{thm:gauge}). The correction term of $F_q$
translates the dual coordinates by $2q-1$ and changes neither the metric
$x^{q-2}$, nor the connections, nor the divergence, which is the
$\beta$-divergence with $\beta=q$. On every transport polytope,
\[
  \textstyle\sum_{ij}F_q(P_{ij})=-\tfrac1q S_q(P)+\text{const},
\]
so $F_q$-regularized transport at strength $\gamma$ is Tsallis-regularized
transport at temperature $\gamma/q$.
\item \emph{Support and Pythagoras} (\cref{thm:support,thm:pyth}). Optimal
plans have full support for $q\le1$ and may be sparse only for $q>1$. For every
feasible plan, the excess cost equals $\gamma$ times its $\beta$-divergence
from the optimum plus a non-negative slack supported on the zero set of the
optimum.
\item \emph{Arithmetic versus differential} (\cref{prop:invisible}).
Chip-firing is a lattice translation in mixture coordinates. If the firing
lattice is extended to the real degree-zero subspace, linear equivalence
becomes trivial, so the Jacobian is not determined by this real-linearized
relation alone. It is encoded as a lattice on the mixture side and as torsion
on the dual side.
\item \emph{The Riemann--Roch transport defect}
(\cref{thm:defect,thm:chipmove,thm:AJ}). The defect $\delta_{\mathcal C}(D)$, the
worst-case cost of approaching an arbitrary target by chip-firing, satisfies
\[
  \frac1d\le\delta_{\TV}(D)\le\min\Bigl(1,\frac gd\Bigr)\quad(g\ge1,\ d\ge1),
\]
and vanishes identically exactly on trees. Moreover $d\,\delta_{W_1}(D)$ is
an eccentricity in a chip-move graph, bounded below by the diameter $\rho(G)$
of a Cayley graph of the Jacobian and by weighted Abel--Jacobi word lengths,
and above by $\diam(G)\,g$.
\item \emph{Topology--information crossover}
(\cref{thm:crossover,thm:phase}). The regularized price $\Gamma_{q,\gamma}$ of
lying in the wrong linear equivalence class tends to the topological defect
as $\gamma\to0$. It lies in the two-sided Riemann--Roch window
\[
  \frac{N_E}{d}-\frac\gamma q\bigl(\ln_q(n^2)-S_q(\mu_E)\bigr)
  \;\le\;\Gamma_{q,\gamma}\;\le\;\frac{\diam(G)\,g}{d}+\frac\gamma q\,H_q(\mu_E,\mu_E),
\]
and its high-temperature slope is a difference of maximal Tsallis coupling
entropies. A positive-gap regime, that is a target with $\Gamma>0$ for some
$\gamma$, exists for divisors of every degree $d\ge1$ if and only if $g\ge1$
(\cref{thm:phase}). The degree condition $d\ge g$ enters only the quantitative
form of this statement, through the Cayley diameter $\rho(G)$.
\end{enumerate}

\begin{enumerate}[label=\textup{(\Roman*)},leftmargin=2.2em,start=6]
\item \emph{Primal--dual structure} (\cref{sec:pd}). The gap splits as
$\Gamma=\mathsf T+\mathsf I$, with $\mathsf T\ge0$ topological and $\mathsf I\le0$
purely informational (\cref{thm:TI}). Dual potentials certify exactly the convex
relaxation over the convex hull of $\lin D$; the remainder is an integrality gap
(\cref{thm:minimax}). The Serre-dual rank refines the defect estimate
(\cref{thm:serre}). Sparsity is a dual-boundary phenomenon tied to the failure
of essential smoothness for $q>1$ (\cref{prop:legendre}). The support and
gap-sign transitions therefore have different mechanisms; that they occur at
different parameters is observed in exact computations, not proved. Low-temperature limits select
$q$-dependent points of an optimal face whose dimension is the cycle number of
its support (\cref{thm:face}).
\item \emph{Simulation at scale} (\cref{sec:sim}). On honeycomb and
Erd\H{o}s--Rényi networks with $10^2$--$10^5$ nodes we report numerical
observations corresponding to the theorems above (\cref{tab:synthesis}). The
Riemann--Roch bounds become informative only at particle densities of order
$g/n\approx\tfrac12$.
\end{enumerate}

Result (III) explains why the Hessian geometry generated by $F_q$, which does
not itself involve the lattice $L\Z^V$, carries no information on linear
equivalence; results (IV) and (V) show where topology reappears: in a
transport cost minimized over a lattice coset, and in its competition with
entropy. Along the way we show that a natural candidate inequality bounding
$W_2^2(\mu_D,\mu_K)$ from below by $\MobIdx(D)/\deg D$ fails
(\cref{prop:naive}). We also record several
consequences for the experiments of~\cite{JSAI2026}
(\cref{rem:jsai,rem:sparse}).

\subsection{Core contributions}\label{ssec:core}
The paper centres on three problems that arise from optimizing over a linear
equivalence class.
\begin{enumerate}[label=\textup{(\Alph*)}]
\item \emph{The Riemann--Roch transport defect.} The definition of
$\delta_{\mathcal C}$, the two-sided bounds through $r(D)$, $r(K-D)$ and the
vertex multiplicities $m_v(D)$, and the equality criterion
(\cref{thm:defect,thm:serre,thm:vertexrank}).
\item \emph{The Jacobian and transport distance.} The chip-move metric, and the
Cayley-diameter and Abel--Jacobi bounds for the $W_1$-defect
(\cref{thm:chipmove,thm:AJ}).
\item \emph{The topology--information crossover.} The limits and window of the
regularized gap, the existence of a positive-gap regime iff $g\ge1$ (for
every $d\ge1$), and the
decomposition $\Gamma=\mathsf T+\mathsf I$ together with its integrality
interpretation (\cref{thm:crossover,thm:phase,thm:TI,thm:minimax}).
\end{enumerate}
The remaining results are supporting. The gauge invariance of $F_q$ and the
support theorem apply standard convex analysis; the optimal-face results apply
standard linear programming; the network experiments are observational. Within
the core, several statements are short consequences of known theory once the
right quantity is defined, for instance the upper bound in (A) and the
existence statement in (C), and we say so where they occur. The quantitative
questions that would go beyond this, notably the finite-temperature behaviour
of $\mathsf T$ and a characterization of equality in (A), are listed in
\cref{sec:conclusion}.

\Cref{tab:status} records the status of every main statement. We use four
categories: \textbf{P} (proved here; a new formulation), \textbf{K} (proved
here, but a short consequence of known theory, applied to the quantities of this
paper), \textbf{O} (numerical observation, on finitely many graphs or networks)
and \textbf{U} (open).

\begin{table}[!htbp]
\centering\small
\renewcommand{\arraystretch}{1.12}
\begin{tabular}{>{\raggedright\arraybackslash}p{6.4cm}c>{\raggedright\arraybackslash}p{6.4cm}}
\toprule
Statement & Status & Remark \\
\midrule
Gauge invariance (\cref{thm:gauge}) & \textbf{K} & affine invariance of Hessian geometry \\
Support of optimal plans (\cref{thm:support,prop:legendre}) & \textbf{K} & boundary behaviour of $F_q'$ \\
Pythagorean relation (\cref{thm:pyth}) & \textbf{K} & Bregman projection with KKT slack \\
Real linearization (\cref{prop:invisible}) & \textbf{K} & rank of $L$; Pontryagin duality \\
Chip-move metric (\cref{thm:chipmove}) & \textbf{K} & total unimodularity \\
Failure of the candidate inequality (\cref{prop:naive}) & \textbf{P} & explicit counterexamples \\
Riemann--Roch transport estimate (\cref{thm:defect}) & \textbf{P} & upper bound short from rank; lower bound uses $\Jac$ \\
Abel--Jacobi bounds (\cref{thm:AJ}) & \textbf{P} & Cayley lower bound needs $d\ge g$ \\
Crossover limits and window (\cref{thm:crossover}) & \textbf{P} & standard regularization estimates, applied to $\Gamma$ \\
Positive-gap regime iff $g\ge1$ (\cref{thm:phase}) & \textbf{P} & short; every $d\ge1$; $\rho$-form needs $d\ge g$ \\
Serre-dual form (\cref{thm:serre}) & \textbf{P} & short from rank and Riemann--Roch \\
Vertex multiplicities, equality criterion (\cref{thm:vertexrank}) & \textbf{P} & criterion sufficient, not necessary \\
When the Serre form is sharper (\cref{prop:serresharp}) & \textbf{P} & short; exact condition $r(D)\ge1$, $r(K-D)\ge0$ \\
Equality in genus $\le1$; cycles (\cref{prop:smallgenus}) & \textbf{P} & complete for $g\le1$, partial for $g=2$ \\
Dual problem and minimax steps (\cref{lem:dual,thm:minimax}) & \textbf{K} & Fenchel--Young, Rockafellar, Sion \\
Decomposition $\Gamma=\mathsf T+\mathsf I$ (\cref{thm:TI}) & \textbf{P} & definitional split; limits proved \\
Minimax and three levels (\cref{thm:minimax,cor:threelevels}) & \textbf{P} & Sion's theorem; gaps defined here \\
Optimal faces and selection (\cref{thm:face}) & \textbf{K} & LP facts; classical for $q=1$ \\
Equality in the rank bound $(d-r(D))/d$, $1769$ of $1784$ classes (\cref{ssec:pdnum}) & \textbf{O} & exhaustive, nine graphs \\
Distinct gap-sign and support transitions (\cref{ssec:pdnum}) & \textbf{O} & three graphs \\
Crossover temperatures, $\gamma^\ast(2)=16$ (\cref{tab:phase}) & \textbf{O} & one graph, numerical \\
Scaling laws, regimes at scale (\cref{sec:sim}) & \textbf{O} & empirical fits \\
Characterization of equality; finite-temperature $\mathsf T$; closed form for $\gamma^\ast$ & \textbf{U} & \cref{sec:conclusion} \\
\bottomrule
\end{tabular}
\caption{Status of the main statements: \textbf{P} proved, new formulation;
\textbf{K} proved, short consequence of known theory; \textbf{O} numerical
observation; \textbf{U} open.}
\label{tab:status}
\end{table}

\subsection{Relation to previous work and what is new}\label{ssec:prior}
\Cref{tab:prior} separates what this paper takes from the literature from what
it contributes. ``New'' means new to the best of our knowledge; we have not
carried out an exhaustive literature search for every item.

\begin{table}[!htbp]
\centering\small
\renewcommand{\arraystretch}{1.15}
\begin{tabular}{>{\raggedright\arraybackslash}p{3.0cm}>{\raggedright\arraybackslash}p{5.3cm}>{\raggedright\arraybackslash}p{5.6cm}}
\toprule
Topic & Known (used here) & Contribution of this paper \\
\midrule
Dual geometry of $F_q$ &
Affine terms leave Hessian metrics and Bregman divergences unchanged
\cite{AmariNagaoka2000}; Tsallis-regularized transport \cite{Muzellec2017};
sparse transport for $q=2$ \cite{Blondel2018}; Bregman projections
\cite{Dessein2018}. &
Explicit identification of the $F_q$ problem of \cite{KYA2022,JSAI2026} with
Tsallis transport at temperature $\gamma/q$ (\cref{thm:gauge}); its
consequences for \cite{JSAI2026} (\cref{rem:jsai,rem:sparse}). Not claimed as
deep. \\
Chip-firing and transport &
Chip-firing, sandpile group and Jacobian
\cite{BLS1991,Dhar1990,Biggs1999,BakerNorine2007}; structure of the sandpile group
and reduced divisors as energy minimizers
\cite{Lorenzini2008,BakerShokrieh2013,PPW2013}; min-cost flow, total
unimodularity and integrality gaps \cite{AhujaMagnantiOrlin1993,Schrijver2003}. &
Chip-move metric (\cref{thm:chipmove}); triviality of the real-linearized
relation, in transport terms (\cref{prop:invisible}); Cayley-diameter and
Abel--Jacobi bounds (\cref{thm:AJ}). We are not aware of these bounds in the
chip-firing literature. Their relation to the invariant-factor structure of the
sandpile group \cite{Lorenzini2008} and to the energy pairing of
\cite{BakerShokrieh2013} remains to be clarified. \\
Riemann--Roch and rank &
Rank, reduced divisors, $r(D)\ge\deg D-g$, Serre duality
\cite{BakerNorine2007,CorryPerkinson2018}. &
Transport defect and its bounds (\cref{thm:defect,thm:serre,thm:vertexrank}),
Abel--Jacobi bounds (\cref{thm:AJ}), failure of the candidate
inequality (\cref{prop:naive}). \\
Topological gap &
--- &
Definition of $\Gamma_{q,\gamma}$, its limits and window
(\cref{thm:crossover}), existence iff $g\ge1$ (\cref{thm:phase}), the
decomposition $\mathsf T+\mathsf I$ and the integrality interpretation
(\cref{thm:TI,thm:minimax}). \\
Degeneracy and supports &
Legendre-type functions \cite{Rockafellar1970}; strict complementarity
\cite{GoldmanTucker1956}; faces of transportation polytopes. &
Application to $q$-selection of optimal faces (\cref{thm:face}) and the
dissociation of support and gap-sign transitions (\cref{ssec:pdnum}). \\
Network experiments &
Particle transport model of \cite{JSAI2026}; random graphs \cite{ER1960}. &
Controlled comparison at matched mean degree and matched genus; observational,
not a theorem (\cref{sec:sim}). \\
\bottomrule
\end{tabular}
\caption{What is used from the literature and what is contributed here.}
\label{tab:prior}
\end{table}

We make the differences concrete, field by field. Before doing so we record
what our literature check covered. We searched for combinations of
``chip-firing'', ``divisor'', ``linear equivalence'' or ``graph Riemann--Roch''
with ``optimal transport'', ``Wasserstein'' or ``earth mover's distance''.
We found no quantity equivalent to $\delta_{\mathcal C}$, no transport-cost bound
in terms of $r(D)$ or $r(K-D)$, and no treatment of the convex relaxation of a
linear system as an integrality gap of regularized transport. Work on the
complexity and approximation of the rank, and on rotor-routing, concerns
$r(D)$ itself and not transport costs. This was a keyword search, not an
exhaustive one, and the claims of novelty below are subject to it.

\paragraph{Chip-firing and reduced divisors.} From
\cite{BakerNorine2007,CorryPerkinson2018} we use the rank, reduced divisors
(computed by Dhar's burning algorithm \cite{Dhar1990}), the Riemann--Roch
formula and the estimate $r(D)\ge\deg D-g$. These works study whether a
configuration can be moved to a target; they do not attach a cost to the
remaining discrepancy. Our new ingredient is the quantity
$\delta_C(D)$, the worst-case residual transport cost after optimal firing,
and the observation that Riemann--Roch and Serre duality bound it
(\cref{thm:defect,thm:serre}). The lower bound of \cref{thm:vertexrank} uses
only the standard fact that $D_v$ maximizes the coefficient at $v$ over $|D|$
(\cref{lem:reducedmax}).

\paragraph{Jacobian and sandpile group.} From
\cite{BLS1991,Biggs1999,Lorenzini2008,BakerShokrieh2013} we use that $\Jac(G)$
is finite of order $\kappa(G)$ and generated by the classes $[v-w]$. The
energy pairing of \cite{BakerShokrieh2013} measures reduced divisors in the
Laplacian quadratic form; our bounds in \cref{thm:AJ} measure classes instead in
a word metric (edge or Abel--Jacobi generators) and relate it to $W_1$. We do
not know whether the two can be compared in general.

\paragraph{Integer min-cost flow.} Total unimodularity of network matrices and
the resulting integrality of min-cost flows \cite{AhujaMagnantiOrlin1993,Schrijver2003}
give \cref{thm:chipmove} directly. What is new is only the reading of this fact in
chip-firing terms: the number of chip moves equals $d\,W_1$, and the integrality gap of
\cref{thm:minimax} is the part of the transport defect that real potentials
cannot certify.

\paragraph{Regularized optimal transport.} From
\cite{Muzellec2017,Blondel2018,Dessein2018} we use the Tsallis- and
$\beta$-divergence-regularized problems, the sparsity of the quadratic case and
Bregman projection. \Cref{thm:gauge} only identifies the problem of
\cite{KYA2022,JSAI2026} with this family; it is not deep. The new elements are
the combination with linear systems, that is the gap $\Gamma_{q,\gamma}$ and its
limits (\cref{thm:crossover,thm:phase,thm:TI}), and the observation that support
transitions and gap-sign transitions have different causes
(\cref{ssec:pdnum}). \Cref{thm:face} is an application of the Goldman--Tucker
theorem; only its parts (iii)--(v), on the selection by $q\ne1$, go beyond what
is standard for $q=1$.

\subsection{Organization}
\Cref{sec:divisors} recalls divisor theory and the Baker--Norine theorem.
\Cref{sec:seven} introduces the seven-vertex family that serves as a test bed
throughout. \Cref{sec:dual,sec:gauge,sec:support} develop the dual geometry of
$F_q$. \Cref{sec:chip} relates chip-firing to transport and proves the
invisibility of the Jacobian. \Cref{sec:defect} establishes the transport
defect and its Abel--Jacobi bounds, and \cref{sec:crossover} couples the two
geometries. \Cref{sec:pd} analyses the primal--dual structure of the gap.
\Cref{sec:sim} compares honeycomb and Erd\H{o}s--Rényi networks by simulation
and interprets the results theorem by theorem. \Cref{sec:conclusion} lists
open problems.
\Cref{app:exercises,app:computation} contain worked examples and the
computational details, including how every numerical result can be
reproduced from the supplementary archive.

\section{Divisors, rank and the Riemann--Roch theorem}\label{sec:divisors}

Let $G=(V,E)$ be a finite connected multigraph without loops, with
$n=\lvert V\rvert$ vertices, graph distance $\dist_G$ (unit edge lengths) and
diameter $\diam(G)$.

\begin{definition}
A \emph{divisor} is an element $D=\sum_vD(v)\,v$ of $\Div(G)=\Z^V$, of degree
$\deg D=\sum_vD(v)$; it is \emph{effective}, $D\ge0$, if $D(v)\ge0$ for all
$v$, and $\Div^d_+(G)$ denotes the effective divisors of degree $d$. The
\emph{Laplacian} is $(Lf)(v)=\sum_{u\sim v}(f(v)-f(u))$, with edges counted
with multiplicity. \emph{Firing} $v$ replaces $D$ by $D-L\one_v$, and
$D\sim D'$ (\emph{linear equivalence}) if $D-D'\in L(\Z^V)$. The \emph{linear
system} of $D$ is $\lin D=\{D'\ge0:D'\sim D\}$. The \emph{rank} $r(D)$ is $-1$
if $\lin D=\varnothing$, and otherwise the largest $k$ such that
$\lin{D-E}\neq\varnothing$ for all $E\in\Div^k_+(G)$. The \emph{canonical
divisor} is $K=\sum_v(\deg_G(v)-2)\,v$, and the \emph{mobility index} is
$\MobIdx(D)=\deg D+1-g$.
\end{definition}

Since the columns of $L$ sum to zero, firing preserves degree. The
\emph{Picard group} $\Pic(G)=\Div(G)/L(\Z^V)$ is therefore graded by degree. Its degree-zero part is the
\emph{Jacobian} $\Jac(G)$. In game-theoretic terms, $r(D)\ge k$ means that an
adversary may remove any $k$ chips and the configuration can still be brought
out of debt by firing. The rank measures a reconfiguration capacity, not a
count of transport operations. The handshake identity gives
$\deg K=2\lvert E\rvert-2\lvert V\rvert=2g-2$.

\begin{theoremBN}[\cite{BakerNorine2007}]
For every divisor $D$ on $G$, $\;r(D)-r(K-D)=\MobIdx(D)$.
\end{theoremBN}

\begin{corollary}\label{cor:RR}
Let $\deg D=d$. Then \textup{(i)} $r(D)\ge d-g$; \textup{(ii)} if $d>2g-2$,
then $r(K-D)=-1$ and $r(D)=d-g$; \textup{(iii)} $r(K)=g-1$.
\end{corollary}

\begin{proof}
(i) holds since $r(K-D)\ge-1$. For (ii), $\deg(K-D)<0$, and since equivalence
preserves degree, $\lin{K-D}=\varnothing$. For (iii), take $D=0$.
\end{proof}

Part (ii) is a \emph{configuration independence}: in the non-special range
$d\ge2g-1$ the rank depends on the degree and the genus only.

\begin{proposition}[Kirchhoff, Biggs, Baker--Norine]\label{prop:jac}
$\Jac(G)$ is finite of order $\kappa(G)$, the number of spanning trees. Hence
each degree carries exactly $\kappa(G)$ linear equivalence classes.
Moreover $\kappa(G)=1$ if $g=0$, and $\kappa(G)\ge2$ if $g\ge1$.
\end{proposition}

\begin{proof}
The first statement is the matrix-tree theorem $\kappa(G)=\det L_q$ in the form
of~\cite[Sect.~1]{Biggs1999}. A tree is its own unique spanning tree. If
$g\ge1$, take a spanning tree $T$ and an edge $e\notin T$; the unique cycle of
$T+e$ contains an edge $f\neq e$, and $T+e-f$ is a second spanning tree.
\end{proof}

\begin{corollary}[Trees]\label{cor:trees}
On a tree, divisors of equal degree are linearly equivalent, and $r(D)=\deg D$
whenever $\deg D\ge-1$.
\end{corollary}

\begin{theorem}[Reduced divisors {\cite[Prop.~3.1]{BakerNorine2007}}]\label{thm:dhar}
Fix $v\in V$. A divisor $D'$ is \emph{$v$-reduced} if $D'(u)\ge0$ for $u\ne v$ and,
for every non-empty $S\subseteq V\setminus\{v\}$, firing $S$ makes some
vertex of $S$ negative. Every divisor $D$ is linearly equivalent to a unique
$v$-reduced divisor $D_v$, and $\lin D\ne\varnothing$ if and only if $D_v(v)\ge0$.
\end{theorem}

Reduced divisors and Dhar's burning algorithm~\cite{Dhar1990} give the
standard procedure for computing ranks; see
\cite[Prop.~3.1]{BakerNorine2007} and \cite{CorryPerkinson2018,Klivans2018}.
For the computations below we use the exact integer class invariant described
in \cref{app:computation}.

\section{A seven-vertex test bed}\label{sec:seven}

We label vertices $v_1,\dots,v_7$ and consider the cycle $C_7$; the complete
binary tree $T_7$ with root $v_1$; $G'=C_7+v_1v_4$; $G''=C_7+v_1v_4+v_1v_5$; the
star $S_7$ with centre $v_1$; the path $P_7$; and the complete graph $K_7$.

\begin{figure}[!htbp]
\centering
\begin{tikzpicture}[scale=0.88]
  \begin{scope}
    \foreach \i in {1,...,7}{\coordinate (A\i) at ({90+(\i-1)*360/7}:1.55cm);}
    \foreach \i in {1,...,6}{\pgfmathtruncatemacro{\j}{\i+1}\draw[gedge] (A\i)--(A\j);}
    \draw[gedge] (A7)--(A1);
    \node[vhub] at (A1) {$v_1$};
    \foreach \i in {2,...,7}{\node[vertex] at (A\i) {$v_{\i}$};}
    \node[anno] at (0,-2.25) {(a) $C_7$:\; $g=1$,\; $K=0$};
  \end{scope}
  \begin{scope}[xshift=5.4cm,yshift=1.0cm]
    \node[vertex] (r) at (0,0) {$v_1$};
    \node[vhub] (a) at (-1.25,-1.25) {$v_2$};
    \node[vhub] (b) at (1.25,-1.25) {$v_3$};
    \node[vleaf] (c) at (-1.9,-2.5) {$v_4$};
    \node[vleaf] (d) at (-0.6,-2.5) {$v_5$};
    \node[vleaf] (e) at (0.6,-2.5) {$v_6$};
    \node[vleaf] (f) at (1.9,-2.5) {$v_7$};
    \draw[gedge] (r)--(a) (r)--(b) (a)--(c) (a)--(d) (b)--(e) (b)--(f);
    \node[anno] at (0,-3.25) {(b) $T_7$:\; $g=0$,\; $\deg K=-2$};
  \end{scope}
  \begin{scope}[xshift=10.8cm]
    \foreach \i in {1,...,7}{\coordinate (B\i) at ({90+(\i-1)*360/7}:1.55cm);}
    \foreach \i in {1,...,6}{\pgfmathtruncatemacro{\j}{\i+1}\draw[gedge] (B\i)--(B\j);}
    \draw[gedge] (B7)--(B1);
    \draw[chord] (B1)--(B4);
    \node[vhub] at (B1) {$v_1$};
    \node[vhub] at (B4) {$v_4$};
    \foreach \i in {2,3,5,6,7}{\node[vertex] at (B\i) {$v_{\i}$};}
    \node[anno] at (0,-2.25) {(c) $G'$:\; $g=2$,\; $K=v_1+v_4$};
  \end{scope}
\end{tikzpicture}
\caption{Three members of the test family. Blue vertices have $K(v)>0$ (hubs),
red vertices $K(v)<0$ (terminals); the dashed edge is the chord $v_1v_4$.}
\label{fig:graphs}
\end{figure}
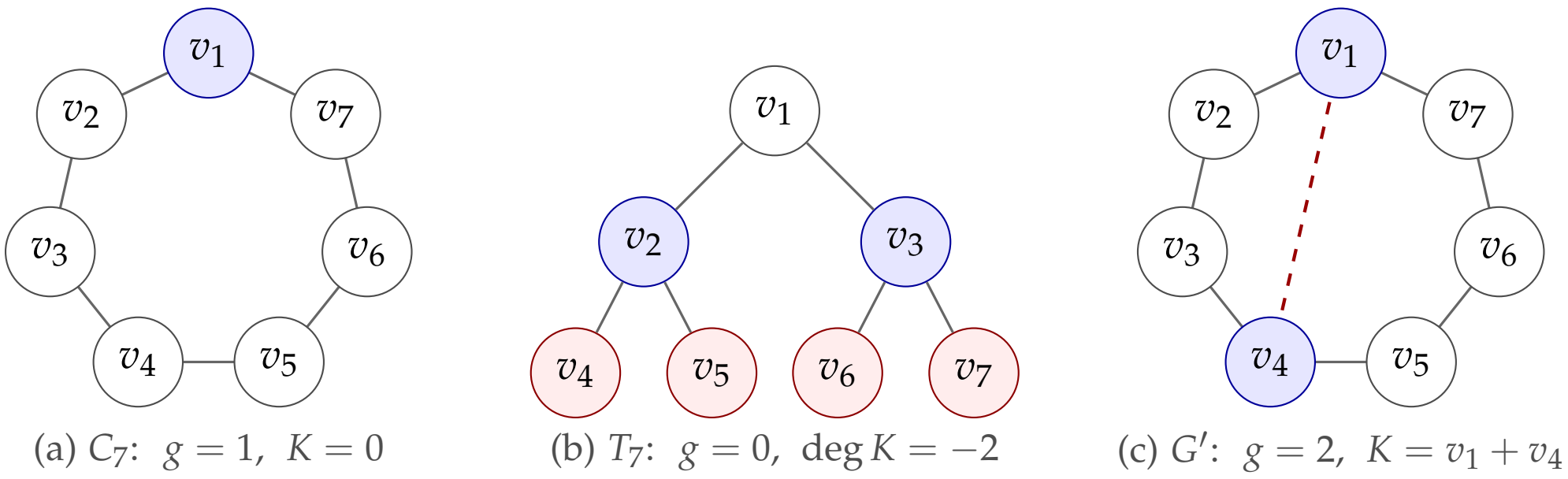

\begin{table}[!htbp]
\centering\small
\begin{tabular}{lccccl}
\toprule
Graph & $\lvert E\rvert$ & $g$ & $\kappa(G)$ & $\diam$ & Canonical divisor $K$ \\
\midrule
$T_7$  & 6  & 0  & 1      & 4 & $v_2+v_3-v_4-v_5-v_6-v_7$ \\
$S_7$  & 6  & 0  & 1      & 2 & $4v_1-(v_2+\dots+v_7)$ \\
$P_7$  & 6  & 0  & 1      & 6 & $-v_1-v_7$ \\
$C_7$  & 7  & 1  & 7      & 3 & $0$ \\
$G'$   & 8  & 2  & 19     & 3 & $v_1+v_4$ \\
$G''$  & 9  & 3  & 40     & 3 & $2v_1+v_4+v_5$ \\
$K_7$  & 21 & 15 & 16807  & 1 & $4(v_1+\dots+v_7)$ \\
\bottomrule
\end{tabular}
\caption{Invariants of the seven-vertex family; in each row $\deg K=2g-2$.}
\label{tab:family}
\end{table}

\begin{example}[Rank and genus]\label{ex:rank}
On $C_7$ we have $K=0$; for $D=3v_1$, \cref{cor:RR}(ii) gives $r(3v_1)=2$. On
$T_7$, \cref{cor:trees} gives $r(3v_1)=3$. On $G'$, $K=v_1+v_4$ and
$r(4v_1)=\MobIdx(4v_1)-1=2$, while $r(K)=g-1=1$. On $C_7$ the divisors $4v_1$ and
$v_1+v_2+v_3+v_4$ both have rank $3$, yet they are not linearly equivalent:
in $\Jac(C_7)\cong\Z/7\Z$ the class of a degree-$d$ divisor is the sum of the
indices of its chips modulo $7$, which is $4$ and $3$ respectively. At fixed
degree four, each independent cycle lowers the rank by one: $r(4v_1)=4,3,2$ on
$T_7,C_7,G'$. All of these values were confirmed independently of the theorem
(\cref{app:computation}).
\end{example}

\begin{remark}[The canonical divisor as curvature]\label{rem:curv}
With the combinatorial curvature $\kappa_v=1-\tfrac12\deg_G(v)$ one has
$K(v)=-2\kappa_v$, and $\deg K=2g-2$ becomes the discrete Gauss--Bonnet formula
$\sum_v\kappa_v=1-g=\chi(G)$, in analogy with $\deg K_X=-\chi(X)$ for a Riemann
surface. Vertices with $K(v)<0$ (leaves) are \emph{terminals}, those with
$K(v)=0$ are \emph{transit} vertices, and those with $K(v)>0$ are \emph{hubs}.
This combinatorial curvature should be distinguished from transport curvatures
such as Ollivier's~\cite{Ollivier2009}.
\end{remark}

\section{The generalized entropy and its Legendre dual}\label{sec:dual}

Fix $q>0$ and put $f_q(x)=\frac{x^q+q(1-x)-1}{q(q-1)}$ and
$F_q=f_q-(2q-1)(1-x)$ on $x\ge0$, with $F_1(x)=x\log x$. We write
$\ln_s x=\frac{x^{1-s}-1}{1-s}$ and $\exp_s y=[1+(1-s)y]_+^{1/(1-s)}$.

\begin{proposition}[Dual structure]\label{prop:dual}
\leavevmode
\begin{enumerate}[label=\textup{(\roman*)}]
\item $F_q'(x)=\ln_{2-q}x+(2q-1)$ and $F_q''(x)=x^{q-2}$.
\item For $q\neq1$ the Legendre dual is
\[
  F_q^*(y)=\frac{\bigl[1+(q-1)(y-2q+1)\bigr]_+^{q/(q-1)}-1}{q}+2q-1,
\]
on the domain where the bracket is positive if $q<1$, and on $\R$ if $q\ge1$;
moreover $F_1^*(y)=e^{y-1}$ and $(F_q^*)'(y)=\exp_{2-q}(y-2q+1)$.
\item The Hessian metric is $g(\eta)=\eta^{q-2}$ in the mixture coordinate
$\eta=x$, and $g^*=\eta^{2-q}$ in the dual coordinate $\theta=F_q'(\eta)$.
\item The Bregman divergence is the $\beta$-divergence~\cite{Basu1998} with
$\beta=q$:
\[
  \Dq(P\,\|\,Q)=\sum_{i}\frac{P_i^{\,q}+(q-1)Q_i^{\,q}-qP_iQ_i^{\,q-1}}{q(q-1)}
  \quad(q\neq1),
\]
and $\mathrm D_1(P\,\|\,Q)=\sum_i\bigl(P_i\log\frac{P_i}{Q_i}-P_i+Q_i\bigr)$.
\end{enumerate}
\end{proposition}

\begin{proof}
(i) is direct. For (ii), $F_q^*(y)=f_q^*(y-2q+1)+2q-1$ since $F_q-f_q$ is
affine, and $f_q^*(s)=\int_0^s[1+(q-1)t]^{1/(q-1)}dt$ because $f_q^*(0)=-f_q(1)=0$;
for $q>1$ and $1+(q-1)s\le0$ the supremum is attained at $x=0$ with value
$-f_q(0)=-1/q$, which the positive part reproduces. (iii) uses
$(F_q^*)''=1/(F_q''\circ(F_q')^{-1})$. (iv) The affine part cancels in
$F_q(P)-F_q(Q)-F_q'(Q)(P-Q)$.
\end{proof}

The pair $(\eta,\theta)$ consists of the two affine coordinates of a dually
flat structure on the \emph{open} orthant $\R^N_{>0}$~\cite{AmariNagaoka2000},
for every $q>0$. The boundary $\{x_i=0\}$ is not part of this manifold, and
there the parameter matters. For $0<q\le1$ the dual coordinate $F_q'(x)$ tends to $-\infty$ as
$x\downarrow0$, so the boundary is pushed to infinity in the $\theta$-chart. This
does \emph{not} mean that the boundary is at infinite Riemannian distance: the
length of $[0,x]$ in the metric $t^{q-2}$ is
$\int_0^x t^{q/2-1}\,dt=\tfrac2q\,x^{q/2}<\infty$ for every $q>0$. The metric
coefficient itself blows up at $0$ for every $q<2$, so it is the derivative
$F_q'$, not the metric, that separates $q\le1$ from $q>1$. For $q>1$, $F_q'$ extends continuously to $x=0$ with a finite
value, $F_q$ is not essentially smooth there, and the closed cone is not a dually
flat manifold in the same sense. Statements about plans with zero entries
(\cref{sec:support,sec:pd}) are therefore statements of convex analysis on the
closed cone, not of the Hessian geometry on the open orthant. Sets that are affine in
$\eta$ are \emph{m-flat}, and sets that are affine in $\theta$ are
\emph{e-flat}. The exponent $q/(q-1)$ in the Legendre formula is the algebraic
conjugate exponent of $q$. It is the H\"older conjugate of $q$ only for $q>1$;
for $0<q<1$ it is negative. The case $q=2$ is self-dual and Euclidean.

\section{Gauge invariance}\label{sec:gauge}

For probability vectors $a\in\R^m$, $b\in\R^n$, a cost $C$ and $\gamma>0$, set
\[
  J_{q,\gamma}(P)=\langle C,P\rangle+\gamma\sum_{i,j}F_q(P_{ij}),\qquad
  \Wq(a,b)=\min_{P\in\Pi(a,b)}J_{q,\gamma}(P),
\]
and let $\mathcal T_{q,\varepsilon}(a,b)=\min_{\Pi(a,b)}\bigl[\langle C,P\rangle-\varepsilon S_q(P)\bigr]$
be Tsallis-regularized transport~\cite{Tsallis1988,Muzellec2017}, where
$S_q(P)=\frac{1-\sum P_{ij}^{\,q}}{q-1}$. Let $N=mn$, and let
$F^{\rm base}_q(x)=\frac{x^q-x}{q-1}$ be the power-law baseline
of~\cite{JSAI2026}.

\emph{Conventions, used throughout the paper.}
\begin{itemize}
\item $q>0$ always. For $q=1$ every $q$-dependent quantity is its continuous
limit: $F_1(x)=x\log x$, $\ln_1=\log$, and $S_1(\pi)=-\sum\pi_i\log\pi_i$. The
formulas for $q\neq1$ converge to these as $q\to1$.
\item $S_q$ is defined for any probability vector on finitely many cells, with
zero entries contributing $0$. On $N$ cells it satisfies
$0\le S_q\le\ln_q N$, with equality on the right exactly for the uniform vector,
for every $q>0$.
\item For a regularization strength $\gamma$ we always write $\varepsilon=\gamma/q$
for the corresponding Tsallis temperature, and
$H_q(\mu,\nu)=\max_{\pi\in\Pi(\mu,\nu)}S_q(\pi)$.
\item All inequalities stated below for $S_q$, $\ln_q$ and $H_q$ hold for every
$q>0$, including $q=1$; no separate case distinction is needed.
\end{itemize}

\begin{theorem}[Gauge invariance]\label{thm:gauge}
\leavevmode
\begin{enumerate}[label=\textup{(\roman*)}]
\item The correction term $-(2q-1)(1-x)$ changes neither the metric, nor the
e- and m-connections, nor $\Dq$; it acts as the translation
$\theta\mapsto\theta+(2q-1)$ of the dual coordinate.
\item For $P\in\Pi(a,b)$,
\[
  \sum_{i,j}F_q(P_{ij})=-\frac{S_q(P)}{q}+c_{q,N},\qquad
  \sum_{i,j}F^{\rm base}_q(P_{ij})=-S_q(P),
\]
where $c_{q,N}=(N-1)\bigl(\frac1q-(2q-1)\bigr)$.
\item Hence $\Wq=\mathcal T_{q,\gamma/q}+\gamma c_{q,N}$. The $F_q$ problem at
strength $\gamma$, the Tsallis problem at temperature $\gamma/q$ and the
power-law problem at strength $\gamma/q$ share the same unique optimal plan.
\end{enumerate}
\end{theorem}

\begin{proof}
(i) Affine functions have vanishing second and third derivatives and cancel
in Bregman divergences. (ii) Using $\sum P_{ij}=1$, we get
$\sum f_q(P_{ij})=\frac{\sum P^q_{ij}-1+(q-1)(N-1)}{q(q-1)}=\frac{N-1-S_q(P)}q$
and $\sum(2q-1)(1-P_{ij})=(2q-1)(N-1)$; the baseline is analogous. (iii)
follows.
\end{proof}

\begin{remark}[Consequences for~\cite{JSAI2026}]\label{rem:jsai}
(a) At $q=1$ both regularizers are $x\log x$, so with an exact solver the two
tables of total distances in~\cite{JSAI2026} must agree at $q=1$; the reported
values $51$ and $57$ point to algorithmic effects. One plausible source is
that a gradient step shifts every entry by the constant $-\eta\gamma(2q-1)$,
which diagonal scaling does not remove. (b) The $F_q$ experiment at
$(q,\gamma)$ is the power-law experiment at $(q,\gamma/q)$, so the improvement
observed for $q>1$ is consistent with a falling effective temperature
$\gamma/q$. (c) In \emph{unbalanced} transport~\cite{Chizat2018,LMS2018} the
total mass is free, and the correction term becomes a genuine chemical
potential $\gamma(2q-1)\sum P_{ij}$.
\end{remark}

Numerically, on random $6\times6$ instances with $q=0.5$, $0.8$, $1$, $1.2$, $2$,
$3$ and $5$, the three optimal plans of \cref{thm:gauge}(iii) agree to within
$10^{-7}$ and the value identity holds to within $2\times10^{-8}$.

\section{Support and the Pythagorean relation}\label{sec:support}

\begin{theorem}[Sparsity only for $q>1$]\label{thm:support}
Let $a,b>0$ and let $P^\star$ minimize $J_{q,\gamma}$ on $\Pi(a,b)$.
\begin{enumerate}[label=\textup{(\roman*)}]
\item If $q\le1$, then $P^\star_{ij}>0$ for all $i,j$.
\item If $q>1$, there are potentials $\alpha,\beta$ with
\[
  P^\star_{ij}=\Bigl[1+(q-1)\Bigl(\tfrac{\alpha_i+\beta_j-C_{ij}}{\gamma}-(2q-1)\Bigr)\Bigr]_+^{1/(q-1)},
\]
and $P^\star_{ij}=0$ exactly when
$\alpha_i+\beta_j-C_{ij}\le\gamma\bigl(2q-1-\frac1{q-1}\bigr)$.
\end{enumerate}
\end{theorem}

\begin{proof}
(i) Here $F_q'(0^+)=-\infty$. If $P^\star_{ij}=0$, choose $l\neq j$ and $k\neq i$
with $P^\star_{il},P^\star_{kj}>0$. Along the feasible direction
$e_{ij}+e_{kl}-e_{il}-e_{kj}$ the one-sided derivative of $J_{q,\gamma}$ is $-\infty$,
because the entries $(i,l)$ and $(k,j)$ contribute finite terms. This
contradicts optimality. (ii) For $q>1$, $F_q$ is $C^1$ on $[0,\infty)$ with
$F_q'(0)=2q-1-\frac1{q-1}$. The KKT conditions
$C_{ij}+\gamma F_q'(P^\star_{ij})=\alpha_i+\beta_j+s_{ij}$, with $s\ge0$ and
$sP^\star=0$, invert through \cref{prop:dual}(ii).
\end{proof}

\begin{remark}\label{rem:sparse}
The derivative blow-up $F_q'(0^+)=-\infty$ for $0<q\le1$ is therefore a
\emph{boundary barrier} that keeps plans dense (although, as noted in
\cref{sec:dual}, the Riemannian distance to the boundary is finite), whereas
$q>1$ permits exact zeros, as
in quadratic regularization at $q=2$~\cite{Blondel2018}. On a random
$6\times6$ instance, the number of zero entries (out of $36$) was
$0,0,0,11,24,25,25$ for $q=0.5,0.8,1,1.2,2,3,5$. This reverses the reading
in~\cite[Sect.~3.3, App.~A.2]{JSAI2026}.
\end{remark}

For fixed $(C,\gamma)$ the optimal plans form the e-flat family
$\{F_q'(P_{ij})=(\alpha_i+\beta_j-C_{ij})/\gamma\}$, which meets the m-flat
polytope $\Pi(a,b)$ at $P^\star$. The orthogonality of these two foliations
takes the following exact form (cf.~\cite{Dessein2018} for Bregman projections
in regularized transport). The divergence $\Dq(P\,\|\,P^\star)$ is well defined
for every $P\in\Pi(a,b)$, including plans with zero entries. For $0<q\le1$ this
uses $P^\star>0$ (\cref{thm:support}(i)), so $F_q'(P^\star_{ij})$ is finite. For
$q>1$, $F_q$ is $C^1$ on $[0,\infty)$ and $F_q'(0)$ is finite, so zero entries of
either argument are allowed. In both cases $F_q(0)$ is finite.

\begin{theorem}[Pythagorean relation]\label{thm:pyth}
For every $P\in\Pi(a,b)$,
\[
  J_{q,\gamma}(P)-J_{q,\gamma}(P^\star)=\gamma\,\Dq(P\,\|\,P^\star)+\sum_{i,j}s_{ij}P_{ij},
\]
where the KKT slack $s\ge0$ is supported on $\{P^\star=0\}$. For $q\le1$ the
slack vanishes.
\end{theorem}

\begin{proof}
By definition of $\Dq$, the left-hand side equals
$\gamma\Dq(P\|P^\star)+\sum(C_{ij}+\gamma F_q'(P^\star_{ij}))(P_{ij}-P^\star_{ij})$.
Insert the KKT relation: the potentials drop out because $P$ and $P^\star$ have
the same marginals, and $\sum s_{ij}P^\star_{ij}=0$.
\end{proof}

Numerically the identity holds to within $10^{-5}$ for $q=\tfrac12$ and $q=1$;
for $q=2$ the slack is strictly positive (on a $5\times5$ instance with $16$
zeros, the excess cost $0.433$ compares with $\gamma\Dq=0.018$).

\section{Chip-firing as transport, and the invisibility of the Jacobian}\label{sec:chip}

For probability measures $\mu,\nu$ on $V$ let $W_p^p$ be the Wasserstein cost
(the $p$-th power of the $p$-Wasserstein distance; throughout the paper all
statements are made for $W_p^p$, never for $W_p$ itself, and $W_1=W_1^1$)
for $\dist_G^p$ and $\TV(\mu,\nu)=\tfrac12\sum_v\lvert\mu(v)-\nu(v)\rvert$. With this
normalization (the factor $\tfrac12$ included), $\TV$ takes values in $[0,1]$ and
\begin{equation}\label{eq:TVmin}
\TV(\mu,\nu)=1-\sum_v\min\bigl(\mu(v),\nu(v)\bigr).
\end{equation}
For divisors of the same degree $d$ we write $\mu_D=D/d$; then
$\TV(\mu_{D'},\mu_E)=\frac1{2d}\sum_v\lvert D'(v)-E(v)\rvert$ is the number of chips
in which $D'$ and $E$ differ, divided by $d$.

\begin{lemma}[Discreteness]\label{lem:discrete}
Let $D',E\in\Div^d_+(G)$ with $D'\ne E$. Then
$\TV(\mu_{D'},\mu_E)\ge1/d$ and $W_p^p(\mu_{D'},\mu_E)\ge1/d$ for every $p\ge1$.
\end{lemma}

\begin{proof}
The vector $x=D'-E$ is a non-zero integer vector with $\sum_vx(v)=0$. Its
positive part and its negative part therefore have the same total, and that
total is a positive integer, hence at least $1$. So $\sum_v\lvert x(v)\rvert\ge2$
and $\TV(\mu_{D'},\mu_E)=\frac1{2d}\sum_v\lvert x(v)\rvert\ge1/d$. The bound for
$W_p^p$ follows from $\TV\le W_p^p$ (\cref{lem:compare} below).
\end{proof}

\begin{lemma}[Chip-firing as an integer flow]\label{lem:flow}
If $D'=D-Lf$ with $f\in\Z^V$, then $F(u,v)=f(u)-f(v)$ is an antisymmetric
integer flow whose net outflow at $v$ is $D(v)-D'(v)$. Consequently
$W_1(\mu_D,\mu_{D'})\le\frac1d\sum_{uv\in E}\lvert f(u)-f(v)\rvert$.
\end{lemma}

\begin{proof}
The net outflow is $\sum_{u\sim v}(f(v)-f(u))=(Lf)(v)$. For the bound, decompose
$F$ into unit paths and cycles~\cite{AhujaMagnantiOrlin1993}, discard the
cycles, and couple the endpoints of each path.
\end{proof}

\begin{lemma}[Comparison]\label{lem:compare}
$\TV\le W_1\le W_p^p\le\diam(G)^p\,\TV$ for all $p\ge1$.
\end{lemma}

\begin{proof}
Off-diagonal mass is at least $\TV$ and $\dist_G\ge1$ off the diagonal;
$\dist_G\le \dist_G^p$ because $\dist_G$ is integer-valued; and the maximal coupling moves
exactly $\TV$ at unit cost at most $\diam(G)^p$.
\end{proof}

A divisor is the mixture coordinate $\eta=D$ of $\mu_D$, and firing a script $f$
is the translation $\eta\mapsto\eta-Lf$, an m-geodesic move.

\begin{proposition}[Arithmetic versus differential]\label{prop:invisible}
$L(\R^V)$ is the hyperplane of degree-zero vectors. Thus, after the firing
lattice $L\Z^V$ is extended to $L\R^V$, any two configurations of equal degree
are equivalent, and the real-linearized relation is trivial. In particular the
Jacobian $\Jac(G)\cong\Z_0^V/L\Z^V$ is not determined by this real-linearized
relation, and a geometric structure on configurations that is defined without
reference to the lattice $L\Z^V$, such as the Hessian geometry of $F_q$, carries
no information on linear equivalence. Dually, the potentials $h$ with
$Lh\in\Z^V$, taken modulo $\Z^V+\R\one$, form the Pontryagin dual
$\widehat{\Jac(G)}$, a finite subgroup of the torus of potentials modulo
integers.
\end{proposition}

\begin{proof}
For connected $G$, $\ker L=\R\one$, so $\operatorname{im}L$ has dimension $n-1$
and lies in the degree-zero hyperplane. For the dual statement,
$\langle h,Lf\rangle=\langle Lh,f\rangle$ is an integer for all $f\in\Z^V$ exactly
when $Lh\in\Z^V$.
\end{proof}

Topology is therefore encoded arithmetically, as a lattice in mixture
coordinates and as torsion among dual potentials. We do not claim that no
geometric structure can detect it: a structure built from the lattice itself
can. Our point is narrower. The Hessian geometry of the regularizer does not
involve the lattice, so in this paper the Jacobian is detected through a
transport cost minimized over a lattice coset. This is
the role of the defect.

\begin{theorem}[Chip-move metric]\label{thm:chipmove}
For $D',E\in\Div^d_+(G)$, the number $d\,W_1(\mu_{D'},\mu_E)$ is the minimal
number of single-chip moves along edges transforming $E$ into $D'$ through
effective configurations.
\end{theorem}

\begin{proof}
The transport problem with integer supplies is a min-cost flow problem with a
totally unimodular constraint matrix, so an integral optimal flow
exists~\cite{AhujaMagnantiOrlin1993,Schrijver2003}. It is acyclic and decomposes into unit
paths; carrying one chip along each path, one edge at a time, keeps every
configuration effective. Conversely, $k$ moves define a flow of cost at most $k$.
\end{proof}

\section{The Riemann--Roch transport defect}\label{sec:defect}

\subsection{A candidate inequality and its failure}
A tempting reading of~\eqref{eq:RR} is
\begin{equation}\tag{N}\label{eq:naive}
  W_2^2(\mu_D,\mu_K)\ge\MobIdx(D)/\deg D,
\end{equation}
which suggests itself as a guiding principle.

\begin{proposition}\label{prop:naive}
\textup{(i)} $\mu_K=K/\deg K$ is a probability measure if and only if $g\ge2$
and $G$ has no leaves. \textup{(ii)} Whenever $\mu_K$ is defined,
\eqref{eq:naive} fails at $D=K$, where the left side is $0$ and the right side
is $\tfrac12$. \textup{(iii)} For every probability measure $\nu$ with rational
weights there are effective $D$ with $W_2^2(\mu_D,\nu)=0$ and $\MobIdx(D)/\deg D$
arbitrarily close to $1$.
\end{proposition}

\begin{proof}
(i) $\mu_K\ge0$ with positive total mass requires $\deg K=2g-2>0$ and
$\deg_G(v)\ge2$ for all $v$. (ii) $\MobIdx(K)/\deg K=(g-1)/(2g-2)$. (iii) Take
$D=N\nu$ for a large integer $N\ge g$; then the ratio is $1-(g-1)/N$.
\end{proof}

On $G'$, for instance, $K=v_1+v_4$ and $W_2^2(\mu_K,\mu_K)=0<\tfrac12$. The
theorem does not obstruct proximity to a fixed target; it governs how far
chip-firing can \emph{move} a configuration.

\subsection{The defect and the Riemann--Roch estimate}

\begin{definition}
For effective $D$ of degree $d\ge1$ and $\mathcal C\in\{\TV,W_1,W_p^p\}$, the
\emph{transport defect} is
$\delta_{\mathcal C}(D)=\max_{E\in\Div^d_+}\min_{D'\in\lin D}\mathcal C(\mu_{D'},\mu_E)$.
For $\mathcal C=W_p^p$ with $p>1$ this is a cost, not a distance: $W_p^p$ is the
$p$-th power of the Wasserstein distance and does not satisfy the triangle
inequality. No statement below uses the triangle inequality for $W_p^p$.
\end{definition}

\begin{theorem}[Riemann--Roch transport estimate]\label{thm:defect}
Let $D$ be effective of degree $d\ge1$.
\begin{enumerate}[label=\textup{(\alph*)}]
\item \textup{(Every $d\ge1$.)} For every $E\in\Div^d_+$ there is $D'\in\lin D$ with
$D'\ge E_1$ for some effective $E_1\le E$ of degree $\max(0,d-g)$. Hence
$\delta_{\TV}(D)\le\min(1,g/d)$ and $\delta_{W_p^p}(D)\le\diam(G)^p\,g/d$.
\item \textup{(Every $d\ge1$.)} If $g=0$, then $\delta_{\TV}(D)=0$.
\item \textup{(Every $d\ge1$.)} If $g\ge1$, then
$\delta_{\TV}(D)\ge1/d$ and $\delta_{W_p^p}(D)\ge1/d$; in particular
$\delta_{\TV}(D)>0$. For $g=1$, $\delta_{\TV}(D)=1/d$.
\end{enumerate}
Consequently, for every $d\ge1$, $\delta_{\TV}(D)=0$ if and only if $g=0$.
\end{theorem}

\begin{proof}
(a) An effective divisor has $r(D)\ge0$, and \cref{cor:RR}(i) gives
$r(D)\ge d-g$. Hence $s=\max(0,d-g)$ satisfies $s\le r(D)\le d$. Fix
$E\in\Div^d_+$. Choose $E_1\le E$ effective of degree $s$ (possible because
$s\le d=\deg E$). Since $r(D)\ge s$, the definition of rank gives
$F\in\lin{D-E_1}$. Put $D'=F+E_1$; then $D'\ge0$ and $D'\sim D$, so $D'\in\lin D$.
Because $D'\ge E_1$ and $E\ge E_1$, we have $\min(D'(v),E(v))\ge E_1(v)$ for
every $v$, so $\sum_v\min(D'(v),E(v))\ge s$. By \eqref{eq:TVmin},
$\TV(\mu_{D'},\mu_E)\le1-s/d=\min(1,g/d)$. As $E$ was arbitrary,
$\delta_{\TV}(D)\le\min(1,g/d)$. The $W_p^p$ bound follows from
$W_p^p\le\diam(G)^p\,\TV$ (\cref{lem:compare}), applied to the same $D'$.
(b) On a tree every $E\in\Div^d_+$ lies in $\lin D$ (\cref{cor:trees}), so one
may take $D'=E$.
(c) We need one effective target $E\in\Div^d_+$ with $E\not\sim D$. If
$g\ge1$, then $\kappa(G)\ge2$, since a connected graph with a cycle has at least
two spanning trees, so $\Jac(G)$ is non-trivial. It is generated by the classes
$[v-u]$ of the edges $uv$, so some edge has $u\not\sim v$. The two effective
divisors $(d-1)u+v$ and $d\,u$ differ by $v-u$, so they are not equivalent, and
at least one of them is not equivalent to $D$. Call it $E$. (Choosing $E$ in an arbitrary class different from $[D]$ would
require $d\ge g$ to guarantee an effective representative; this choice needs
only $d\ge1$.)
Every $D'\in\lin D$ satisfies $D'\sim D\not\sim E$, hence $D'\ne E$, and
\cref{lem:discrete} gives $\TV(\mu_{D'},\mu_E)\ge1/d$ and
$W_p^p(\mu_{D'},\mu_E)\ge1/d$. Taking the minimum over the finite set $\lin D$
and then the maximum over targets gives both lower bounds. For $g=1$ the upper
bound of (a) is $1/d$, so $\delta_{\TV}(D)=1/d$.
\end{proof}

Thus the upper bound comes from rank: at least $\max(0,d-g)$ chips of any
target can be matched, so at most $g$ chips remain constrained. The lower bound
is a separation argument of a different kind: integral divisors of equal degree
in different classes cannot be closer than one chip, that is $1/d$ in
total-variation units.

\subsection{Abel--Jacobi bounds}
Three different quantities appear in this subsection:
\begin{itemize}
\item $W_1(\mu,\nu)$, a transport cost between probability measures;
\item $d\,W_1(\mu_{D'},\mu_E)$, which by \cref{thm:chipmove} is the minimal number of
single-chip moves between the integral configurations $D'$ and $E$;
\item $\rho(G)$, a diameter in the Jacobian.
\end{itemize}
Each chip move along an edge $uv$ changes the class by $\pm[v-u]\in\Jac(G)$, so
move counts are word lengths for the \emph{edge generators}
$\{\pm[v-u]:uv\in E(G)\}$. The quantities $N_w$ below are instead weighted word
lengths for the \emph{Abel--Jacobi generators} $\{[v-w]:v\in V\}$ based at a
fixed vertex $w$, with weight $\dist_G(w,v)$.

\begin{definition}
Let $\rho(G)$ be the diameter of the Cayley graph of $\Jac(G)$ with generators
$\{\pm[v-u]:uv\in E(G)\}$. For $w\in V$ and $c\in\Jac(G)$, let $N_w(c)$ be the
minimum of $\sum_vF(v)\,\dist_G(w,v)$ over effective $F$ with $\sum_vF(v)[v-w]=c$.
\end{definition}

\begin{theorem}[Abel--Jacobi bounds]\label{thm:AJ}
Let $D$ be effective of degree $d\ge1$.
\begin{enumerate}[label=\textup{(\roman*)}]
\item \textup{(Every $d\ge1$.)} We have
\[\max_{w}N_w\bigl([D-d\,w]\bigr)\le d\,\delta_{W_1}(D)\le\diam(G)\,g,\]
and $N_w(c)\le g\max_v\dist_G(w,v)$ for every $c$ (Jacobi inversion).
\item \textup{(Requires $d\ge g$.)} If $d\ge\max(g,1)$, then $\rho(G)\le d\,\delta_{W_1}(D)$.
\end{enumerate}
\end{theorem}

\begin{proof}
\emph{Lower bound in (i).} Fix $w$ and take the target $E=d\,w$, which is
effective for every $d\ge1$. For $D'\in\lin D$ the only coupling of $\mu_{D'}$ with
the point mass $\delta_w=\mu_{dw}$ is the product, so
\[
  d\,W_1(\mu_{D'},\mu_{dw})=\sum_vD'(v)\,\dist_G(w,v).
\]
On the other hand $\sum_vD'(v)[v-w]=[D'-d\,w]=[D-d\,w]$, because $D'\sim D$. So
$F=D'$ is admissible in the definition of $N_w([D-d\,w])$, and
$d\,W_1(\mu_{D'},\mu_{dw})\ge N_w([D-d\,w])$. Minimizing over $D'\in\lin D$ and using
$\delta_{W_1}(D)\ge\min_{D'}W_1(\mu_{D'},\mu_{dw})$ gives
$d\,\delta_{W_1}(D)\ge N_w([D-d\,w])$ for every $w$.

\emph{Upper bound in (i).} By \cref{lem:compare} and \cref{thm:defect}(a),
$\delta_{W_1}(D)\le\diam(G)\,\delta_{\TV}(D)\le\diam(G)\min(1,g/d)$, so
$d\,\delta_{W_1}(D)\le\diam(G)\,g$.

\emph{Jacobi inversion.} Let $c\in\Jac(G)$ and choose a divisor $A$ of degree $0$
in the class $c$. The divisor $A+g\,w$ has degree $g$, so by \cref{cor:RR}(i) it is
equivalent to an effective $F$ of degree $g$. Then
$\sum_vF(v)[v-w]=[F-g\,w]=c$, and
$\sum_vF(v)\dist_G(w,v)\le g\max_v\dist_G(w,v)$.

\emph{(ii).} By \cref{thm:chipmove}, $d\,W_1(\mu_{D'},\mu_E)$ is the least number of
single-chip moves from $E$ to $D'$. Each move along an edge $uv$ changes the
class of the configuration by $\pm[v-u]$, so a move sequence from $E$ to
$D'\in\lin D$ is a word in the edge generators representing $[D-E]$, and its
length is at least the Cayley distance of $[D-E]$ from $0$. Since the Cayley
graph is vertex-transitive, some class $c$ has distance $\rho(G)$ from $0$.
Because $d\ge g$, the class $[D]-c$ of degree $d$ contains an effective $E$
(\cref{cor:RR}(i)); this is the only use of $d\ge g$. For this $E$,
$d\min_{D'}W_1(\mu_{D'},\mu_E)\ge\rho(G)$.
\end{proof}

\begin{remark}[$N_w$ is a transport cost]\label{rem:Nw}
$N_w(c)$ is the cheapest way of producing the class $c$ by moving chips out of
a single vertex $w$. If $F$ has degree $k$, then $\sum_vF(v)\dist_G(w,v)$ is
exactly $k\,W_1(\mu_F,\delta_w)$, and the condition $\sum_vF(v)[v-w]=c$ says
$[F-k\,w]=c$. Hence
\[
  N_w(c)=\min_{k\ge0}\ \min\bigl\{k\,W_1(\mu_F,\delta_w):F\in\Div^k_+(G),\ [F-k\,w]=c\bigr\}.
\]
The lower bound of (i) is the restriction of this minimum to $k=d$ and to
$F\in\lin D$, which explains the direction of the inequality. Two facts make
the correspondence exact. The target $d\,w$ has a unique coupling, and the
weights $\dist_G(w,v)$ are the costs of that coupling. For a general target $E$
no single base point exists, which is why the general lower bound uses the
unweighted edge generators and $\rho(G)$. The values in \cref{tab:AJ} were
computed by Dijkstra's algorithm on the weighted Cayley graph
(\cref{app:computation}).
\end{remark}

\begin{table}[!htbp]
\centering\small
\setlength{\tabcolsep}{3.3pt}\renewcommand{\arraystretch}{1.12}
\begin{tabular}{llcccccccccc}
\toprule
Graph & $D$ & $d$ & $g$ & $r(D)$ & $r(K\!-\!D)$ & $\lvert\lin D\rvert$
 & $\delta_{\TV}$ & $g/d$ & $\delta_{W_1}$ & $\diam\cdot g/d$ & $\delta_{W_2^2}$ \\
\midrule
$T_7$  & $4v_1$ & 4 & 0 & 4 & $-1$ & 210 & 0 & 0 & 0 & 0 & 0 \\
$C_7$  & $4v_1$ & 4 & 1 & 3 & $-1$ & 30  & 1/4 & 1/4 & 3/4 & 3/4 & 3/4 \\
$C_7$  & $v_1{+}v_2{+}v_3{+}v_4$ & 4 & 1 & 3 & $-1$ & 30 & 1/4 & 1/4 & 3/4 & 3/4 & 3/4 \\
$G'$   & $4v_1$ & 4 & 2 & 2 & $-1$ & 9   & 1/2 & 1/2 & 5/4 & 3/2 & 5/2 \\
$G''$  & $5v_1$ & 5 & 3 & 2 & $-1$ & 11  & 3/5 & 3/5 & 6/5 & 9/5 & 12/5 \\
$G'$   & $K$ & 2 & 2 & 1 & $0$ & 4 & 1/2 & 1 & 3/2 & 3 & 5/2 \\
\bottomrule
\end{tabular}
\caption{Ranks, linear systems and exact transport defects. The bound $g/d$ is
attained in every non-special case shown.}
\label{tab:defect}
\end{table}

\begin{table}[!htbp]
\centering\small
\begin{tabular}{lcccccl}
\toprule
$G$ & $g$ & $\kappa$ & $\diam$ & $\rho(G)$ & $\max_{w,c}N_w(c)$ &
$d\,\delta_{W_1}(dv_1)$, $d=1,\dots,10$ \\
\midrule
$C_7$  & 1  & 7     & 3 & 3 & 3  & $3,3,3,3,3,3,3,3,3,3$ \\
$G'$   & 2  & 19    & 3 & 3 & 5  & $2,4,4,5,5,5,5,5,5,5$ \\
$G''$  & 3  & 40    & 3 & 4 & 7  & $2,4,6,5,6,5,5,6,6,6$ \\
$K_7$  & 15 & 16807 & 1 & 6 & 15 & --- \\
\bottomrule
\end{tabular}
\caption{Jacobian invariants and exact $W_1$-defects computed through
\cref{thm:chipmove}.}
\label{tab:AJ}
\end{table}

On $C_7$ the bounds of \cref{thm:AJ} coincide, $\rho=\diam\cdot g=3$, so
$\delta_{W_1}(D)=3/d$ for every $D$. On $G'$ and $G''$ effectivity matters: the
defect stabilizes at $5$ or oscillates between $5$ and $6$. The
concentrated-target bound $\max_wN_w([D-dw])$ is exact for $7$ of the $10$
degrees on $G'$ and $9$ of the $10$ on $G''$. On $K_7$ the Jacobi-inversion
bound $15=g$ is attained by $\max N_w$, although $\delta_{\TV}(15v_1)=7/15$ and
$\delta_{\TV}(20v_1)=2/5$ stay well below $g/d$.

\section{The topology--information crossover}\label{sec:crossover}

We now let the regularized geometry act on configurations. Fix $q>0$,
$\gamma>0$, effective $D,E$ of the same degree $d\ge1$, the cost $c=\dist_G$ and
$n=\lvert V\rvert\ge2$, and define the
\emph{regularized topological gap}
\[
  \Gamma_{q,\gamma}(D,E)=\min_{D'\in\lin D}\Wq(\mu_{D'},\mu_E)-\Wq(\mu_E,\mu_E),
\]
the regularized price of lying in the class of $D$ rather than in that of the
target. By \cref{thm:gauge} the constant $\gamma c_{q,N}$ cancels, so we may
work with $\mathcal T_{q,\varepsilon}$ and $\varepsilon=\gamma/q$. Let
$N_E=\min_{D'\in\lin D}d\,W_1(\mu_{D'},\mu_E)$ be the chip-move distance and
$H_q(\mu,\nu)=\max_{\pi\in\Pi(\mu,\nu)}S_q(\pi)$.

Throughout this section the graph $G$, and hence its genus and Jacobian, is
fixed; only the regularization parameters $(q,\gamma)$ vary. To avoid any
suggestion that the graph itself changes, we fix the following terminology.

\begin{definition}[Gap regime and transitions]\label{def:transitions}
Fix $G$, $D$ and $E$.
\begin{enumerate}[label=\textup{(\roman*)}]
\item The \emph{positive-gap regime} is $\{(q,\gamma):\Gamma_{q,\gamma}(D,E)>0\}$, the
parameters at which the class of the target is strictly preferred.
\item A \emph{gap-sign transition} is a boundary point of the positive-gap
regime, i.e.\ a parameter at which $\Gamma_{q,\gamma}(D,E)=0$ and $\Gamma$ changes
sign. We write $\gamma^\ast(q)=\inf\{\gamma>0:\Gamma_{q,\gamma}(D,E)\le0\}$ for the
crossover temperature at fixed $q$ (with $\gamma^\ast=\infty$ if the set is
empty), and $q_{\rm gap}(\gamma)$ for the value of $q$ at which $\Gamma$ changes
sign at fixed $\gamma$, when it exists.
\item A \emph{support transition} is a change of $\operatorname{supp}P^\star_{q,\gamma}$,
the support of an optimal regularized plan; it is studied in \cref{sec:pd}.
\end{enumerate}
\end{definition}

None of these is a change of the topology of $G$. ``Topological'' refers only
to the origin of the positive gap, which is the integrality of linear
equivalence.

\begin{remark}[What ``crossover'' does and does not assert]\label{rem:crossoverword}
Three statements must be kept apart.
\begin{enumerate}[label=(\alph*)]
\item As $\gamma\to0$ the gap tends to the chip-move defect $N_E/d$
(\cref{thm:crossover}(i)). If $N_E>0$, the gap is positive on the whole interval
$(0,\gamma_c(q))$ (\cref{thm:crossover}(ii)), so $\gamma^\ast(q)\ge\gamma_c(q)>0$.
\item As $\gamma\to\infty$ the sign of $\Gamma$ is the sign of the entropic slope of
\cref{thm:crossover}(iii). If the slope is negative, then $\gamma^\ast(q)<\infty$.
\item Neither limit says anything about what happens in between. We do not
claim that $\Gamma_{q,\gamma}$ is monotone in $\gamma$, nor that it changes sign
only once.
\end{enumerate}
The word ``crossover'' and the symbol $\gamma^\ast(q)$ therefore refer only to
the infimum in \cref{def:transitions}(ii), the first temperature at which the
gap is no longer positive. Numerically we evaluated $\Gamma$ on $16$
temperatures in $[0.03,30]$ for each of $11$ values of $q$ on $C_7$, $G'$ and
$G''$. In these $33$ scans $\Gamma$ changed sign at most once. The reported
$\gamma^\ast$ is the root found by bisection in the first bracketing interval. A
further pair of sign changes between two grid points cannot be excluded by such
a scan.
\end{remark}

\begin{theorem}[Crossover]\label{thm:crossover}
Let $\varepsilon=\gamma/q$.
\begin{enumerate}[label=\textup{(\roman*)}]
\item \textup{(Low temperature.)} $\lim_{\gamma\to0}\Gamma_{q,\gamma}(D,E)=N_E/d$.
\item \textup{(Riemann--Roch window.)} For all $\gamma>0$ (and every $d\ge1$)
\[
  \frac{N_E}{d}-\varepsilon\bigl(\ln_q(n^2)-S_q(\mu_E)\bigr)
  \le\Gamma_{q,\gamma}(D,E)\le
  \frac{\diam(G)\,g}{d}+\varepsilon\,H_q(\mu_E,\mu_E).
\]
In particular $\Gamma_{q,\gamma}(D,E)>0$ whenever
$\gamma<\gamma_c(q)=\dfrac{q\,N_E}{d\,(\ln_q(n^2)-S_q(\mu_E))}$. The denominator is
positive because $S_q(\mu_E)\le\ln_q n<\ln_q(n^2)$, and $\gamma_c(q)>0$ if and only
if $N_E>0$, that is $E\notin\lin D$.
\item \textup{(High temperature.)} With the normalization by $\varepsilon=\gamma/q$,
\[
  \lim_{\gamma\to\infty}\frac{\Gamma_{q,\gamma}(D,E)}{\varepsilon}
  =H_q(\mu_E,\mu_E)-\max_{D'\in\lin D}H_q(\mu_{D'},\mu_E);
\]
for $q=1$ this equals $S(\mu_E)-\max_{D'\in\lin D}S(\mu_{D'})$.
\end{enumerate}
\end{theorem}

\begin{proof}
Write $\mathcal T=\mathcal T_{q,\varepsilon}$, so that
$\mathcal T(\mu,\nu)=\min_{\pi\in\Pi(\mu,\nu)}\{\langle c,\pi\rangle-\varepsilon S_q(\pi)\}$
with $c=\dist_G$. By \cref{thm:gauge}(iii),
$\Wq=\mathcal T+\gamma c_{q,N}$ with the same constant in both terms of $\Gamma$, hence
\begin{equation}\label{eq:gapT}
  \Gamma_{q,\gamma}(D,E)=\min_{D'\in\lin D}\mathcal T(\mu_{D'},\mu_E)-\mathcal T(\mu_E,\mu_E).
\end{equation}
We use four elementary inequalities, valid for all probability measures
$\mu,\nu$ on $V$. They rest on $0\le S_q(\pi)\le\ln_q(n^2)$ for every coupling, a
probability vector on $n^2$ cells, and on $c\ge0$.
\begin{align}
  \mathcal T(\mu,\nu)&\le W_1(\mu,\nu)
  &&\text{(evaluate at a $W_1$-optimal $\pi$; $-\varepsilon S_q\le0$)},\tag{T1}\label{eq:a}\\
  \mathcal T(\mu,\nu)&\ge W_1(\mu,\nu)-\varepsilon\ln_q(n^2)
  &&\text{($\langle c,\pi\rangle\ge W_1$ and $S_q\le\ln_q(n^2)$)},\tag{T2}\label{eq:b}\\
  \mathcal T(\mu_E,\mu_E)&\le-\varepsilon S_q(\mu_E)
  &&\text{(diagonal coupling: cost $0$, entropy $S_q(\mu_E)$)},\tag{T3}\label{eq:c}\\
  \mathcal T(\mu,\nu)&\ge-\varepsilon H_q(\mu,\nu)
  &&\text{($\langle c,\pi\rangle\ge0$ and $S_q(\pi)\le H_q(\mu,\nu)$)}.\tag{T4}\label{eq:d}
\end{align}
(i) By \eqref{eq:a} and \eqref{eq:b},
$\lvert\mathcal T-W_1\rvert\le\varepsilon\ln_q(n^2)\to0$ uniformly. Since
$W_1(\mu_E,\mu_E)=0$ and $\min_{D'\in\lin D}W_1(\mu_{D'},\mu_E)=N_E/d$ by
definition of $N_E$, \eqref{eq:gapT} gives the limit.

(ii) Lower bound. By \eqref{eq:b}, the first term of \eqref{eq:gapT} is at
least $N_E/d-\varepsilon\ln_q(n^2)$. By \eqref{eq:c}, the subtracted term
satisfies $-\mathcal T(\mu_E,\mu_E)\ge+\varepsilon S_q(\mu_E)$. Adding,
$\Gamma\ge N_E/d-\varepsilon\ln_q(n^2)+\varepsilon S_q(\mu_E)$, which is the stated
bound. The right-hand side is positive exactly when
$\varepsilon<N_E/(d(\ln_q(n^2)-S_q(\mu_E)))$, i.e.\ $\gamma<\gamma_c(q)$.
Upper bound. By \eqref{eq:a} and \cref{thm:defect}(a) with \cref{lem:compare},
the first term is at most $\min_{D'}W_1(\mu_{D'},\mu_E)\le\diam(G)\,g/d$. By
\eqref{eq:d}, $-\mathcal T(\mu_E,\mu_E)\le\varepsilon H_q(\mu_E,\mu_E)$. Adding gives the
stated bound.

(iii) Choosing $\pi$ to maximize $S_q$ gives
$\mathcal T\le\max c-\varepsilon H_q$, and \eqref{eq:d} gives $\mathcal T\ge-\varepsilon H_q$.
Hence $\lvert\mathcal T(\mu,\nu)/\varepsilon+H_q(\mu,\nu)\rvert\le\max c/\varepsilon$,
uniformly over the finitely many pairs $(\mu_{D'},\mu_E)$ and $(\mu_E,\mu_E)$.
Dividing \eqref{eq:gapT} by $\varepsilon$ and letting $\varepsilon=\gamma/q\to\infty$ gives
$\Gamma/\varepsilon\to-\max_{D'}H_q(\mu_{D'},\mu_E)+H_q(\mu_E,\mu_E)$. For $q=1$
the product coupling maximizes Shannon entropy among couplings, so
$H_1(\mu,\nu)=S(\mu)+S(\nu)$.
\end{proof}

\begin{remark}[Normalizations in \cref{thm:crossover}]\label{rem:crossnorm}
Every inequality and limit of the theorem uses the same five conventions.
\begin{enumerate}[label=(\arabic*)]
\item $\Gamma$ is a difference of two values of $\Wq$. The plan-independent
constant $\gamma c_{q,N}$ of \cref{thm:gauge} appears in both and cancels, so
$\Gamma$ is the same whether computed with $F_q$ or with Tsallis entropy
(\eqref{eq:gapT}).
\item The Tsallis temperature is $\varepsilon=\gamma/q$, and all bounds are linear in
$\varepsilon$.
\item $S_q$ is evaluated on couplings, which are probability vectors on $n^2$
cells, whence $\ln_q(n^2)$; on the marginal $\mu_E$ it is evaluated on $n$ cells.
\item $H_q(\mu,\nu)=\max_{\pi\in\Pi(\mu,\nu)}S_q(\pi)$ is attained, since $\Pi(\mu,\nu)$ is
compact and $S_q$ continuous. For $q=1$ it is $S(\mu)+S(\nu)$, by the product coupling.
\item The limit in (iii) is taken after division by $\varepsilon$. In terms of
$\gamma$ it reads $\Gamma/\gamma\to\frac1q\bigl(H_q(\mu_E,\mu_E)-\max_{D'}H_q(\mu_{D'},\mu_E)\bigr)$.
It uses only the bound $\lvert\mathcal T/\varepsilon+H_q\rvert\le\max c/\varepsilon$, uniform over
the finitely many pairs involved, and the maximum over the finite set
$\lin D$ exists.
\end{enumerate}
All these conventions are continuous at $q=1$, so the case $q=1$ needs no
separate argument.
\end{remark}

\begin{remark}[Which upper bound to use]\label{rem:whichbound}
The upper side $\diam(G)\,g/d$ of the window is not sharp in general. For a
\emph{special} divisor ($r(K-D)\ge0$), \cref{thm:serre} replaces $g$ by
$g-1-r(K-D)$, a genuine improvement by $(r(K-D)+1)\diam(G)/d$. For a
non-special divisor the two coincide. Where $r(K-D)$ is available, the refined
bound should be used.
\end{remark}

\begin{theorem}[A positive-gap regime exists if and only if $g\ge1$]\label{thm:phase}
Let $D$ be effective of degree $d\ge1$, let $q>0$, and put $\varepsilon=\gamma/q$.
\begin{enumerate}[label=\textup{(\alph*)}]
\item If $g=0$, then $\Gamma_{q,\gamma}(D,E)\le0$ for every target $E$ and every
$\gamma>0$.
\item If $g\ge1$, there is a target $E$ with
\[
  \Gamma_{q,\gamma}(D,E)\ge\frac1d-\varepsilon\ln_q(n^2)>0
  \qquad\text{for all }\gamma<\frac{q}{d\,\ln_q(n^2)}.
\]
\item If $g\ge1$ and moreover $d\ge g$, the target can be chosen with
\[
  \Gamma_{q,\gamma}(D,E)\ge\frac{\rho(G)}d-\varepsilon\ln_q(n^2)>0
  \qquad\text{for all }\gamma<\frac{q\,\rho(G)}{d\,\ln_q(n^2)}.
\]
\end{enumerate}
In particular, for every $d\ge1$ the positive-gap regime of some pair $(D,E)$ is
non-empty if and only if $g\ge1$.
\end{theorem}

\begin{proof}
(a) By \cref{cor:trees}, $E\in\lin D$, so the minimum in \eqref{eq:gapT} is at
most $\mathcal T(\mu_E,\mu_E)$.
(b) The proof of \cref{thm:defect}(c) gives an effective $E\not\sim D$ for every
$d\ge1$. Every $D'\in\lin D$ differs from $E$, so by \cref{lem:discrete} and
\cref{lem:compare}, $d\,W_1(\mu_{D'},\mu_E)\ge d\,\TV(\mu_{D'},\mu_E)\ge1$; hence
$N_E\ge1$. Insert this and $S_q(\mu_E)\ge0$ into the lower bound of
\cref{thm:crossover}(ii), which holds for every $d\ge1$.
(c) Since $d\ge g$, every class of degree $d$ contains an effective divisor
(\cref{cor:RR}(i)); choose $E$ with $[D-E]$ at Cayley distance $\rho(G)$ from
$0$. By \cref{thm:chipmove}, a sequence of chip moves from $E$ to $\lin D$ is a
word in the edge generators for $[D-E]$, so $N_E\ge\rho(G)$. Conclude as in (b).
\end{proof}

\begin{remark}[Existence versus persistence]\label{rem:existpersist}
\Cref{thm:phase} and \cref{thm:crossover} answer two different questions.
\emph{Existence} of a positive gap is purely topological: for every $d\ge1$ it
is equivalent to $g\ge1$, and it uses only that the Jacobian is non-trivial. The
hypothesis $d\ge g$, which makes every class effective, enters only the
quantitative form \cref{thm:phase}(c) (\cref{rem:scope}). \emph{Persistence}, the temperature up to which
the gap stays positive, is not topological. Its bounds involve the chip-move
distance $N_E$, the diameter, and the entropies $S_q(\mu_E)$ and $H_q$, and
\cref{thm:TI} below isolates the informational part exactly.
\end{remark}

\begin{figure}[!htbp]
\centering
\begin{tikzpicture}
\begin{axis}[width=0.78\textwidth,height=6.2cm,xmode=log,
  xlabel={$\gamma$},ylabel={$\Gamma_{q,\gamma}(4v_1,E)$},
  ymin=-3,ymax=1.4,xmin=0.01,xmax=5,
  legend style={font=\footnotesize,at={(0.02,0.03)},anchor=south west},
  tick label style={font=\footnotesize},label style={font=\small},
  grid=major,grid style={black!12}]
\addplot[blue!70!black,thick,mark=*,mark size=1.2pt] coordinates {(0.01,1.2100) (0.02,1.1701) (0.05,1.0510) (0.1,0.8555) (0.15,0.6650) (0.2,0.4782) (0.3,0.1166) (0.4,-0.2372) (0.5,-0.5823) (0.7,-1.2645) (1.0,-2.2735) (1.5,-3.9444)};
\addplot[green!45!black,thick,mark=square*,mark size=1.2pt] coordinates {(0.01,1.2396) (0.02,1.2292) (0.05,1.1980) (0.1,1.1460) (0.15,1.0941) (0.2,1.0427) (0.3,0.9433) (0.4,0.8499) (0.5,0.7619) (0.7,0.5974) (1.0,0.3669) (1.5,0.0012) (2.0,-0.3554) (3.0,-1.0587) (5.0,-2.4533)};
\addplot[red!70!black,thick,mark=triangle*,mark size=1.6pt] coordinates {(0.01,1.2484) (0.02,1.2469) (0.05,1.2422) (0.1,1.2344) (0.15,1.2266) (0.2,1.2187) (0.3,1.2031) (0.4,1.1875) (0.5,1.1719) (0.7,1.1406) (1.0,1.0937) (1.5,1.0156) (2.0,0.9375) (3.0,0.8229) (5.0,0.6875)};
\addplot[black!60,dashed,thick] coordinates {(0.01,0.7431) (0.02,0.7361) (0.05,0.7153) (0.1,0.6805) (0.15,0.6449) (0.2,0.6087) (0.3,0.5395) (0.4,0.4795) (0.5,0.4301) (0.7,0.3582) (1.0,0.2952) (1.5,0.2466) (2.0,0.2253) (3.0,0.1929) (5.0,0.1631)};
\addplot[black,thin,forget plot] coordinates {(0.01,0) (5,0)};
\legend{{$G'$, $q=\tfrac12$},{$G'$, $q=1$},{$G'$, $q=2$},{$C_7$, $q=1$}}
\end{axis}
\end{tikzpicture}
\caption{Regularized topological gap for $D=4v_1$ against a worst target
($E=2v_5+2v_6$ on $G'$, $E=v_1+v_2+v_4+v_7$ on $C_7$). The curves start at
$N_E/d$ ($5/4$ and $3/4$). On $G'$ every curve eventually crosses zero (for
$q=2$ near $\gamma\approx16$, beyond the plotted range). On $C_7$ the
high-temperature slope vanishes and no crossover was observed.}
\label{fig:gap}
\end{figure}
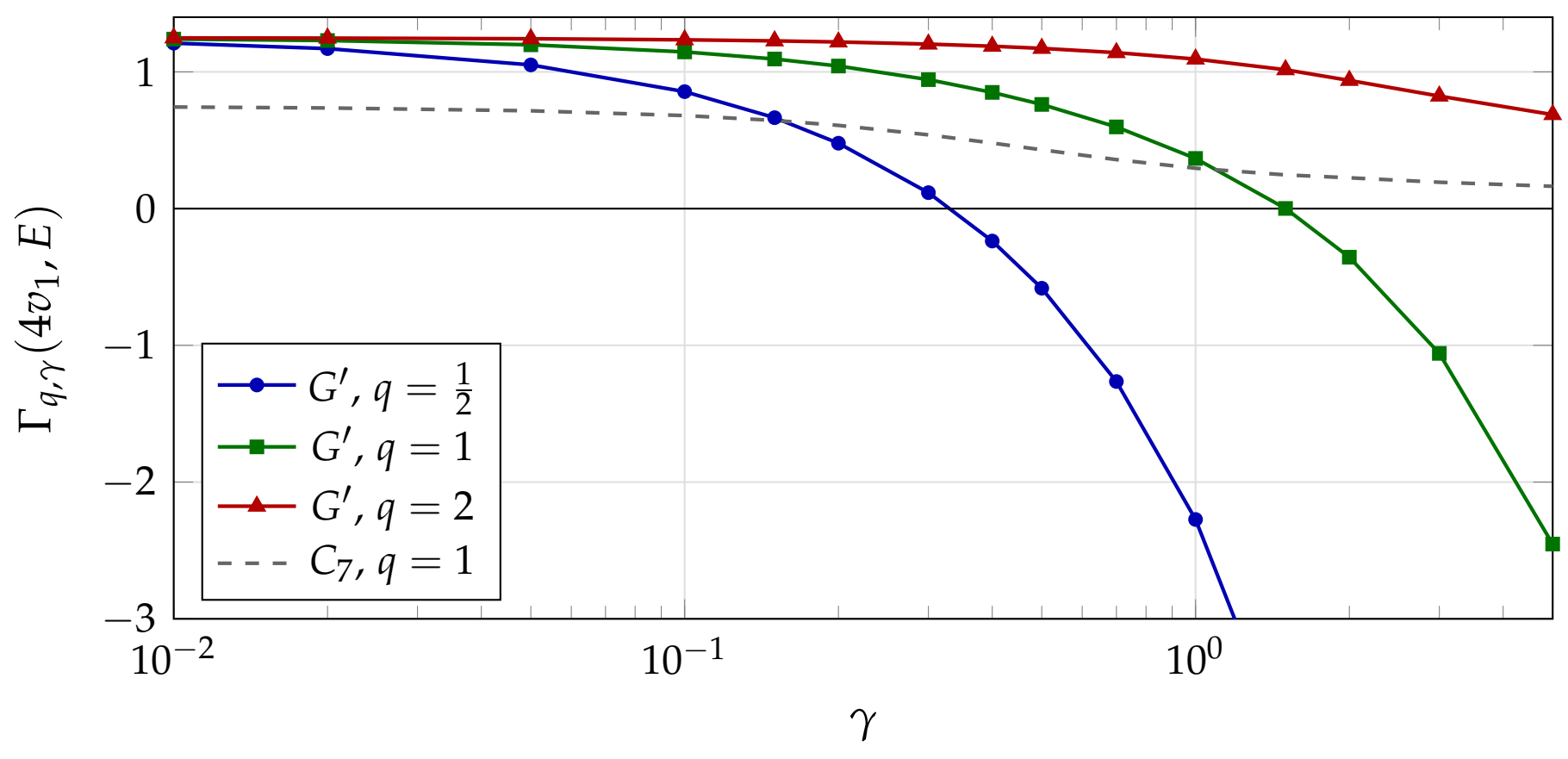

\begin{table}[!htbp]
\centering\small
\begin{tabular}{ccccc}
\toprule
$q$ & $\gamma_c(q)$ (sufficient) & $\gamma^\ast(q)$ (exact) &
high-$T$ slope & $\ln_q4-\ln_q8$ \\
\midrule
0.50 & 0.056 & 0.333  & $-1.6573$ & $-1.6569$ \\
0.75 & 0.161 & 0.746  & $-1.0703$ & $-1.0703$ \\
1.00 & 0.391 & 1.502  & $-0.6931$ & $-\log 2$ \\
1.25 & 0.844 & 2.85   & $-0.4500$ & $-0.4500$ \\
1.50 & 1.661 & 5.19   & $-0.2929$ & $-0.2929$ \\
2.00 & 5.213 & 16.0   & $-0.1250$ & $-1/8$ \\
3.00 & 30.05 & 126.0  & $-0.0234$ & $-3/128$ \\
\bottomrule
\end{tabular}
\caption{Crossover on $G'$ ($g=2$), $D=4v_1$, $E=2v_5+2v_6$. The slope of
\cref{thm:crossover}(iii) equals $\ln_q4-\ln_q8$: the target spreads over two
vertices, while $\lin{4v_1}$ contains a configuration spread over four.}
\label{tab:phase}
\end{table}

Three conclusions follow. First, $q$ is an inverse-temperature multiplier: by
\cref{thm:gauge} the effective temperature is $\gamma/q$, and $\ln_q(n^2)$
decreases in $q$; both effects protect topology, and $\gamma^\ast(q)$ grows by
a factor of about $380$ between $q=\tfrac12$ and $q=3$. Second, the crossover
is entropic rather than metric. Its occurrence is decided by the comparison of
coupling-entropy capacities in \cref{thm:crossover}(iii); on $C_7$ the linear
system contains a configuration as spread as the target, the slope vanishes,
and the gap remained positive for all $\gamma\le5$. Third, the
``phase-transition-like'' behaviour reported in~\cite{JSAI2026} acquires a
precise counterpart: a gap-sign transition (\cref{def:transitions}) separating a topology-dominated
regime, in which the Jacobian class is visible, from an information-dominated
one. The sufficient window $\gamma_c$ is conservative by a factor of $3$ to $6$.

\section{Primal--dual structure of the topological gap}\label{sec:pd}

Three dualities have appeared so far: the Legendre duality $(\eta,\theta)$ of
$F_q$ (\cref{sec:dual}), the Kantorovich duality between plans and potentials
(\cref{sec:support}), and the Serre duality $D\leftrightarrow K-D$ of the
Riemann--Roch theorem (\cref{sec:divisors}), together with the Pontryagin
duality of the Jacobian (\cref{prop:invisible}). In this section we show that
the three phenomena singled out in the preceding analysis are each governed by
one of these dualities: the size of the topological gap, the change of
support of optimal plans, and the degeneracy of optimal transport. The
picture that emerges is that \emph{topology enters through the integrality of
the primal problem, while sparsity is a boundary phenomenon on the dual side}.

Throughout, $\mathcal T_{q,\varepsilon}$ and $\Wq$ are as in
\cref{sec:gauge}, $\varepsilon=\gamma/q$, $n=\lvert V\rvert$, the cost is
$c=\dist_G$, and $E$ is an effective divisor of degree $d\ge1$. We write
$F_q^*(t)=\sup_{x\ge0}\bigl(tx-F_q(x)\bigr)\in(-\infty,+\infty]$ for the convex
conjugate of $F_q$ on its domain $[0,\infty)$.

\begin{lemma}[Dual problem]\label{lem:dual}
For probability vectors $\mu,\nu$ on $V$,
\begin{equation}\label{eq:dual}
  \Wq(\mu,\nu)=\sup_{\alpha,\beta\in\R^V}\Phi(\mu;\alpha,\beta),\qquad
  \Phi(\mu;\alpha,\beta)=\langle\alpha,\mu\rangle+\langle\beta,\nu\rangle
  -\gamma\sum_{i,j}F_q^*\Bigl(\frac{\alpha_i+\beta_j-c_{ij}}{\gamma}\Bigr),
\end{equation}
and $P^\star_{ij}=(F_q^*)'\bigl((\alpha_i+\beta_j-c_{ij})/\gamma\bigr)$ at optimal
potentials whenever these exist.
\end{lemma}

\begin{proof}
\emph{Weak duality.} Let $P\in\Pi(\mu,\nu)$ and put
$t_{ij}=(\alpha_i+\beta_j-c_{ij})/\gamma$. The Fenchel--Young inequality
$F_q(x)\ge t x-F_q^*(t)$, multiplied by $\gamma$, gives
$\gamma F_q(P_{ij})\ge(\alpha_i+\beta_j-c_{ij})P_{ij}-\gamma F_q^*(t_{ij})$. Summing
over $(i,j)$ and adding $\langle c,P\rangle$, the terms $c_{ij}P_{ij}$ cancel, and
the marginal constraints turn $\sum_{ij}(\alpha_i+\beta_j)P_{ij}$ into
$\langle\alpha,\mu\rangle+\langle\beta,\nu\rangle$. Hence $J_{q,\gamma}(P)\ge\Phi(\mu;\alpha,\beta)$.

\emph{No duality gap.} Suppose first $\mu,\nu>0$. The primal problem minimizes a
finite convex function over $[0,\infty)^{V\times V}$ subject to linear equality
constraints, and the feasible plan $\mu\otimes\nu$ has all entries positive, so
it lies in the relative interior of the domain. By
\cite[Thm.~28.2 and Sect.~30]{Rockafellar1970} the Lagrangian dual, which is
the supremum of $\Phi$ computed in the first step, has the same value. If some
$\mu_i=0$ or $\nu_j=0$, every feasible plan vanishes in that row or column.
Apply the previous case to the problem restricted to the supports, and let
$\alpha_i\to-\infty$ (respectively $\beta_j\to-\infty$) for the removed indices.
Since $F_q^*(t)\to-F_q(0)$ as $t\to-\infty$, each removed cell contributes
$\gamma F_q(0)$ to the dual objective in the limit, which is what the zero entry
contributes to $J_{q,\gamma}$. The two values therefore agree. The formula for
$P^\star$ is the equality case of Fenchel--Young.
\end{proof}

\subsection{Serre duality refines the Riemann--Roch estimate}

\begin{theorem}[Rank and Serre-dual form of the defect]\label{thm:serre}
Let $D$ be effective of degree $d\ge1$. Then
\[
  \delta_{\TV}(D)\le\frac{d-r(D)}{d}=\frac{g-1-r(K-D)}{d},\qquad
  \delta_{W_p^p}(D)\le\diam(G)^p\,\frac{g-1-r(K-D)}{d},
\]
and for every target $E$ and all $q,\gamma>0$,
\[
  \Gamma_{q,\gamma}(D,E)\le\diam(G)\,\frac{g-1-r(K-D)}{d}+\frac\gamma q\,H_q(\mu_E,\mu_E).
\]
\end{theorem}

\begin{proof}
An effective divisor has $r(D)\ge0$. Given $E\in\Div^d_+$, choose an effective
$E_1\le E$ of degree $r(D)$ and $F\in\lin{D-E_1}$, and put $D'=F+E_1\in\lin D$.
Then $\sum_v\min(D'(v),E(v))\ge r(D)$, so $\TV(\mu_{D'},\mu_E)\le(d-r(D))/d$.
The Riemann--Roch theorem gives $d-r(D)=g-1-r(K-D)$. The $W_p^p$ bound
follows from \cref{lem:compare}, and the bound on $\Gamma$ follows as in the
proof of \cref{thm:crossover}(ii).
\end{proof}

When $D$ is non-special ($r(K-D)=-1$), \cref{thm:serre} reduces to
\cref{thm:defect}(a) and \cref{thm:crossover}(ii). When $D$ is special, the
\emph{dual} rank $r(K-D)\ge0$ lowers the bound $g/d$ by $(r(K-D)+1)/d$, and no
assumption $d\ge g$ is needed. The improvement must be compared with the
bound actually available, $\min(1,g/d)$, which may be capped at $1$. The
following proposition says exactly when the improvement is real.

\begin{proposition}[When the Serre form is sharper]\label{prop:serresharp}
Let $D$ be effective of degree $d\ge1$, and put $u_0=\min(1,g/d)$ and
$u=(d-r(D))/d=(g-1-r(K-D))/d$. Then $u\le u_0$, and $u<u_0$ if and only if
$r(D)\ge1$ and $r(K-D)\ge0$. In that case
\[
  u_0-u=\min\Bigl(\frac{r(D)}{d},\ \frac{r(K-D)+1}{d}\Bigr).
\]
\end{proposition}

\begin{proof}
$u<1$ if and only if $r(D)\ge1$, and $1-u=r(D)/d$. By Riemann--Roch,
$g/d-u=(r(K-D)+1)/d$, which is positive if and only if $r(K-D)\ge0$. Since
$u_0$ is the smaller of $1$ and $g/d$, the difference $u_0-u$ is the smaller of
the two differences.
\end{proof}

For example, on $G'$ the divisor $D=K$ has $d=2$, $r(K)=g-1=1$ and
$r(K-D)=r(0)=0$. Thus $u=\tfrac12<u_0=1$, and the bound $\tfrac12$ is attained.
Examples with $r(K-D)\ge1$ occur from genus four on. On $K_{3,3}$ ($g=4$), the
divisor $D=3v$ has $r(D)=1$ and $r(K-D)=1$, so $u=\tfrac23$ while $u_0=1$. On
$K_5$ ($g=6$), $D=4v$ has $r(D)=1$ and $r(K-D)=2$, so $u=\tfrac34$ while $u_0=1$;
on the Petersen graph ($g=6$) the same values occur for $D=2v+2w$ with $vw$ an
edge. In all three examples $\delta_{\TV}(D)=u$, so the
refined bound is attained and the unrefined bound is not.

\Cref{tab:serresharp} counts these cases among the $1784$ classes of
\cref{ssec:pdnum}. The refined bound is strictly sharper for $75$ classes, and it
is attained for $68$ of them. The $7$ exceptions show that the refined bound
need not be sharp either: on $K_{3,3}$ six classes of degree $5$, for example the class of $4v+w$
with $v,w$ on the same side, have $r(D)=2$, $r(K-D)=0$ and $u=\tfrac35$, but $\delta_{\TV}(D)=\tfrac25$, which
equals the vertex-multiplicity lower bound of \cref{thm:vertexrank}.

\begin{table}[!htbp]
\centering\small
\begin{tabular}{l c c c c c c}
\toprule
Graph & $g$ & classes & special & $u<u_0$ & attained & $\max r(K-D)$ \\
 & & & ($r(K-D)\ge0$) & & ($\delta_{\TV}=u$) & among $u<u_0$ \\
\midrule
$C_7$, $T_7$ & $1$, $0$ & $40$ & $0$ & $0$ & --- & --- \\
$G'$ & $2$ & $83$ & $8$ & $1$ & $1$ & $0$ \\
Honeycomb (2 hex.) & $2$ & $115$ & $11$ & $1$ & $1$ & $0$ \\
$G''$ & $3$ & $149$ & $37$ & $8$ & $8$ & $0$ \\
$K_4$ & $3$ & $62$ & $19$ & $5$ & $5$ & $0$ \\
$K_{3,3}$ & $4$ & $239$ & $104$ & $29$ & $23$ & $1$ \\
$K_5$ & $6$ & $221$ & $221$ & $26$ & $25$ & $2$ \\
Petersen & $6$ & $875$ & $875$ & $5$ & $5$ & $2$ \\
\midrule
Total & & $1784$ & $1275$ & $75$ & $68$ & \\
\bottomrule
\end{tabular}
\caption{Sharpness of the Serre-dual form (\cref{prop:serresharp}). ``Special''
counts classes with $r(K-D)\ge0$; on $K_5$ and on the Petersen graph every class
of degree $\le5<g$ is special. ``$u<u_0$'' counts classes for which
$(g-1-r(K-D))/d<\min(1,g/d)$. ``Attained'' counts those among them with
$\delta_{\TV}(D)=(g-1-r(K-D))/d$. Degrees $d\le5$ ($d\le4$ for the Petersen graph and
the honeycomb patch).}
\label{tab:serresharp}
\end{table}

\emph{What the Serre-dual form adds.} The Riemann--Roch theorem enters
\cref{thm:serre} in two distinct ways. The proof uses only the definition of
rank, which shows that $d-r(D)$ chips are the most that can be constrained. The
theorem then identifies this number with $g-1-r(K-D)$. Thus every effective
divisor in $\lin{K-D}$ (there are such exactly when $D$ is special) frees one
further chip, and the amount of freedom is measured on the Serre-dual side.
The inequality itself is a direct consequence of the definitions; the new
content is that the transport defect is controlled by the \emph{dual} rank, and
the lower bound and equality criterion below.

\begin{theorem}[Vertex multiplicities and an equality criterion]\label{thm:vertexrank}
For an effective divisor $D$ of degree $d\ge1$ and $v\in V$ let
$m_v(D)=\max_{D'\in\lin D}D'(v)$. Then $r(D)\le\min_v m_v(D)$ and
\[
  1-\frac{\min_v m_v(D)}{d}\;\le\;\delta_{\TV}(D)\;\le\;1-\frac{r(D)}{d}=\frac{g-1-r(K-D)}{d}.
\]
In particular, if $r(D)=\min_vm_v(D)$, then
$\delta_{\TV}(D)=(g-1-r(K-D))/d$ exactly. This holds whenever $r(D)=0$, in which case
$\delta_{\TV}(D)=1$.
\end{theorem}

\begin{proof}
If $r(D)\ge k$ then $\lin{D-kv}\neq\varnothing$, so some $D'\in\lin D$ has
$D'(v)\ge k$; hence $r(D)\le m_v(D)$ for every $v$. For the target
$E=d\,v\in\Div^d_+$ and any $D'\in\lin D$,
$\TV(\mu_{D'},\mu_E)=1-\min(D'(v),d)/d=1-D'(v)/d$, whose minimum over $\lin D$ is
$1-m_v(D)/d$. Maximizing over $v$ gives the lower bound, and the upper bound is
\cref{thm:serre}. If $r(D)=0$, there is a vertex $v$ with $\lin{D-v}=\varnothing$,
so $m_v(D)=0=r(D)$.
\end{proof}

The multiplicities $m_v(D)$ are computable by Dhar's algorithm, by the
following standard property of reduced divisors. We include a proof to fix the
conventions: firing $v$ maps $D\mapsto D-L\one_v$, and reducedness is as in
\cref{sec:divisors}.

\begin{lemma}\label{lem:reducedmax}
Let $D$ be effective and let $D_v$ be its $v$-reduced representative. Then
$D_v\in\lin D$ and $m_v(D)=D_v(v)$.
\end{lemma}

\begin{proof}
Since $\lin D\neq\varnothing$, \cref{thm:dhar} gives $D_v(v)\ge0$, so $D_v$ is
effective and $D_v\in\lin D$. Let $D'\in\lin D$ and write $D'=D_v-Lf$ with
$f\in\Z^V$. Let $S$ be the set where $f$ attains its maximum. Suppose $v\notin S$.
For $u\in S$, every term of $(Lf)(u)=\sum_{w\sim u}(f(u)-f(w))$ is non-negative, and
each edge from $u$ to $V\setminus S$ contributes at least $1$. Hence
$D'(u)\le D_v(u)-\#\{\text{edges from }u\text{ to }V\setminus S\}$, and by
reducedness some $u\in S$ has $D'(u)<0$, contradicting $D'\ge0$. Therefore
$v\in S$, so $(Lf)(v)\ge0$ and $D'(v)\le D_v(v)$.
\end{proof}

The next proposition settles the equality question completely in genus at most
one and partly in genus two. It also shows that the criterion of
\cref{thm:vertexrank} can fail although equality holds.

\begin{proposition}[Equality in small genus]\label{prop:smallgenus}
Let $D$ be effective of degree $d\ge1$.
\begin{enumerate}[label=\textup{(\roman*)}]
\item If $g=0$, then $r(D)=m_v(D)=d$ for every $v$, and $\delta_{\TV}(D)=0=(d-r(D))/d$.
\item If $g=1$, then $r(D)=d-1$ and $\delta_{\TV}(D)=1/d=(d-r(D))/d$. Thus the upper
bound is attained for every class.
\item On the cycle $C_n$ ($n\ge3$), $m_v(D)=d$ if $D\sim d\,v$ and $m_v(D)=d-1$
otherwise. Hence the criterion $r(D)=\min_vm_v(D)$ fails for exactly one class
of degree $d$ when $n\mid d$, namely the class of $d\,v$, which does not depend
on $v$. It holds for every class when $n\nmid d$.
\item If $g\ge1$ and $d=1$, then $r(D)=0$ and $\delta_{\TV}(D)=1$. If $g=2$, $d=2$
and $D\not\sim K$, then $r(D)=0$ and $\delta_{\TV}(D)=1$.
\end{enumerate}
\end{proposition}

\begin{proof}
(i) On a tree all divisors of degree $d$ are equivalent (\cref{cor:trees}), so
$d\,v\in\lin D$ for every $v$, and $r(D)=d$.

(ii) Since $d\ge1=2g-1$, $D$ is non-special and $r(D)=d-g=d-1$ by
\cref{cor:RR}(ii). The value $\delta_{\TV}(D)=1/d$ is \cref{thm:defect}(c).

(iii) Number the vertices $0,\dots,n-1$ in cyclic order. The map
$\mathrm{AJ}(x)=\sum_i i\,x(i)\bmod n$ on degree-zero divisors vanishes on
$L\Z^V$, because firing $i$ changes the sum by $(i-1)+(i+1)-2i\equiv0\pmod n$.
It is surjective, and $\lvert\Jac(C_n)\rvert=\kappa(C_n)=n$, so it identifies
$\Jac(C_n)$ with $\Z/n$. Since $r(D)=d-1$, every $m_v(D)\ge d-1$, and
$m_v(D)=d$ exactly when $d\,v\in\lin D$. Now
$\mathrm{AJ}(d\,v-d\,u)=d(v-u)\bmod n$. Taking $u=v+1$ shows that the classes of
$d\,v$, $v\in V$, all coincide if $n\mid d$, and that at least two of them
differ otherwise. In the second case $D$ is equivalent to at most one of two
distinct classes, so some $m_u(D)=d-1=r(D)$. In the first case the class of
$d\,v$ has $m_v=d$ for all $v$, so $\min_vm_v=d>r(D)$, while every other class
has $\min_vm_v=d-1$.

(iv) If $d=1$ and $r(D)\ge1$, then $D-u$ is equivalent to an effective divisor
of degree $0$, i.e.\ $D\sim u$, for every $u$. Then $u\sim w$ for all $u,w$,
and since the classes $[u-w]$ generate $\Jac(G)$, $\kappa(G)=1$ and $G$ is a
tree. So $r(D)=0$ when $g\ge1$. If $g=2$ and $d=2$, Riemann--Roch gives
$r(D)=1+r(K-D)$ with $\deg(K-D)=0$, so $r(K-D)=0$ if $D\sim K$ and $-1$ otherwise.
In both statements $r(D)=0$, and \cref{thm:vertexrank} gives $\delta_{\TV}(D)=1$.
\end{proof}

By (iii), the class of $3v$ on $C_3$ satisfies $\delta_{\TV}=\tfrac13=(d-r)/d$
although $\min_vm_v=3>r=2$. It is a second example, besides $4v_1$ on $K_4$, of
equality without the criterion. A brute-force check over all classes of degree
$d\le n+2$ on $C_3,\dots,C_6$ ($122$ classes) confirms (ii) and (iii) and finds the
criterion failing exactly in the $4$ predicted classes (supplementary script
\texttt{cycles.py}).

\emph{Proved results, computational evidence and conjecture.} We separate
three levels of evidence.
\begin{enumerate}[label=(\alph*)]
\item \emph{Proved:} the two-sided bound of \cref{thm:vertexrank}, and equality in
the upper bound whenever $r(D)=\min_vm_v(D)$, in particular whenever $r(D)=0$;
equality for every class when $g\le1$, and for $d=1$, or for $g=2$, $d=2$, $D\not\sim K$
(\cref{prop:smallgenus}).
\item \emph{Exhaustive computation on finitely many graphs.} All quantities
are integers or rationals with denominator $d$ and were compared in exact
rational arithmetic; the Riemann--Roch identity was checked for every class.
The full list of classes, with $r(D)$, $r(K-D)$, all $m_v(D)$, $\delta_{\TV}(D)$
and both bounds, is in the supplementary file \texttt{classes\_all.csv}. For
every linear equivalence class of effective divisors of degree at most $5$ on $C_7$, $G'$,
$G''$, $T_7$, $K_4$, $K_{3,3}$ and $K_5$, and of degree at most $4$ on the
Petersen graph and a two-hexagon honeycomb patch ($1784$ classes in total):
\begin{itemize}
\item the criterion $r(D)=\min_vm_v(D)$ holds for $1768$ classes;
\item equality in the \emph{rank bound}, $\delta_{\TV}(D)=(d-r(D))/d$, holds for
$1769$, namely those $1768$
and the single class $4v_1$ on $K_4$;
\item $\delta_{\TV}$ equals one of the two bounds of \cref{thm:vertexrank} for
$1779$ classes;
\item the five classes attaining neither bound are $5v_1$, $4v_1+v_j$ ($j=2,3,4$)
on $K_4$, and $5v_1$ on $K_5$.
\item the $1233$ classes with $g\ge1$ and $d<g$ all satisfy $\delta_{\TV}(D)\ge1/d$, in
agreement with \cref{thm:defect}(c);
\item for comparison, the coarser bound $\min(1,g/d)$ of \cref{thm:defect}(a) is
attained for only $1701$ classes; the two counts must not be confused;
\end{itemize}
The $15$ classes without equality include the non-special class $K+v_1$ on
$G''$ ($d=5=2g-1$, $\delta_{\TV}=\tfrac25<\tfrac35$). There the lower bound is
attained ($\min_vm_v=3$), and the bound $g/d$ is \emph{not} attained
throughout the non-special range of $G''$. All matched-genus graphs of
\cref{tab:matched} attain equality. Sorted by genus: equality holds for all
$40$ classes of genus $\le1$, as proved; for all $198$ classes on the two
genus-two graphs ($G'$ and the honeycomb patch), which goes beyond what
\cref{prop:smallgenus}(iv) proves; and the first failures occur in genus three
($G''$, $K_4$).
\item \emph{Open:} the criterion is sufficient but, by the class $4v_1$ on
$K_4$, not necessary. Characterizing equality, and the classes attaining
neither bound, remains open (\cref{sec:conclusion}). We state no conjecture
beyond the finite evidence in (b).
\end{enumerate}

\begin{remark}[Scope of the degree assumptions]\label{rem:scope}
\Cref{tab:hyp} lists the degree hypotheses of the core results. Three
thresholds occur. $d\ge1$ suffices for everything that uses only $r(D)\ge0$,
the definitions, or the existence of \emph{one} effective divisor outside
$[D]$, which holds for every $d\ge1$ when $g\ge1$ (proof of
\cref{thm:defect}(c)). $d\ge g$ is needed exactly where a result uses that
\emph{every} class of degree $d$ contains an effective divisor: the Cayley
bound of \cref{thm:AJ}(ii) and the quantitative form \cref{thm:phase}(c). $d\ge2g-1$ (the non-special
range) is where $r(D)=d-g$ holds exactly (\cref{cor:RR}(ii)), so that the bounds
of \cref{thm:serre} and \cref{thm:defect}(a) coincide. The condition $d\ge g$
cannot be dropped: when $g\ge1$ and $0\le d<g$, subtracting an effective divisor
of degree $g-1-d$ from a divisor of degree $g-1$ and rank $-1$, which exists by
\cite[Sect.~3]{BakerNorine2007}, gives a class of degree $d$ with no effective
representative. The Cayley lower bound of \cref{thm:AJ}(ii) does fail for $d<g$: on $G''$ at
$d=1$ one has $d\,\delta_{W_1}=2<\rho=4$, and on $G'$ at $d=1$ one has $2<\rho=3$.
\end{remark}

\begin{table}[!htbp]
\centering\small
\renewcommand{\arraystretch}{1.12}
\begin{tabular}{>{\raggedright\arraybackslash}p{6.2cm}>{\centering\arraybackslash}p{2.3cm}>{\raggedright\arraybackslash}p{5.2cm}}
\toprule
Result & Degree hypothesis & Where the hypothesis enters \\
\midrule
\Cref{cor:RR}(ii): $r(D)=d-g$ & $d\ge2g-1$ & $\deg(K-D)<0$ \\
\Cref{thm:defect}(a),(b): $\delta_{\TV}\le\min(1,g/d)$; trees & $d\ge1$ & $r(D)\ge\max(0,d-g)$ \\
\Cref{thm:defect}(c): $\delta_{\TV}\ge1/d$ ($g\ge1$) & $d\ge1$ & one effective divisor outside $[D]$ \\
\Cref{thm:AJ}(i): concentrated targets, upper bound & $d\ge1$ & targets $E=d\,w$ always effective \\
\Cref{thm:AJ}(ii): $\rho(G)\le d\,\delta_{W_1}$ & $d\ge\max(g,1)$ & every class has an effective divisor \\
\Cref{thm:crossover}: limits and window & $d\ge1$ & upper side via \cref{thm:defect}(a) \\
\Cref{thm:phase}(a): $g=0\Rightarrow\Gamma\le0$ & $d\ge1$ & --- \\
\Cref{thm:phase}(b): positive gap for $g\ge1$ & $d\ge1$ & one effective divisor outside $[D]$ \\
\Cref{thm:phase}(c): gap $\ge\rho(G)/d-\varepsilon\ln_q(n^2)$ & $d\ge\max(g,1)$ & choice of $E$ in a far class \\
\Cref{thm:serre,thm:vertexrank} & $d\ge1$ & $r(D)\ge0$ \\
\Cref{thm:TI,thm:minimax,cor:threelevels} & $d\ge1$ & --- \\
\bottomrule
\end{tabular}
\caption{Degree hypotheses of the core results. In the non-special range
$d\ge2g-1$, $(d-r(D))/d=g/d$, so the refined and unrefined upper bounds agree.}
\label{tab:hyp}
\end{table}

\subsection{Decomposition of the gap: topology versus information}

\begin{definition}
For effective $D,E$ of degree $d$ put
\begin{align*}
  \mathsf T_{q,\gamma}(D,E)&=\min_{D'\in\lin D}\Wq(\mu_{D'},\mu_E)-\min_{E'\in\Div^d_+}\Wq(\mu_{E'},\mu_E),\\
  \mathsf I_{q,\gamma}(E)&=\min_{E'\in\Div^d_+}\Wq(\mu_{E'},\mu_E)-\Wq(\mu_E,\mu_E).
\end{align*}
\end{definition}

The identity $\Gamma=\mathsf T+\mathsf I$ below is definitional, a telescoping
sum, and we do not present it as a structure theorem. Its content lies in the
properties of the two terms, each derived separately: the sign of $\mathsf T$ from
the inclusion $\lin D\subseteq\Div^d_+$, the sign of $\mathsf I$ from
$E\in\Div^d_+$, the independence of $\mathsf I$ from $D$, and the limits from
(T1)--(T4). It is a tool for reading the gap, used in \cref{rem:existpersist}
and \cref{ssec:pdnum}.

\begin{theorem}[Topology--information decomposition]\label{thm:TI}
$\Gamma_{q,\gamma}(D,E)=\mathsf T_{q,\gamma}(D,E)+\mathsf I_{q,\gamma}(E)$, where
\begin{enumerate}[label=\textup{(\roman*)}]
\item $\mathsf T\ge0$; $\mathsf T\equiv0$ if $g=0$; and
$\lim_{\gamma\to0}\mathsf T_{q,\gamma}(D,E)=N_E/d$;
\item $\mathsf I\le0$; $\mathsf I$ does not depend on $D$; and
$-\frac\gamma q\bigl(\ln_q(n^2)-S_q(\mu_E)\bigr)\le\mathsf I_{q,\gamma}(E)\le0$, so
$\mathsf I\to0$ as $\gamma\to0$;
\item $\lim_{\gamma\to\infty}\frac q\gamma\,\mathsf I_{q,\gamma}(E)=H_q(\mu_E,\mu_E)-\max_{E'\in\Div^d_+}H_q(\mu_{E'},\mu_E)$
and
$\lim_{\gamma\to\infty}\frac q\gamma\,\mathsf T_{q,\gamma}(D,E)=\max_{E'\in\Div^d_+}H_q(\mu_{E'},\mu_E)-\max_{D'\in\lin D}H_q(\mu_{D'},\mu_E)\ge0$.
\end{enumerate}
In particular, $\Gamma_{q,\gamma}(D,E)>0$ if and only if $\mathsf T_{q,\gamma}(D,E)>\lvert\mathsf I_{q,\gamma}(E)\rvert$:
a crossover is a temperature at which the topological term $\mathsf T$ and the
entropic term $\mathsf I$ balance.
\end{theorem}

\begin{proof}
The identity is a telescoping sum. $\mathsf T\ge0$ because
$\lin D\subseteq\Div^d_+$, with equality of the two sets on trees
(\cref{cor:trees}). $\mathsf I\le0$ because $E\in\Div^d_+$. Since
$W_1-\varepsilon\ln_q(n^2)\le\mathcal T_{q,\varepsilon}\le W_1$ and
$\min_{E'}W_1(\mu_{E'},\mu_E)=0$, the limit in (i) and the lower bound in (ii)
follow as in the proof of \cref{thm:crossover}. The limits in (iii) follow
from $\mathcal T_{q,\varepsilon}/\varepsilon\to-H_q$ uniformly over the finite
families involved.
\end{proof}

Thus the two forces that compete in \cref{sec:crossover} separate into two
terms of definite sign. $\mathsf T$ is purely topological: it vanishes on
trees and at low temperature equals the chip-move defect. $\mathsf I$ is
purely informational: it is blind to $D$ and hence to the Jacobian, and it is
negative because entropy rewards configurations that are more spread than the
target.

\subsection{Integrality: what dual certificates certify}

\Cref{prop:invisible} showed that over $\R$ chip-firing connects all
configurations of equal degree. The dual problem~\eqref{eq:dual} makes this
quantitative.

\begin{theorem}[Minimax and integrality gap]\label{thm:minimax}
Let $q>0$ and $\gamma>0$. Fix a probability measure $\nu$ on $V$ and an effective
$D$ of degree $d\ge1$, and let
$\mathcal C_D=\operatorname{conv}\{\mu_{D'}:D'\in\lin D\}$. Then
\[
  \min_{D'\in\lin D}\Wq(\mu_{D'},\nu)\;\ge\;\min_{\mu\in\mathcal C_D}\Wq(\mu,\nu)
  =\sup_{\alpha,\beta}\Bigl\{\tfrac1d\min_{D'\in\lin D}\langle\alpha,D'\rangle+\langle\beta,\nu\rangle
  -\gamma\sum_{i,j}F_q^*\Bigl(\tfrac{\alpha_i+\beta_j-c_{ij}}{\gamma}\Bigr)\Bigr\},
\]
and the inner minimum has the firing-script form
\[
  \min_{D'\in\lin D}\langle\alpha,D'\rangle=\langle\alpha,D\rangle-\max\bigl\{\langle L\alpha,f\rangle:\ f\in\Z^V,\ D-Lf\ge0\bigr\},
\]
in which potentials and scripts are paired by the Dirichlet form
\[
  \langle L\alpha,f\rangle=\sum_{uv\in E}(\alpha(u)-\alpha(v))(f(u)-f(v)).
\]
If
the scripts are allowed to be real, $f\in\R^V$, the inner minimum becomes
$d\min_v\alpha(v)$, independent of $D$, and the right-hand side becomes
$\min_{\mu}\Wq(\mu,\nu)$ over all probability measures.
\end{theorem}

\begin{proof}
We check each step, including the factor $\tfrac1d$.

\emph{Step 1 (the inequality).} Every $D'\in\lin D$ gives the point
$\mu_{D'}=D'/d\in\mathcal C_D$, so the minimum over the finite set $\lin D$ is a
minimum over a subset of $\mathcal C_D$ and cannot be smaller.

\emph{Step 2 (convexity and existence).} If $P_0\in\Pi(\mu_0,\nu)$ and
$P_1\in\Pi(\mu_1,\nu)$ are optimal, then $(1-t)P_0+tP_1\in\Pi((1-t)\mu_0+t\mu_1,\nu)$,
and convexity of $J_{q,\gamma}$ gives convexity of $\mu\mapsto\Wq(\mu,\nu)$. It is
continuous, by Berge's maximum theorem, because $J_{q,\gamma}$ is continuous and the
compact set $\Pi(\mu,\nu)$ depends continuously on $\mu$. So the minimum over the
compact set $\mathcal C_D$ is attained.

\emph{Step 3 (exchange of $\min$ and $\sup$).} By \cref{lem:dual},
\[\min_{\mu\in\mathcal C_D}\Wq(\mu,\nu)=\min_{\mu\in\mathcal C_D}\sup_{\alpha,\beta}\Phi(\mu;\alpha,\beta).\]
Restrict $(\alpha,\beta)$ to the convex set $Y$ on which every $F_q^*$ term is
finite; this does not change the supremum. On $\mathcal C_D\times Y$, $\Phi$ is
affine and continuous in $\mu$, and concave and upper semicontinuous in
$(\alpha,\beta)$, because $F_q^*$ is convex and lower semicontinuous. Since
$\mathcal C_D$ is compact and convex, Sion's theorem~\cite{Sion1958} gives
$\min_{\mu}\sup_{\alpha,\beta}\Phi=\sup_{\alpha,\beta}\min_{\mu}\Phi$.

\emph{Step 4 (the factor $\tfrac1d$).} For fixed $(\alpha,\beta)$, only the term
$\langle\alpha,\mu\rangle$ depends on $\mu$. A linear function on the polytope
$\mathcal C_D$ attains its minimum at a vertex, and every vertex has the form
$\mu_{D'}=D'/d$ with $D'\in\lin D$. Hence
$\min_{\mu\in\mathcal C_D}\langle\alpha,\mu\rangle=\tfrac1d\min_{D'\in\lin D}\langle\alpha,D'\rangle$.
The factor $\tfrac1d$ is thus the normalization $\mu_{D'}=D'/d$, and nothing else.

\emph{Step 5 (firing scripts).} $D'\in\lin D$ means $D'=D-Lf$ with $f\in\Z^V$ and
$D-Lf\ge0$. Since $L$ is symmetric,
$\langle\alpha,D'\rangle=\langle\alpha,D\rangle-\langle L\alpha,f\rangle$, which gives
the firing-script form. Expanding $(L\alpha)(v)=\sum_{u\sim v}(\alpha(v)-\alpha(u))$
and collecting the two endpoints of every edge gives the Dirichlet form.

\emph{Step 6 (real scripts).} Let
$\mathcal P_D^{\R}=\{D-Lf:\ f\in\R^V,\ D-Lf\ge0\}$. If $x\ge0$ and $\sum_vx(v)=d$,
then $x-D$ has degree $0$, so $x-D\in L(\R^V)$ by \cref{prop:invisible}, that is
$x\in\mathcal P_D^{\R}$. Conversely every element of $\mathcal P_D^{\R}$ is
non-negative of degree $d$. Hence $\mathcal P_D^{\R}=d\,\Delta_V$, and
$\min_{x\in d\Delta_V}\langle\alpha,x\rangle=d\min_v\alpha(v)$, attained at
$x=d\,\delta_v$. After the factor $\tfrac1d$ of Step~4 this is
$\min_v\alpha(v)=\min_{\mu\in\Delta_V}\langle\alpha,\mu\rangle$. Steps~3--4 applied
to the compact convex set $\Delta_V$ show that the resulting supremum equals
$\min_{\mu\in\Delta_V}\Wq(\mu,\nu)$.
\end{proof}

\begin{remark}[Three objects, not two]\label{rem:threeobjects}
Step~6 shows that relaxing the integrality of firing scripts is \emph{not} the
same as taking the convex hull. Three sets of probability measures appear:
\begin{enumerate}[label=(\arabic*)]
\item the finite set $\{\mu_{D'}:D'\in\lin D\}$, i.e.\ the lattice coset
$D+L\Z^V$ intersected with the non-negative orthant, divided by $d$;
\item its convex hull $\mathcal C_D$, the \emph{integer hull} of that coset in
the simplex. It still depends on the class $[D]$;
\item the simplex $\Delta_V=\tfrac1d\,\mathcal P_D^{\R}$, which is what real
scripts give. It forgets the coset entirely.
\end{enumerate}
In general $(1)\subsetneq(2)\subsetneq(3)$, and relaxing scripts jumps
from (1) directly to (3).

A simple example shows the difference. Let $G$ be $2$-edge-connected, for
instance the cycle $C_n$, and let $D=v$ with $d=1$. Then $\lin D=\{v\}$. Indeed,
if $u-v=Lf$ with $u\ne v$, let $S$ be the set where $f$ is maximal. If
$S\ne V$, then $\sum_{w\in S}(Lf)(w)\ge\lvert\partial S\rvert\ge2$, while
$\sum_{w\in S}(u-v)(w)\le1$, a contradiction. So $f$ is constant and $u=v$.
Hence $\mathcal C_D=\{\delta_v\}$ is a single point, whereas $\Delta_V$ is the
whole simplex. For $\nu=\delta_u$ with $u\ne v$ and small $\gamma$,
$m_{\Z}=m_{\mathcal C}\approx\dist_G(u,v)$ while $m_{\R}\approx0$. In this example
the whole topological effect is a class gap, and the integrality gap is zero.
The opposite situation, a vanishing class gap and a positive integrality gap,
occurs on $C_7$ at $d=4$ (\cref{tab:levels}).
\end{remark}

The structure is as follows. (1) The dual problem gives exactly the optimal
value of the convex relaxation over $\mathcal C_D$. (2) The additional cost of
restricting to the integral representatives is the \emph{integrality gap}
$\min_{\lin D}\Wq-\min_{\mathcal C_D}\Wq\ge0$, defined as the difference between
the discrete problem and its convex relaxation. (3) This gap cannot be recovered
from the optimal value of the relaxation. The relaxation itself still depends on
the class: $\mathcal C_D$ is the convex hull of the linear system $\lin D$, so
arithmetic information enters the dual through the inner minimum over $\lin D$.
What the dual cannot certify is the integrality gap in (2). Only if the
integrality of firing scripts is dropped, so that $\mathcal C_D$ is replaced by
the whole simplex, does the dependence on $[D]$ disappear altogether. The arithmetic
information originates in the primal lattice $D+L\Z^V$. The dual sees part of it,
through the convex hull $\mathcal C_D$ and the torsion $\widehat{\Jac(G)}$ of
\cref{prop:invisible}, and \cref{cor:threelevels} makes the split precise.

\begin{corollary}[Three levels of relaxation]\label{cor:threelevels}
Fix $\nu$ and an effective $D$ of degree $d$. Put
\[
  m_{\Z}=\min_{D'\in\lin D}\Wq(\mu_{D'},\nu),\qquad
  m_{\mathcal C}=\min_{\mu\in\mathcal C_D}\Wq(\mu,\nu),\qquad
  m_{\R}=\min_{\mu\in\Delta_V}\Wq(\mu,\nu),
\]
where $\Delta_V$ is the simplex of all probability measures on $V$. Then
$m_{\Z}\ge m_{\mathcal C}\ge m_{\R}$, and we call
$\Delta_{\rm int}=m_{\Z}-m_{\mathcal C}\ge0$ the \emph{integrality gap} and
$\Delta_{\rm cls}=m_{\mathcal C}-m_{\R}\ge0$ the \emph{class gap}.
\begin{enumerate}[label=\textup{(\roman*)}]
\item Dual potentials certify exactly $m_{\mathcal C}$ (\cref{thm:minimax}). Lower
bounds obtained from them can therefore detect $\Delta_{\rm cls}$, but never
$\Delta_{\rm int}$.
\item $\Delta_{\rm cls}$ depends on $[D]$ only through the polytope $\mathcal C_D$.
It vanishes if and only if $\mathcal C_D$ contains a minimizer of
$\Wq(\cdot,\nu)$ over $\Delta_V$.
\item Replacing integral firing scripts by real ones replaces $\mathcal C_D$ by
$\Delta_V$, and both gaps vanish.
\end{enumerate}
\end{corollary}

\begin{proof}
The inequalities follow from $\mu(\lin D)\subseteq\mathcal C_D\subseteq\Delta_V$, and (i)
is \cref{thm:minimax}. For (ii), $\Wq(\cdot,\nu)$ is continuous and $\mathcal C_D$
is compact, so $m_{\mathcal C}$ is attained, and $m_{\mathcal C}=m_{\R}$ exactly when
some point of $\mathcal C_D$ attains $m_{\R}$. (iii) is the last statement of
\cref{thm:minimax}.
\end{proof}

Thus the arithmetic information carried by the class splits into two parts.
The class gap is retained by the convex relaxation and is visible to the dual.
The integrality gap is lost under convexification. Which part dominates
depends on the graph. In the computations of \cref{ssec:pdnum} (with $\nu=\mu_E$
the worst target) the class gap vanishes on $C_7$, so the whole topological
effect is an integrality gap there; on $G'$ both parts are present
(\cref{tab:levels}).

\Cref{thm:minimax} and \cref{cor:threelevels} accomplish two different tasks,
which we keep apart. The first is a \emph{representation}: the optimal value over
$\mathcal C_D$ equals a supremum over dual potentials. This is exact and is
proved. The second is an \emph{evaluation}: when are the two gaps positive, and
how large are they? The corollary only defines them and gives their signs. The
next proposition gives the quantitative information we can prove, namely exact
criteria for vanishing and exact low-temperature values.

\begin{proposition}[When the gaps vanish, and their low-temperature values]\label{prop:gapsvanish}
Fix an effective $D$ of degree $d\ge1$, and write $\Delta_{\rm int}$ and
$\Delta_{\rm cls}$ for the gaps of \cref{cor:threelevels}, as functions of
$(\nu,q,\gamma)$.
\begin{enumerate}[label=\textup{(\roman*)}]
\item $\Delta_{\rm int}(\nu,q,\gamma)=0$ if and only if some $\mu_{D'}$, $D'\in\lin D$,
minimizes $\Wq(\cdot,\nu)$ over $\mathcal C_D$. In particular
$\Delta_{\rm int}\equiv0$ when $\lin D$ has one element, for instance when
$D=v$ on a $2$-edge-connected graph (\cref{rem:threeobjects}).
\item $\Delta_{\rm cls}(\nu,q,\gamma)=0$ for all $\nu$, $q$ and $\gamma$ if and only if
$\mathcal C_D=\Delta_V$, if and only if $d\,v\in\lin D$ for every $v\in V$. This
holds on every tree. On $C_n$ it holds exactly for the class of $d\,v$ when
$n\mid d$ (\cref{prop:smallgenus}(iii)). Otherwise, for every $q$, there are $\nu$
and $\gamma_0>0$ with $\Delta_{\rm cls}(\nu,q,\gamma)>0$ for all $\gamma<\gamma_0$.
\item Let $\nu=\mu_E$ with $E\in\Div^d_+$, and put
$\Delta^0_{\rm cls}=\min_{\mu\in\mathcal C_D}W_1(\mu,\mu_E)$. As $\gamma\to0$,
\[
  \Delta_{\rm cls}\to\Delta^0_{\rm cls},\qquad
  \Delta_{\rm int}\to\frac{N_E}d-\Delta^0_{\rm cls}.
\]
$\Delta^0_{\rm cls}=0$ if and only if $\mu_E\in\mathcal C_D$. In that case the whole
low-temperature gap $N_E/d$ is an integrality gap, and it is positive if and
only if $E\notin\lin D$.
\end{enumerate}
\end{proposition}

\begin{proof}
(i) $m_{\Z}\ge m_{\mathcal C}$, with equality exactly when the minimum over $\mathcal C_D$
is attained at one of the points $\mu_{D'}$.

(ii) If $\mathcal C_D=\Delta_V$, then $m_{\mathcal C}=m_{\R}$ for every $\nu$. Each $\delta_v$ is
an extreme point of $\Delta_V$, so $\delta_v\in\mathcal C_D\subseteq\Delta_V$ forces $\delta_v$
to be one of the generating points $\mu_{D'}$, that is $d\,v\in\lin D$; and if
all $d\,v$ lie in $\lin D$, then $\mathcal C_D\supseteq\operatorname{conv}\{\delta_v\}=\Delta_V$.
Suppose now $\delta_v\notin\mathcal C_D$ for some $v$, and take $\nu=\delta_v$. On the
compact set $\mathcal C_D$ the continuous function $\mu\mapsto\mu(v)$ is $<1$, so
$\eta=1-\max_{\mathcal C_D}\mu(v)>0$. Every $\mu\in\mathcal C_D$ has $W_1(\mu,\delta_v)\ge
1-\mu(v)\ge\eta$. With $\mathcal T$ as in the proof of \cref{thm:crossover}, (T2)
gives $m_{\mathcal C}\ge\eta-\varepsilon\ln_q(n^2)+\gamma c_{q,N}$, and (T1) at $\mu=\delta_v$
gives $m_{\R}\le0+\gamma c_{q,N}$. Hence $\Delta_{\rm cls}\ge\eta-\varepsilon\ln_q(n^2)>0$ for
$\gamma<q\,\eta/\ln_q(n^2)$. The statements on trees and cycles follow from
\cref{cor:trees} and \cref{prop:smallgenus}(iii).

(iii) By (T1)--(T2), $\lvert\Wq(\mu,\nu)-\gamma c_{q,N}-W_1(\mu,\nu)\rvert\le\varepsilon\ln_q(n^2)$
uniformly in $\mu$. Minima over the fixed sets $\lin D$, $\mathcal C_D$ and
$\Delta_V$ therefore converge, after subtracting $\gamma c_{q,N}$, to the
corresponding minima of $W_1(\cdot,\mu_E)$. These are $N_E/d$, $\Delta^0_{\rm cls}$ and
$0$ (at $\mu=\mu_E$). Since $W_1$ is a metric and $\mathcal C_D$ is compact,
$\Delta^0_{\rm cls}=0$ exactly when $\mu_E\in\mathcal C_D$. Finally $N_E=0$ exactly when
$E\in\lin D$.
\end{proof}

The two graphs of \cref{tab:levels} illustrate (iii). We computed
$\Delta^0_{\rm cls}$ exactly by a linear program over convex combinations of
$\lin D$ (script \texttt{lowtemp\_split.py}). On $C_7$, $\mu_E\in\mathcal C_D$, so
$\Delta^0_{\rm cls}=0$ and $\Delta_{\rm int}\to N_E/d=\tfrac34$; the tabulated values
$0.729$ ($q=1$) and $0.748$ ($q=2$) at $\gamma=0.03$ approach this limit. On $G'$,
$N_E/d=\tfrac54$ splits exactly as $\Delta^0_{\rm cls}=\tfrac78$ and
$\Delta_{\rm int}\to\tfrac38$, in agreement with the tabulated $0.844$--$0.870$
and $0.375$. Criteria for the sign of the two gaps at \emph{finite} temperature,
beyond (i) and (ii), remain open (\cref{sec:conclusion}).

\subsection{Sparsity is a boundary phenomenon of the dual}

\begin{proposition}[Essential smoothness and support]\label{prop:legendre}
Let $q>0$. The function $F_q$ is finite and continuous on $\operatorname{dom}F_q=[0,\infty)$
and differentiable on the interior $(0,\infty)$. It is essentially smooth in
the sense of \cite[Sect.~26]{Rockafellar1970}, i.e.\ $\lvert F_q'(x)\rvert\to\infty$ as
$x\downarrow0$, the only boundary point of its domain, if and only if
$0<q\le1$. Being strictly convex on $(0,\infty)$, it is then of Legendre type.
For $q>1$, $F_q'$ extends continuously to $x=0$ with the finite value
$F_q'(0)=2q-1-\frac1{q-1}$. Consequently:
\begin{enumerate}[label=\textup{(\roman*)}]
\item for $0<q\le1$ the dual map $(F_q^*)'$ is a bijection from the interior of
$\operatorname{dom}F_q^*$ onto $(0,\infty)$, and every optimal plan is
positive wherever $\mu_i\nu_j>0$;
\item for $q>1$ one has $\operatorname{dom}F_q^*=\R$ and $(F_q^*)'=0$ on
$(-\infty,F_q'(0)]$, so that
$\operatorname{supp}P^\star=\{(i,j):(\alpha_i+\beta_j-c_{ij})/\gamma>F_q'(0)\}$
is cut out by the boundary of the dual domain.
\end{enumerate}
Hence $q_{\rm supp}(\gamma)=\inf\{q:\ P^\star_{q,\gamma}\text{ has a zero entry with }\mu_i\nu_j>0\}$
satisfies $q_{\rm supp}(\gamma)\ge1$ for every $\gamma$.
\end{proposition}

\begin{proof}
By \cref{prop:dual}(i), $F_q'(0^+)=-\infty$ if $q\le1$ and
$F_q'(0)=2q-1-\frac1{q-1}$ is finite if $q>1$. (i) is the standard
characterization of Legendre-type functions~\cite[Thm.~26.5]{Rockafellar1970},
combined with \cref{thm:support}(i). (ii) follows from
\cref{prop:dual}(ii) and \cref{thm:support}(ii).
\end{proof}

The support transition is therefore a statement about where the dual
coordinate $\theta=(\alpha_i+\beta_j-c_{ij})/\gamma$ meets the edge of the
domain of $F_q^*$. The gap-sign transition, by \cref{thm:TI,thm:minimax},
is a statement about the primal lattice. The two need not coincide, and
\cref{ssec:pdnum} shows that they do not.

\emph{Where the support changes: the KKT threshold.} For $q>1$,
\cref{prop:legendre}(ii) and the formula for $P^\star$ in \cref{lem:dual} give
an exact test for each cell: at optimal potentials,
\begin{equation}\label{eq:kktzero}
  P^\star_{ij}=0\iff\alpha_i+\beta_j-c_{ij}\le\gamma\,F_q'(0)
  =\gamma\Bigl(2q-1-\frac1{q-1}\Bigr).
\end{equation}
A support transition happens when the potentials make one of these
inequalities an equality. \Cref{prop:legendre} only says that zeros are
\emph{possible} for $q>1$. Whether, and at which $\gamma$, they occur depends on
the data. The following example computes the threshold in closed form.

\begin{example}[A $2\times2$ support threshold]\label{ex:twobytwo}
Let $a=b=(\tfrac12,\tfrac12)$ and $C=\bigl(\begin{smallmatrix}0&c\\c&0\end{smallmatrix}\bigr)$
with $c>0$. Every plan has the form
$P_t=\bigl(\begin{smallmatrix}1/2-t&t\\t&1/2-t\end{smallmatrix}\bigr)$, $0\le t\le\tfrac12$, and
\[
  \tfrac12\tfrac{d}{dt}J_{q,\gamma}(P_t)=c-\gamma\bigl(F_q'(\tfrac12-t)-F_q'(t)\bigr)
  =c-\frac{\gamma}{q-1}\Bigl(\bigl(\tfrac12-t\bigr)^{q-1}-t^{q-1}\Bigr)\quad(q\ne1).
\]
The function $t\mapsto J_{q,\gamma}(P_t)$ is convex.
\begin{itemize}
\item For $q\le1$ the derivative tends to $-\infty$ as $t\downarrow0$, so the
optimal $t^\star$ is positive for every $\gamma>0$. For $q=1$,
$t^\star=\tfrac12\,(1+e^{c/\gamma})^{-1}$.
\item For $q>1$ the derivative at $t=0$ equals $c-\gamma\,2^{1-q}/(q-1)$. Hence
\[
  t^\star=0\iff\gamma\le\gamma_{\rm supp}(q)=c\,(q-1)\,2^{\,q-1}.
\]
For $q=2$, $\gamma_{\rm supp}=2c$ and $t^\star=\tfrac14-\tfrac{c}{2\gamma}$ for $\gamma>2c$.
\end{itemize}
At $\gamma=\gamma_{\rm supp}(q)$ itself the diagonal plan is still optimal
($t^\star=0$); for every $\gamma>\gamma_{\rm supp}(q)$ all four entries are positive.
The support therefore changes across the threshold, from the diagonal (two
cells) for $\gamma\le\gamma_{\rm supp}$ to all four cells for $\gamma>\gamma_{\rm supp}$. In the Tsallis temperature $\varepsilon=\gamma/q$ the threshold is
$c\,(q-1)\,2^{q-1}/q$. It equals $c$ at $q=2$, $\tfrac83c$ at $q=3$, and tends
to $0$ as $q\downarrow1$, in agreement with the absence of zeros at $q=1$. A
numerical minimization confirms these values (supplementary script
\texttt{twobytwo.py}). The example involves no linear equivalence: it shows a
support transition with no gap $\Gamma$ in sight. This is the simplest instance
of the dissociation observed in \cref{ssec:pdnum}.
\end{example}

\subsection{Degeneracy of optimal transport: primal faces and dual components}

The following theorem is an application rather than a core result. Parts (i)
and (ii), including the dimension formula $\dim O=\beta_1(T^\ast)$, are standard
consequences of strict complementarity \cite{GoldmanTucker1956}, and we claim
no novelty for them. What is added is the link in parts (iii)--(v) between the
selection by the regularizer $F_q$ and the cycle structure $\beta_1$ of the
selected support: for $q\le1$ the limit plan keeps every cycle of $T^\ast$, for
$q>1$ it may lose some. This is what the experiments of \cref{ssec:reg} test.

\begin{theorem}[Optimal faces and their selection]\label{thm:face}
Let $a,b>0$, let $O\subseteq\Pi(a,b)$ be the set of optimal plans of the
unregularized problem, and let $T^\ast$ be the support of a plan in the relative
interior of $O$. Regard supports as bipartite graphs on the $m+n$ rows and
columns.
\begin{enumerate}[label=\textup{(\roman*)}]
\item $O=\{P\in\Pi(a,b):\operatorname{supp}P\subseteq T^\ast\}$ and
$\dim O=\beta_1(T^\ast)=\lvert T^\ast\rvert-(m+n)+c(T^\ast)$.
\item The optimal dual potentials, modulo $(\alpha,\beta)\mapsto(\alpha+t,\beta-t)$,
form a set of dimension $c(T^\ast)-1$.
\item For every $q>0$, $P^\star_{q,\gamma}\to P_0(q)=\operatorname{argmin}_{P\in O}\sum_{ij}F_q(P_{ij})$
as $\gamma\to0$.
\item If $q\le1$, then $\operatorname{supp}P_0(q)=T^\ast$, so
$\beta_1(\operatorname{supp}P_0(q))=\dim O$. If $q>1$, then
$\operatorname{supp}P_0(q)\subseteq T^\ast$.
\item For every $q>0$, $\beta_1(\operatorname{supp}P_0(q))\le\beta_1(T^\ast)=\dim O$,
with equality if $q\le1$.
\end{enumerate}
\end{theorem}

\begin{proof}
(i) and (ii). By the Goldman--Tucker theorem~\cite{GoldmanTucker1956} there is a strictly complementary
pair: a relative-interior optimal plan with support $T^\ast$ and optimal
potentials with $\alpha_i+\beta_j=c_{ij}$ exactly on $T^\ast$. Every feasible
plan supported in $T^\ast$ satisfies complementary slackness with these
potentials and is therefore optimal, and every optimal plan is supported in
$T^\ast$; this gives the description of $O$. Its affine hull is cut out by
the marginal equations on the coordinates in $T^\ast$, whose matrix is the
incidence matrix of the bipartite graph $T^\ast$, of rank $(m+n)-c(T^\ast)$; this
gives $\dim O$. The optimal potentials form the set
$\{\alpha_i+\beta_j=c_{ij}\text{ on }T^\ast,\ \le c_{ij}\text{ elsewhere}\}$, which
by strict complementarity has the same dimension as its affine hull, namely
$(m+n)-\operatorname{rank}=c(T^\ast)$; removing the shift leaves $c(T^\ast)-1$.
(iii) Write $\Phi=\sum F_q$ and $P_\gamma=P^\star_{q,\gamma}$. Every limit point of
$P_\gamma$ lies in $O$. For $P_0\in O$, optimality gives
$\langle C,P_\gamma\rangle+\gamma\Phi(P_\gamma)\le\langle C,P_0\rangle+\gamma\Phi(P_0)
\le\langle C,P_\gamma\rangle+\gamma\Phi(P_0)$, hence $\Phi(P_\gamma)\le\Phi(P_0)$, and
every limit point minimizes $\Phi$ on $O$. That minimizer is unique by strict
convexity. (iv) For $q\le1$, let $Q\in O$ have support $T^\ast$. If $P_0(q)$
vanished at some entry of $T^\ast$, the derivative of $\Phi$ at $P_0(q)$ in the
feasible direction $Q-P_0(q)$ would be $-\infty$, contradicting minimality. (v)
Removing edges from a graph on a fixed vertex set never increases its cycle
number, so (v) follows from (iv) and (i).
\end{proof}

\Cref{thm:face} gives the bipartite cycle number $\beta_1$ of a support a
precise meaning, and it involves three distinct objects that should not be
confused:
\begin{itemize}
\item the support $T^\ast$ of a relative-interior optimal plan, whose cycle
number is the dimension of the optimal face (primal degeneracy);
\item the set of optimal dual potentials, of dimension $c(T^\ast)-1$ modulo the
shift (dual degeneracy);
\item the support of the selected limit point $P_0(q)$, which coincides with
$T^\ast$ for $q\le1$ but may be a proper subset for $q>1$, when $P_0(q)$ lies on the
boundary of $O$.
\end{itemize}
By Euler's relation the first two satisfy
$\beta_1(T^\ast)-(c(T^\ast)-1)=\lvert T^\ast\rvert-m-n+1$. For $q>1$ the cycle number of
$\operatorname{supp}P_0(q)$ bounds $\dim O$ from below but need not equal it, so
it is a property of the selection rather than of the face. For $q\le1$ the
support of a regularized plan at positive $\gamma$ is always full, with
$\beta_1=(m-1)(n-1)$. Only the limit $\gamma\to0$ recovers $\dim O$. For Shannon
entropy ($q=1$), the selection of a distinguished point of the optimal face in
the limit $\gamma\to0$ is classical~\cite{CominettiSanMartin1994,Weed2018}; (iii)
and (iv) extend the elementary part of this picture to all $q>0$.

\subsection{Exact computations}\label{ssec:pdnum}

We computed $\Gamma=\mathsf T+\mathsf I$, the support of the optimal plan of the
minimizing configuration $D'\in\lin D$, and its bipartite cycle number on a
grid of $11$ values of $q\in[\tfrac12,3]$ and $16$ values of $\gamma\in[0.03,30]$,
for $C_7$ ($D=4v_1$), $G'$ ($D=4v_1$) and $G''$ ($D=5v_1$), each against its worst
target $E$. All problems were solved in the gauge-reduced Tsallis form of
\cref{thm:gauge}, restricted to the supports of the marginals; this
restriction is exact because $0^q=0$, and every solve met the solver's accuracy
criteria.

\begin{table}[!htbp]
\centering\small
\begin{tabular}{r rrr rrr}
\toprule
 & \multicolumn{3}{c}{$G'$, $q=1$} & \multicolumn{3}{c}{$C_7$, $q=1$} \\
\cmidrule(lr){2-4}\cmidrule(lr){5-7}
$\gamma$ & $\mathsf T$ & $\mathsf I$ & $\Gamma$ & $\mathsf T$ & $\mathsf I$ & $\Gamma$ \\
\midrule
0.03  & 1.219 & 0.000   & 1.219   & 0.729 & 0.000  & 0.729 \\
0.30  & 0.943 & 0.000   & 0.943   & 0.539 & 0.000  & 0.539 \\
0.75  & 0.577 & $-0.022$ & 0.555  & 0.344 & 0.000  & 0.344 \\
1.19  & 0.551 & $-0.328$ & 0.223  & 0.271 & 0.000  & 0.271 \\
1.89  & 0.533 & $-0.812$ & $-0.279$ & 0.229 & 0.000 & 0.229 \\
3.00  & 0.521 & $-1.579$ & $-1.059$ & 0.193 & 0.000 & 0.193 \\
7.54  & 0.508 & $-4.723$ & $-4.215$ & 0.185 & $-0.036$ & 0.149 \\
30.0  & 0.502 & $-20.29$ & $-19.79$ & 0.187 & $-0.056$ & 0.131 \\
\bottomrule
\end{tabular}
\caption{The decomposition of \cref{thm:TI} at $q=1$. On $G'$ the topological
term $\mathsf T$ decreases from $N_E/d=5/4$ to a plateau near $\tfrac12$, while
the informational term $\mathsf I$ is exactly $0$ until $\gamma\approx0.5$ and
then grows linearly in magnitude, as \cref{thm:TI}(iii) predicts. The sign
change occurs at $\gamma^\ast=1.50$. On $C_7$ the informational term stays
below $0.06$ in magnitude and never overtakes $\mathsf T$.}
\label{tab:TI}
\end{table}

\emph{The decomposition.} \Cref{tab:TI} shows the two terms separating as the
theory predicts. A feature not visible in $\Gamma$ alone is that
$\mathsf I_{q,\gamma}(E)=0$ exactly over a whole low-temperature range: as long as
the target itself is the best configuration of its degree, information exerts
no force at all, and the gap is purely topological. The crossover is preceded
by the moment at which $\mathsf I$ leaves zero, at $\gamma\in(0.48,0.75]$ for
$q=1$ on $G'$. On $G''$ the same order holds for every $q$ in the grid. On all
three graphs the high-temperature slope of $\mathsf T$ vanishes, so whether a
crossover occurs is decided by the slope of $\mathsf I$ alone, which is
$\ln_q4-\ln_q8<0$ on $G'$ and $0$ on $C_7$.

\emph{Integrality.} For $q\in\{1,2\}$ and $\gamma\in\{0.03,0.3,1,3\}$ we also
solved the convex relaxation of \cref{thm:minimax} over
$\mathcal C_D=\operatorname{conv}\mu(\lin D)$, and the relaxation over all
measures. On $C_7$ the convex relaxation coincides with the relaxation over all
measures (difference below $2\times10^{-8}$), because $\mu_E\in\mathcal C_D$
although $E\notin\lin D$. Dual potentials therefore certify \emph{no} positive gap
on $C_7$: the entire topological gap, between $0.55$ and $1.65$, is an integrality
gap. On $G'$ the integrality gap is $0.375=3/8$ in the low-temperature limit, at
both $q=1$ and $q=2$, and the convex relaxation retains the remainder, so part
of the topology is visible to dual certificates there.

\begin{table}[!htbp]
\centering\small
\begin{tabular}{c c cc cc}
\toprule
 & & \multicolumn{2}{c}{$G'$} & \multicolumn{2}{c}{$C_7$} \\
\cmidrule(lr){3-4}\cmidrule(lr){5-6}
$q$ & $\gamma$ & $\Delta_{\rm int}$ & $\Delta_{\rm cls}$ & $\Delta_{\rm int}$ & $\Delta_{\rm cls}$ \\
\midrule
1 & 0.03 & 0.375 & 0.844 & 0.729 & 0.000 \\
1 & 0.3  & 0.399 & 0.555 & 0.550 & 0.000 \\
1 & 1    & 0.655 & 0.163 & 0.610 & 0.000 \\
1 & 3    & 1.663 & 0.011 & 1.648 & 0.000 \\
\addlinespace
2 & 0.03 & 0.375 & 0.870 & 0.748 & 0.000 \\
2 & 0.3  & 0.375 & 0.828 & 0.734 & 0.000 \\
2 & 1    & 0.375 & 0.719 & 0.695 & 0.000 \\
2 & 3    & 0.376 & 0.461 & 0.586 & 0.000 \\
\bottomrule
\end{tabular}
\caption{The two gaps of \cref{cor:threelevels} for $D=4v_1$ and the worst
target $\nu=\mu_E$. Entries shown as $0.000$ are below $2\times10^{-8}$ in absolute
value.}
\label{tab:levels}
\end{table}

\begin{figure}[!htbp]
\centering
\begin{tikzpicture}
\begin{axis}[width=0.66\textwidth,height=6.3cm,xmode=log,xmin=0.02,xmax=200,ymin=0.25,ymax=3.3,
  xlabel={$\gamma$},ylabel={$q$},
  legend style={font=\footnotesize,at={(1.03,1.0)},anchor=north west},
  tick label style={font=\footnotesize},label style={font=\small},grid=major,grid style={black!12}]
\addplot[blue!70!black,thick,mark=*,mark size=1.4pt] coordinates
 {(0.2,0.369) (0.333,0.5) (0.35,0.514) (0.5,0.620) (0.746,0.75) (0.75,0.752) (1.0,0.851) (1.5,1.0) (2.0,1.110) (2.847,1.25) (3.0,1.271) (5.0,1.484) (5.194,1.5) (16.0,2.0) (126.0,3.0)};
\addplot[red!70!black,thick,only marks,mark=square*,mark size=1.6pt,error bars/.cd,y dir=both,y explicit] coordinates
 {(0.03,1.05)+-(0,0.05) (0.3,1.05)+-(0,0.05) (0.754,1.175)+-(0,0.075) (1.89,1.375)+-(0,0.125) (18.9,1.75)+-(0,0.25) (30,2.5)+-(0,0.5)};
\addplot[black!50,dashed] coordinates {(0.02,1) (200,1)};
\legend{{$q_{\rm gap}$ on $G'$},{$q_{\rm supp}$ on $C_7$},{$q=1$ (Legendre)}}
\end{axis}
\end{tikzpicture}
\caption{The two transitions. Blue: the gap-sign boundary on $G'$, below
which $\Gamma\le0$ (information wins) and above which $\Gamma>0$, computed both as
$q_{\rm gap}(\gamma)$ and as $\gamma^\ast(q)$. Red: the support boundary on $C_7$,
above which optimal plans have exact zeros (intervals from the grid). On $G'$
plans keep full support for all $q\le3$; on $C_7$, $\Gamma>0$ everywhere. Below the
dashed line no zeros can occur (\cref{prop:legendre}).}
\label{fig:phase}
\end{figure}
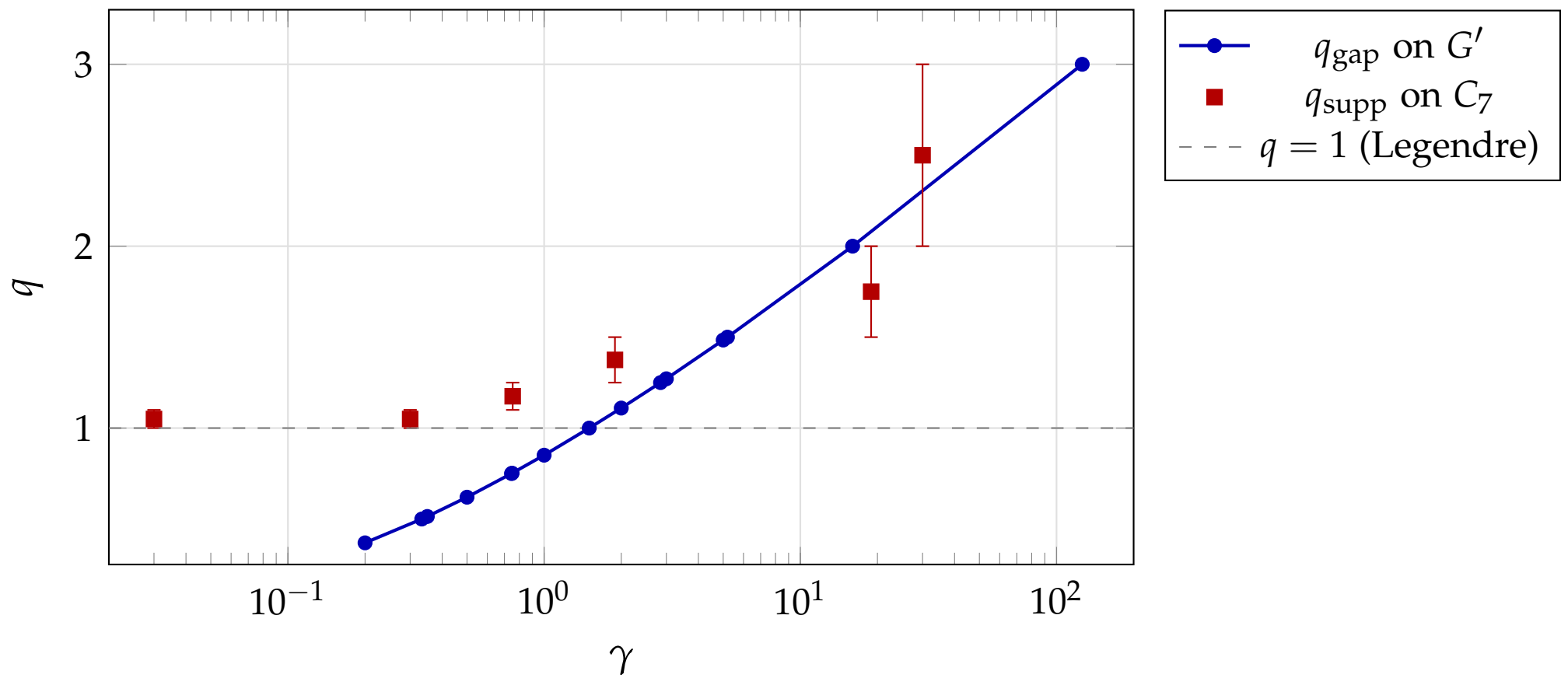

\emph{Two transitions, dissociated.} \Cref{fig:phase} plots the gap-sign
boundary $q_{\rm gap}(\gamma)$ (the $q$ at which $\Gamma$ changes sign at fixed
$\gamma$) and the support boundary $q_{\rm supp}(\gamma)$.
\begin{itemize}
\item On $G'$, $q_{\rm gap}(\gamma)$ increases from $0.37$ at $\gamma=0.2$ through
$1.00$ at $\gamma=1.5$ to $1.48$ at $\gamma=5$, consistently with the inverse
crossover temperatures $\gamma^\ast(q)$ of \cref{tab:phase}. Yet the plan of the
minimizing configuration has \emph{full} support at every grid point with
$q\le3$. On $G''$ the picture is the same. Hence $q_{\rm gap}<1\le q_{\rm supp}$ for
$\gamma<1.5$, and $q_{\rm gap}<q_{\rm supp}$ throughout the grid.
\item On $C_7$ the opposite occurs: $\Gamma>0$ at every grid point, so there is no
gap-sign transition, while exact zeros (entries below $10^{-7}$, which for $q\le1$ are excluded by
\cref{prop:legendre}) appear for every $q>1$ in the grid at small $\gamma$. The support boundary rises from $q_{\rm supp}\in(1,1.1]$ at
$\gamma\le0.3$ to $(2,3]$ at $\gamma=30$, consistent with
$F_q'(0)=2q-1-\frac1{q-1}\to-\infty$ as $q\downarrow1$.
\end{itemize}
At fixed effective temperature $\varepsilon=\gamma/q$ (experiment B), the
gap-sign boundary on $G'$ is $q_{\rm gap}=0.32$, $0.58$, $0.75$, $1.00$, $1.17$,
$1.41$, $1.72$ at $\varepsilon=0.5$, $0.75$, $1$, $1.5$, $2$, $3$, $5$. The
$q$-dependence therefore survives the removal of the factor $1/q$, and is
carried by $\ln_q$ and the shape of $S_q$. In these computations the two
transitions occur at different parameters. This is what the different
mechanisms identified in \cref{prop:legendre} (a dual-boundary event) and in
\cref{thm:TI,thm:minimax} (a primal-lattice event) lead one to expect, but it is
a numerical observation on three graphs, not a general theorem.

\emph{Degeneracy at scale.} For the honeycomb and Erd\H{o}s--Rényi (ER) networks of \cref{sec:sim}
($k=50$ particles, $10$ instances per row) we computed the optimal face $O$ by
one linear program for a relative-interior plan, together with the selections
$P_0(\tfrac12)$ and $P_0(2)$ of \cref{thm:face}(iii).

\begin{table}[!htbp]
\centering\small
\setlength{\tabcolsep}{4pt}
\begin{tabular}{lrrrrrr}
\toprule
Network & $\lvert T^\ast\rvert$ & $\dim O$ & $c(T^\ast)-1$ &
$\beta_1(P_0(\tfrac12))$ & $\beta_1(P_0(2))$ & strict \\
\midrule
honeycomb, $n\approx10^3$ & 130.9 & 56.1 & 24.2 & 56.1 & 48.7 & 8/10 \\
ER, $n\approx10^3$     & 93.3  & 24.8 & 30.5 & 24.8 & 20.5 & 8/10 \\
honeycomb, $n\approx10^4$ & 87.5  & 25.7 & 37.2 & 25.7 & 20.9 & 2/10 \\
ER, $n\approx10^4$     & 107.7 & 33.7 & 25.0 & 33.7 & 20.1 & 8/10 \\
\bottomrule
\end{tabular}
\caption{Primal and dual degeneracy of optimal transport with hop costs (means
over $10$ instances; $\dim O=\beta_1(T^\ast)$; ``strict'' counts instances with
$\operatorname{supp}P_0(2)\subsetneq T^\ast$). Every instance is degenerate ($\dim O>0$). For $q=\tfrac12$
the selection has support exactly $T^\ast$ in all $40$ instances, as
\cref{thm:face}(iv) asserts; for $q=2$ the support is strictly smaller in $26$ of
the $40$.}
\label{tab:face}
\end{table}

\Cref{tab:face} shows that integer hop costs make optimal transport
\emph{highly} degenerate on both families. With $k=50$ the optimal face has
dimension $25$--$56$, and the dual potentials have $24$--$37$ free directions.
This is the precise sense in which the many equidistant targets of
\cref{tab:simeff} matter. The low-temperature limit of $F_q$-regularized
transport is not ``the'' optimal plan but a $q$-dependent selection from a large
face. For $q\le1$ it fills the whole face; for $q=2$ it typically lies on its
boundary, with smaller cycle number.

\section{Honeycomb and Erd\H{o}s--Rényi networks: simulation and interpretation}\label{sec:sim}

The results above were tested on seven-vertex graphs. We now ask what they
predict, and what they do not, for particle transport on large networks of two
contrasting kinds. In a \emph{honeycomb} network every node is linked only to
its geometric neighbours. In an \emph{Erd\H{o}s--Rényi} (ER) network, the random graph
$G(n,p)$~\cite{ER1960}, links join near and far nodes alike. We compare them at
$n\approx10^2,10^3,10^4,10^5$, and then by an exact experiment at matched genus.

\emph{What these experiments test.} Two kinds of computation must be
distinguished.
\begin{enumerate}[label=(\alph*)]
\item The \emph{large-network experiments} (\cref{ssec:struct,ssec:eff,ssec:routes,ssec:reg})
compute ordinary, unregularized and regularized transport between two
configurations of $k\le50$ particles. They test the statements about
regularized plans: the role of $\gamma/q$ and $\ln_q(N)$
(\cref{thm:gauge}), the support (\cref{thm:support,prop:legendre}) and optimal
faces (\cref{thm:face}). They do \emph{not} minimize over a linear equivalence
class. Since the particle configurations have degree $d=k\ll g$, where the
Riemann--Roch bounds are vacuous, they are not tests of the transport defect,
of the gap $\Gamma$, or of \cref{thm:defect,thm:AJ,thm:crossover,thm:phase,thm:serre,thm:vertexrank,thm:TI}.
What they show about those results is only consistency, for instance the
dependence of the metric bounds on $\diam(G)$.
\item The \emph{exact computations} on small graphs (\cref{ssec:pdnum} and
\cref{ssec:matched}) minimize over linear systems. They are direct tests of the
defect and gap theorems.
\end{enumerate}
\Cref{tab:synthesis} records, for each result, which kind of evidence applies.

\subsection{Design}\label{ssec:design}

\emph{Networks.} The honeycomb network is a near-square patch of the hexagonal
lattice. The ER network is the giant component of $G(n_0,p)$ with
$p=3/(n_0-1)$, with $n_0$ chosen so that the giant component has about $n$
nodes; its mean degree ($3.0$--$3.2$) matches the honeycomb value~$3$. Both
families therefore have essentially the same number of edges per node, hence
the same genus density $g/n$ up to boundary and degree effects.

Representative members of both families are drawn in \cref{fig:netexamples}.

\begin{figure}[t]
\centering
\begin{minipage}[b]{0.49\textwidth}\centering
\begin{tikzpicture}[x=1cm,y=1cm]
\draw[black!22,line width=0.35pt] (0.278,0.000)--(0.000,0.482);
\draw[black!22,line width=0.35pt] (0.278,0.000)--(0.835,0.000);
\draw[black!22,line width=0.35pt] (0.000,0.482)--(0.278,0.965);
\draw[black!22,line width=0.35pt] (0.278,0.965)--(0.000,1.447);
\draw[black!22,line width=0.35pt] (0.278,0.965)--(0.835,0.965);
\draw[black!22,line width=0.35pt] (0.000,1.447)--(0.278,1.929);
\draw[black!22,line width=0.35pt] (0.278,1.929)--(0.000,2.412);
\draw[black!22,line width=0.35pt] (0.278,1.929)--(0.835,1.929);
\draw[black!22,line width=0.35pt] (0.000,2.412)--(0.278,2.894);
\draw[black!22,line width=0.35pt] (0.278,2.894)--(0.000,3.376);
\draw[black!22,line width=0.35pt] (0.278,2.894)--(0.835,2.894);
\draw[black!22,line width=0.35pt] (0.000,3.376)--(0.278,3.858);
\draw[black!22,line width=0.35pt] (0.278,3.858)--(0.000,4.341);
\draw[black!22,line width=0.35pt] (0.278,3.858)--(0.835,3.858);
\draw[black!22,line width=0.35pt] (0.000,4.341)--(0.278,4.823);
\draw[black!22,line width=0.35pt] (0.278,4.823)--(0.000,5.305);
\draw[black!22,line width=0.35pt] (0.278,4.823)--(0.835,4.823);
\draw[black!22,line width=0.35pt] (0.000,5.305)--(0.278,5.788);
\draw[black!22,line width=0.35pt] (0.278,5.788)--(0.835,5.788);
\draw[black!22,line width=0.35pt] (0.835,0.000)--(1.114,0.482);
\draw[black!22,line width=0.35pt] (1.114,0.482)--(0.835,0.965);
\draw[black!22,line width=0.35pt] (1.114,0.482)--(1.671,0.482);
\draw[black!22,line width=0.35pt] (0.835,0.965)--(1.114,1.447);
\draw[black!22,line width=0.35pt] (1.114,1.447)--(0.835,1.929);
\draw[black!22,line width=0.35pt] (1.114,1.447)--(1.671,1.447);
\draw[black!22,line width=0.35pt] (0.835,1.929)--(1.114,2.412);
\draw[black!22,line width=0.35pt] (1.114,2.412)--(0.835,2.894);
\draw[black!22,line width=0.35pt] (1.114,2.412)--(1.671,2.412);
\draw[black!22,line width=0.35pt] (0.835,2.894)--(1.114,3.376);
\draw[black!22,line width=0.35pt] (1.114,3.376)--(0.835,3.858);
\draw[black!22,line width=0.35pt] (1.114,3.376)--(1.671,3.376);
\draw[black!22,line width=0.35pt] (0.835,3.858)--(1.114,4.341);
\draw[black!22,line width=0.35pt] (1.114,4.341)--(0.835,4.823);
\draw[black!22,line width=0.35pt] (1.114,4.341)--(1.671,4.341);
\draw[black!22,line width=0.35pt] (0.835,4.823)--(1.114,5.305);
\draw[black!22,line width=0.35pt] (1.114,5.305)--(0.835,5.788);
\draw[black!22,line width=0.35pt] (1.114,5.305)--(1.671,5.305);
\draw[black!22,line width=0.35pt] (0.835,5.788)--(1.114,6.270);
\draw[black!22,line width=0.35pt] (1.114,6.270)--(1.671,6.270);
\draw[black!22,line width=0.35pt] (1.949,0.000)--(1.671,0.482);
\draw[black!22,line width=0.35pt] (1.949,0.000)--(2.506,0.000);
\draw[black!22,line width=0.35pt] (1.671,0.482)--(1.949,0.965);
\draw[black!22,line width=0.35pt] (1.949,0.965)--(1.671,1.447);
\draw[black!22,line width=0.35pt] (1.949,0.965)--(2.506,0.965);
\draw[black!22,line width=0.35pt] (1.671,1.447)--(1.949,1.929);
\draw[black!22,line width=0.35pt] (1.949,1.929)--(1.671,2.412);
\draw[black!22,line width=0.35pt] (1.949,1.929)--(2.506,1.929);
\draw[black!22,line width=0.35pt] (1.671,2.412)--(1.949,2.894);
\draw[black!22,line width=0.35pt] (1.949,2.894)--(1.671,3.376);
\draw[black!22,line width=0.35pt] (1.949,2.894)--(2.506,2.894);
\draw[black!22,line width=0.35pt] (1.671,3.376)--(1.949,3.858);
\draw[black!22,line width=0.35pt] (1.949,3.858)--(1.671,4.341);
\draw[black!22,line width=0.35pt] (1.949,3.858)--(2.506,3.858);
\draw[black!22,line width=0.35pt] (1.671,4.341)--(1.949,4.823);
\draw[black!22,line width=0.35pt] (1.949,4.823)--(1.671,5.305);
\draw[black!22,line width=0.35pt] (1.949,4.823)--(2.506,4.823);
\draw[black!22,line width=0.35pt] (1.671,5.305)--(1.949,5.788);
\draw[black!22,line width=0.35pt] (1.949,5.788)--(1.671,6.270);
\draw[black!22,line width=0.35pt] (1.949,5.788)--(2.506,5.788);
\draw[black!22,line width=0.35pt] (2.506,0.000)--(2.785,0.482);
\draw[black!22,line width=0.35pt] (2.785,0.482)--(2.506,0.965);
\draw[black!22,line width=0.35pt] (2.785,0.482)--(3.342,0.482);
\draw[black!22,line width=0.35pt] (2.506,0.965)--(2.785,1.447);
\draw[black!22,line width=0.35pt] (2.785,1.447)--(2.506,1.929);
\draw[black!22,line width=0.35pt] (2.785,1.447)--(3.342,1.447);
\draw[black!22,line width=0.35pt] (2.506,1.929)--(2.785,2.412);
\draw[black!22,line width=0.35pt] (2.785,2.412)--(2.506,2.894);
\draw[black!22,line width=0.35pt] (2.785,2.412)--(3.342,2.412);
\draw[black!22,line width=0.35pt] (2.506,2.894)--(2.785,3.376);
\draw[black!22,line width=0.35pt] (2.785,3.376)--(2.506,3.858);
\draw[black!22,line width=0.35pt] (2.785,3.376)--(3.342,3.376);
\draw[black!22,line width=0.35pt] (2.506,3.858)--(2.785,4.341);
\draw[black!22,line width=0.35pt] (2.785,4.341)--(2.506,4.823);
\draw[black!22,line width=0.35pt] (2.785,4.341)--(3.342,4.341);
\draw[black!22,line width=0.35pt] (2.506,4.823)--(2.785,5.305);
\draw[black!22,line width=0.35pt] (2.785,5.305)--(2.506,5.788);
\draw[black!22,line width=0.35pt] (2.785,5.305)--(3.342,5.305);
\draw[black!22,line width=0.35pt] (2.506,5.788)--(2.785,6.270);
\draw[black!22,line width=0.35pt] (2.785,6.270)--(3.342,6.270);
\draw[black!22,line width=0.35pt] (3.620,0.000)--(3.342,0.482);
\draw[black!22,line width=0.35pt] (3.620,0.000)--(4.177,0.000);
\draw[black!22,line width=0.35pt] (3.342,0.482)--(3.620,0.965);
\draw[black!22,line width=0.35pt] (3.620,0.965)--(3.342,1.447);
\draw[black!22,line width=0.35pt] (3.620,0.965)--(4.177,0.965);
\draw[black!22,line width=0.35pt] (3.342,1.447)--(3.620,1.929);
\draw[black!22,line width=0.35pt] (3.620,1.929)--(3.342,2.412);
\draw[black!22,line width=0.35pt] (3.620,1.929)--(4.177,1.929);
\draw[black!22,line width=0.35pt] (3.342,2.412)--(3.620,2.894);
\draw[black!22,line width=0.35pt] (3.620,2.894)--(3.342,3.376);
\draw[black!22,line width=0.35pt] (3.620,2.894)--(4.177,2.894);
\draw[black!22,line width=0.35pt] (3.342,3.376)--(3.620,3.858);
\draw[black!22,line width=0.35pt] (3.620,3.858)--(3.342,4.341);
\draw[black!22,line width=0.35pt] (3.620,3.858)--(4.177,3.858);
\draw[black!22,line width=0.35pt] (3.342,4.341)--(3.620,4.823);
\draw[black!22,line width=0.35pt] (3.620,4.823)--(3.342,5.305);
\draw[black!22,line width=0.35pt] (3.620,4.823)--(4.177,4.823);
\draw[black!22,line width=0.35pt] (3.342,5.305)--(3.620,5.788);
\draw[black!22,line width=0.35pt] (3.620,5.788)--(3.342,6.270);
\draw[black!22,line width=0.35pt] (3.620,5.788)--(4.177,5.788);
\draw[black!22,line width=0.35pt] (4.177,0.000)--(4.455,0.482);
\draw[black!22,line width=0.35pt] (4.455,0.482)--(4.177,0.965);
\draw[black!22,line width=0.35pt] (4.455,0.482)--(5.012,0.482);
\draw[black!22,line width=0.35pt] (4.177,0.965)--(4.455,1.447);
\draw[black!22,line width=0.35pt] (4.455,1.447)--(4.177,1.929);
\draw[black!22,line width=0.35pt] (4.455,1.447)--(5.012,1.447);
\draw[black!22,line width=0.35pt] (4.177,1.929)--(4.455,2.412);
\draw[black!22,line width=0.35pt] (4.455,2.412)--(4.177,2.894);
\draw[black!22,line width=0.35pt] (4.455,2.412)--(5.012,2.412);
\draw[black!22,line width=0.35pt] (4.177,2.894)--(4.455,3.376);
\draw[black!22,line width=0.35pt] (4.455,3.376)--(4.177,3.858);
\draw[black!22,line width=0.35pt] (4.455,3.376)--(5.012,3.376);
\draw[black!22,line width=0.35pt] (4.177,3.858)--(4.455,4.341);
\draw[black!22,line width=0.35pt] (4.455,4.341)--(4.177,4.823);
\draw[black!22,line width=0.35pt] (4.455,4.341)--(5.012,4.341);
\draw[black!22,line width=0.35pt] (4.177,4.823)--(4.455,5.305);
\draw[black!22,line width=0.35pt] (4.455,5.305)--(4.177,5.788);
\draw[black!22,line width=0.35pt] (4.455,5.305)--(5.012,5.305);
\draw[black!22,line width=0.35pt] (4.177,5.788)--(4.455,6.270);
\draw[black!22,line width=0.35pt] (4.455,6.270)--(5.012,6.270);
\draw[black!22,line width=0.35pt] (5.012,0.482)--(5.291,0.965);
\draw[black!22,line width=0.35pt] (5.291,0.965)--(5.012,1.447);
\draw[black!22,line width=0.35pt] (5.012,1.447)--(5.291,1.929);
\draw[black!22,line width=0.35pt] (5.291,1.929)--(5.012,2.412);
\draw[black!22,line width=0.35pt] (5.012,2.412)--(5.291,2.894);
\draw[black!22,line width=0.35pt] (5.291,2.894)--(5.012,3.376);
\draw[black!22,line width=0.35pt] (5.012,3.376)--(5.291,3.858);
\draw[black!22,line width=0.35pt] (5.291,3.858)--(5.012,4.341);
\draw[black!22,line width=0.35pt] (5.012,4.341)--(5.291,4.823);
\draw[black!22,line width=0.35pt] (5.291,4.823)--(5.012,5.305);
\draw[black!22,line width=0.35pt] (5.012,5.305)--(5.291,5.788);
\draw[black!22,line width=0.35pt] (5.291,5.788)--(5.012,6.270);
\draw[orange!85!black,line width=1.3pt,opacity=0.85,line cap=round,line join=round] (1.114,1.447)--(0.835,0.965);
\draw[violet!80!black,line width=1.3pt,opacity=0.85,line cap=round,line join=round] (0.278,3.858)--(0.000,3.376);
\draw[teal!80!black,line width=1.3pt,opacity=0.85,line cap=round,line join=round] (1.114,3.376)--(0.835,2.894)--(1.114,2.412)--(1.671,2.412)--(1.949,1.929);
\draw[magenta!80!black,line width=1.3pt,opacity=0.85,line cap=round,line join=round] (0.000,1.447)--(0.278,0.965)--(0.835,0.965)--(1.114,0.482)--(1.671,0.482)--(1.949,0.000)--(2.506,0.000);
\draw[olive!85!black,line width=1.3pt,opacity=0.85,line cap=round,line join=round] (4.177,4.823)--(4.455,4.341)--(4.177,3.858)--(4.455,3.376)--(4.177,2.894)--(4.455,2.412)--(4.177,1.929)--(4.455,1.447);
\draw[cyan!60!black,line width=1.3pt,opacity=0.85,line cap=round,line join=round] (3.342,6.270)--(2.785,6.270)--(2.506,5.788);
\node[transn] at (0.278,0.000) {};
\node[transn] at (0.000,0.482) {};
\node[hubn] at (0.278,0.965) {};
\node[transn] at (0.000,1.447) {};
\node[hubn] at (0.278,1.929) {};
\node[transn] at (0.000,2.412) {};
\node[hubn] at (0.278,2.894) {};
\node[transn] at (0.000,3.376) {};
\node[hubn] at (0.278,3.858) {};
\node[transn] at (0.000,4.341) {};
\node[hubn] at (0.278,4.823) {};
\node[transn] at (0.000,5.305) {};
\node[transn] at (0.278,5.788) {};
\node[transn] at (0.835,0.000) {};
\node[hubn] at (1.114,0.482) {};
\node[hubn] at (0.835,0.965) {};
\node[hubn] at (1.114,1.447) {};
\node[hubn] at (0.835,1.929) {};
\node[hubn] at (1.114,2.412) {};
\node[hubn] at (0.835,2.894) {};
\node[hubn] at (1.114,3.376) {};
\node[hubn] at (0.835,3.858) {};
\node[hubn] at (1.114,4.341) {};
\node[hubn] at (0.835,4.823) {};
\node[hubn] at (1.114,5.305) {};
\node[hubn] at (0.835,5.788) {};
\node[transn] at (1.114,6.270) {};
\node[transn] at (1.949,0.000) {};
\node[hubn] at (1.671,0.482) {};
\node[hubn] at (1.949,0.965) {};
\node[hubn] at (1.671,1.447) {};
\node[hubn] at (1.949,1.929) {};
\node[hubn] at (1.671,2.412) {};
\node[hubn] at (1.949,2.894) {};
\node[hubn] at (1.671,3.376) {};
\node[hubn] at (1.949,3.858) {};
\node[hubn] at (1.671,4.341) {};
\node[hubn] at (1.949,4.823) {};
\node[hubn] at (1.671,5.305) {};
\node[hubn] at (1.949,5.788) {};
\node[transn] at (1.671,6.270) {};
\node[transn] at (2.506,0.000) {};
\node[hubn] at (2.785,0.482) {};
\node[hubn] at (2.506,0.965) {};
\node[hubn] at (2.785,1.447) {};
\node[hubn] at (2.506,1.929) {};
\node[hubn] at (2.785,2.412) {};
\node[hubn] at (2.506,2.894) {};
\node[hubn] at (2.785,3.376) {};
\node[hubn] at (2.506,3.858) {};
\node[hubn] at (2.785,4.341) {};
\node[hubn] at (2.506,4.823) {};
\node[hubn] at (2.785,5.305) {};
\node[hubn] at (2.506,5.788) {};
\node[transn] at (2.785,6.270) {};
\node[transn] at (3.620,0.000) {};
\node[hubn] at (3.342,0.482) {};
\node[hubn] at (3.620,0.965) {};
\node[hubn] at (3.342,1.447) {};
\node[hubn] at (3.620,1.929) {};
\node[hubn] at (3.342,2.412) {};
\node[hubn] at (3.620,2.894) {};
\node[hubn] at (3.342,3.376) {};
\node[hubn] at (3.620,3.858) {};
\node[hubn] at (3.342,4.341) {};
\node[hubn] at (3.620,4.823) {};
\node[hubn] at (3.342,5.305) {};
\node[hubn] at (3.620,5.788) {};
\node[transn] at (3.342,6.270) {};
\node[transn] at (4.177,0.000) {};
\node[hubn] at (4.455,0.482) {};
\node[hubn] at (4.177,0.965) {};
\node[hubn] at (4.455,1.447) {};
\node[hubn] at (4.177,1.929) {};
\node[hubn] at (4.455,2.412) {};
\node[hubn] at (4.177,2.894) {};
\node[hubn] at (4.455,3.376) {};
\node[hubn] at (4.177,3.858) {};
\node[hubn] at (4.455,4.341) {};
\node[hubn] at (4.177,4.823) {};
\node[hubn] at (4.455,5.305) {};
\node[hubn] at (4.177,5.788) {};
\node[transn] at (4.455,6.270) {};
\node[transn] at (5.012,0.482) {};
\node[transn] at (5.291,0.965) {};
\node[hubn] at (5.012,1.447) {};
\node[transn] at (5.291,1.929) {};
\node[hubn] at (5.012,2.412) {};
\node[transn] at (5.291,2.894) {};
\node[hubn] at (5.012,3.376) {};
\node[transn] at (5.291,3.858) {};
\node[hubn] at (5.012,4.341) {};
\node[transn] at (5.291,4.823) {};
\node[hubn] at (5.012,5.305) {};
\node[transn] at (5.291,5.788) {};
\node[transn] at (5.012,6.270) {};
\node[srcn] at (1.114,1.447) {};
\node[srcn] at (0.278,3.858) {};
\node[srcn] at (1.114,3.376) {};
\node[srcn] at (0.000,1.447) {};
\node[srcn] at (4.177,4.823) {};
\node[srcn] at (3.342,6.270) {};
\node[tgtn] at (2.506,0.000) {};
\node[tgtn] at (1.949,1.929) {};
\node[tgtn] at (0.000,3.376) {};
\node[tgtn] at (0.835,0.965) {};
\node[tgtn] at (4.455,1.447) {};
\node[tgtn] at (2.506,5.788) {};
\node[anchor=north,font=\footnotesize] at (2.645,-0.25) {(a) Honeycomb ($n=96$, $g=36$)};
\end{tikzpicture}
\end{minipage}\hfill
\begin{minipage}[b]{0.49\textwidth}\centering
\begin{tikzpicture}[x=1cm,y=1cm]
\draw[black!22,line width=0.35pt] (4.530,0.719)--(4.130,1.388);
\draw[black!22,line width=0.35pt] (4.530,0.719)--(4.919,0.125);
\draw[black!22,line width=0.35pt] (5.961,1.085)--(5.305,1.369);
\draw[black!22,line width=0.35pt] (3.415,4.151)--(4.399,3.906);
\draw[black!22,line width=0.35pt] (3.415,4.151)--(2.923,4.527);
\draw[black!22,line width=0.35pt] (3.415,4.151)--(3.011,3.408);
\draw[black!22,line width=0.35pt] (3.415,4.151)--(3.129,4.677);
\draw[black!22,line width=0.35pt] (2.039,1.195)--(2.356,2.031);
\draw[black!22,line width=0.35pt] (3.468,4.736)--(3.830,5.445);
\draw[black!22,line width=0.35pt] (3.468,4.736)--(3.451,5.106);
\draw[black!22,line width=0.35pt] (3.468,4.736)--(3.801,3.789);
\draw[black!22,line width=0.35pt] (3.468,4.736)--(2.923,4.527);
\draw[black!22,line width=0.35pt] (3.468,4.736)--(2.905,5.552);
\draw[black!22,line width=0.35pt] (2.778,1.969)--(1.656,2.643);
\draw[black!22,line width=0.35pt] (2.778,1.969)--(3.859,2.139);
\draw[black!22,line width=0.35pt] (1.116,3.651)--(1.656,2.643);
\draw[black!22,line width=0.35pt] (1.116,3.651)--(1.616,4.195);
\draw[black!22,line width=0.35pt] (2.269,0.684)--(2.028,0.000);
\draw[black!22,line width=0.35pt] (2.269,0.684)--(2.536,1.462);
\draw[black!22,line width=0.35pt] (3.142,3.870)--(3.110,2.982);
\draw[black!22,line width=0.35pt] (3.142,3.870)--(2.923,4.527);
\draw[black!22,line width=0.35pt] (3.830,5.445)--(3.232,5.440);
\draw[black!22,line width=0.35pt] (1.839,3.707)--(2.772,3.610);
\draw[black!22,line width=0.35pt] (1.839,3.707)--(1.359,2.762);
\draw[black!22,line width=0.35pt] (4.879,2.767)--(4.074,2.950);
\draw[black!22,line width=0.35pt] (3.232,5.440)--(3.129,4.677);
\draw[black!22,line width=0.35pt] (4.859,2.490)--(4.766,3.132);
\draw[black!22,line width=0.35pt] (4.859,2.490)--(4.453,2.624);
\draw[black!22,line width=0.35pt] (4.859,2.490)--(4.101,2.320);
\draw[black!22,line width=0.35pt] (5.305,1.369)--(4.548,1.688);
\draw[black!22,line width=0.35pt] (1.545,3.466)--(1.616,4.195);
\draw[black!22,line width=0.35pt] (1.545,3.466)--(2.424,2.677);
\draw[black!22,line width=0.35pt] (1.545,3.466)--(1.012,3.200);
\draw[black!22,line width=0.35pt] (4.766,3.132)--(4.399,3.906);
\draw[black!22,line width=0.35pt] (4.766,3.132)--(3.976,2.663);
\draw[black!22,line width=0.35pt] (5.275,4.631)--(4.699,4.159);
\draw[black!22,line width=0.35pt] (2.412,4.338)--(2.518,3.336);
\draw[black!22,line width=0.35pt] (2.412,4.338)--(2.923,4.527);
\draw[black!22,line width=0.35pt] (1.267,4.858)--(1.616,4.195);
\draw[black!22,line width=0.35pt] (1.267,4.858)--(1.432,5.601);
\draw[black!22,line width=0.35pt] (2.169,5.884)--(1.432,5.601);
\draw[black!22,line width=0.35pt] (2.169,5.884)--(2.905,5.552);
\draw[black!22,line width=0.35pt] (1.766,1.302)--(1.489,2.048);
\draw[black!22,line width=0.35pt] (1.766,1.302)--(2.536,1.462);
\draw[black!22,line width=0.35pt] (3.981,3.597)--(4.076,4.290);
\draw[black!22,line width=0.35pt] (3.981,3.597)--(4.737,3.407);
\draw[black!22,line width=0.35pt] (3.981,3.597)--(3.011,3.408);
\draw[black!22,line width=0.35pt] (4.556,3.567)--(5.410,3.661);
\draw[black!22,line width=0.35pt] (4.556,3.567)--(3.801,3.789);
\draw[black!22,line width=0.35pt] (4.556,3.567)--(4.074,2.950);
\draw[black!22,line width=0.35pt] (4.556,3.567)--(4.699,4.159);
\draw[black!22,line width=0.35pt] (1.656,2.643)--(0.955,2.277);
\draw[black!22,line width=0.35pt] (1.656,2.643)--(2.099,3.186);
\draw[black!22,line width=0.35pt] (1.656,2.643)--(0.746,2.539);
\draw[black!22,line width=0.35pt] (2.772,3.610)--(3.311,3.156);
\draw[black!22,line width=0.35pt] (2.772,3.610)--(2.471,4.141);
\draw[black!22,line width=0.35pt] (2.772,3.610)--(2.099,3.186);
\draw[black!22,line width=0.35pt] (2.772,3.610)--(2.482,3.732);
\draw[black!22,line width=0.35pt] (2.772,3.610)--(3.631,3.675);
\draw[black!22,line width=0.35pt] (3.322,1.266)--(3.484,0.504);
\draw[black!22,line width=0.35pt] (3.322,1.266)--(3.186,2.068);
\draw[black!22,line width=0.35pt] (3.322,1.266)--(2.845,1.583);
\draw[black!22,line width=0.35pt] (4.453,2.624)--(4.074,2.950);
\draw[black!22,line width=0.35pt] (4.453,2.624)--(4.737,3.407);
\draw[black!22,line width=0.35pt] (4.453,2.624)--(3.859,2.139);
\draw[black!22,line width=0.35pt] (2.043,1.945)--(1.261,1.655);
\draw[black!22,line width=0.35pt] (2.043,1.945)--(2.741,2.381);
\draw[black!22,line width=0.35pt] (3.451,5.106)--(2.923,4.527);
\draw[black!22,line width=0.35pt] (3.451,5.106)--(4.205,5.070);
\draw[black!22,line width=0.35pt] (0.000,2.544)--(0.746,2.539);
\draw[black!22,line width=0.35pt] (2.898,2.808)--(3.110,2.982);
\draw[black!22,line width=0.35pt] (2.898,2.808)--(3.976,2.663);
\draw[black!22,line width=0.35pt] (2.898,2.808)--(2.518,3.336);
\draw[black!22,line width=0.35pt] (1.616,4.195)--(2.471,4.141);
\draw[black!22,line width=0.35pt] (1.616,4.195)--(1.931,4.023);
\draw[black!22,line width=0.35pt] (3.151,6.270)--(2.905,5.552);
\draw[black!22,line width=0.35pt] (5.410,3.661)--(6.100,3.809);
\draw[black!22,line width=0.35pt] (3.110,2.982)--(4.101,2.320);
\draw[black!22,line width=0.35pt] (3.110,2.982)--(2.741,2.381);
\draw[black!22,line width=0.35pt] (3.110,2.982)--(2.482,3.732);
\draw[black!22,line width=0.35pt] (0.955,2.277)--(0.622,1.822);
\draw[black!22,line width=0.35pt] (4.399,3.906)--(4.459,4.277);
\draw[black!22,line width=0.35pt] (3.311,3.156)--(3.186,2.068);
\draw[black!22,line width=0.35pt] (3.311,3.156)--(3.505,2.284);
\draw[black!22,line width=0.35pt] (3.311,3.156)--(3.801,3.789);
\draw[black!22,line width=0.35pt] (3.602,2.502)--(4.074,2.950);
\draw[black!22,line width=0.35pt] (3.602,2.502)--(2.424,2.677);
\draw[black!22,line width=0.35pt] (2.117,2.715)--(1.489,2.048);
\draw[black!22,line width=0.35pt] (2.117,2.715)--(2.356,2.031);
\draw[black!22,line width=0.35pt] (2.117,2.715)--(1.724,3.403);
\draw[black!22,line width=0.35pt] (2.117,2.715)--(3.011,3.408);
\draw[black!22,line width=0.35pt] (4.101,2.320)--(4.829,1.794);
\draw[black!22,line width=0.35pt] (3.329,2.774)--(3.505,2.284);
\draw[black!22,line width=0.35pt] (3.329,2.774)--(3.591,2.958);
\draw[black!22,line width=0.35pt] (3.329,2.774)--(3.011,3.408);
\draw[black!22,line width=0.35pt] (2.471,4.141)--(2.923,4.527);
\draw[black!22,line width=0.35pt] (2.471,4.141)--(2.388,3.233);
\draw[black!22,line width=0.35pt] (2.471,4.141)--(3.129,4.677);
\draw[black!22,line width=0.35pt] (4.459,4.277)--(3.801,3.789);
\draw[black!22,line width=0.35pt] (4.459,4.277)--(4.076,4.290);
\draw[black!22,line width=0.35pt] (1.489,2.048)--(1.261,1.655);
\draw[black!22,line width=0.35pt] (3.186,2.068)--(2.741,2.381);
\draw[black!22,line width=0.35pt] (3.505,2.284)--(2.845,1.583);
\draw[black!22,line width=0.35pt] (3.505,2.284)--(3.976,2.663);
\draw[black!22,line width=0.35pt] (2.099,3.186)--(2.518,3.336);
\draw[black!22,line width=0.35pt] (4.548,1.688)--(3.859,2.139);
\draw[black!22,line width=0.35pt] (4.130,1.388)--(3.859,2.139);
\draw[black!22,line width=0.35pt] (2.845,1.583)--(1.736,1.828);
\draw[black!22,line width=0.35pt] (1.261,1.655)--(0.622,1.822);
\draw[black!22,line width=0.35pt] (3.801,3.789)--(3.591,2.958);
\draw[black!22,line width=0.35pt] (4.074,2.950)--(3.591,2.958);
\draw[black!22,line width=0.35pt] (4.074,2.950)--(3.859,2.139);
\draw[black!22,line width=0.35pt] (4.074,2.950)--(3.631,3.675);
\draw[black!22,line width=0.35pt] (2.741,2.381)--(2.424,2.677);
\draw[black!22,line width=0.35pt] (2.741,2.381)--(2.356,2.031);
\draw[black!22,line width=0.35pt] (2.741,2.381)--(3.591,2.958);
\draw[black!22,line width=0.35pt] (2.741,2.381)--(2.388,3.233);
\draw[black!22,line width=0.35pt] (2.741,2.381)--(2.536,1.462);
\draw[black!22,line width=0.35pt] (4.076,4.290)--(4.205,5.070);
\draw[black!22,line width=0.35pt] (4.076,4.290)--(3.631,3.675);
\draw[black!22,line width=0.35pt] (2.518,3.336)--(1.931,4.023);
\draw[black!22,line width=0.35pt] (2.518,3.336)--(3.011,3.408);
\draw[black!22,line width=0.35pt] (2.518,3.336)--(1.660,3.117);
\draw[black!22,line width=0.35pt] (2.518,3.336)--(3.179,2.445);
\draw[black!22,line width=0.35pt] (1.907,4.487)--(2.482,3.732);
\draw[black!22,line width=0.35pt] (0.622,1.822)--(0.746,2.539);
\draw[black!22,line width=0.35pt] (1.724,3.403)--(2.482,3.732);
\draw[black!22,line width=0.35pt] (4.737,3.407)--(4.699,4.159);
\draw[black!22,line width=0.35pt] (1.012,3.200)--(1.359,2.762);
\draw[black!22,line width=0.35pt] (1.012,3.200)--(1.660,3.117);
\draw[black!22,line width=0.35pt] (2.388,3.233)--(2.482,3.732);
\draw[black!22,line width=0.35pt] (2.482,3.732)--(3.052,4.381);
\draw[black!22,line width=0.35pt] (1.736,1.828)--(1.359,2.762);
\draw[black!22,line width=0.35pt] (1.359,2.762)--(1.660,3.117);
\draw[black!22,line width=0.35pt] (3.859,2.139)--(3.179,2.445);
\draw[black!22,line width=0.35pt] (3.822,4.411)--(3.052,4.381);
\draw[black!22,line width=0.35pt] (3.822,4.411)--(3.631,3.675);
\draw[black!22,line width=0.35pt] (3.822,4.411)--(4.699,4.159);
\draw[black!22,line width=0.35pt] (3.822,4.411)--(3.129,4.677);
\draw[black!22,line width=0.35pt] (2.536,1.462)--(2.776,0.717);
\draw[orange!85!black,line width=1.3pt,opacity=0.85,line cap=round,line join=round] (3.822,4.411)--(3.631,3.675)--(4.074,2.950)--(3.602,2.502);
\draw[violet!80!black,line width=1.3pt,opacity=0.85,line cap=round,line join=round] (3.142,3.870)--(2.923,4.527)--(3.451,5.106);
\draw[teal!80!black,line width=1.3pt,opacity=0.85,line cap=round,line join=round] (1.267,4.858)--(1.616,4.195)--(1.545,3.466);
\draw[magenta!80!black,line width=1.3pt,opacity=0.85,line cap=round,line join=round] (2.039,1.195)--(2.356,2.031)--(2.741,2.381)--(2.536,1.462)--(2.269,0.684);
\draw[olive!85!black,line width=1.3pt,opacity=0.85,line cap=round,line join=round] (2.482,3.732)--(3.110,2.982)--(4.101,2.320)--(4.829,1.794);
\draw[cyan!60!black,line width=1.3pt,opacity=0.85,line cap=round,line join=round] (1.931,4.023)--(2.518,3.336)--(3.011,3.408)--(3.329,2.774)--(3.591,2.958);
\node[transn] at (4.530,0.719) {};
\node[termn] at (5.961,1.085) {};
\node[hubn] at (3.415,4.151) {};
\node[termn] at (2.039,1.195) {};
\node[hubn] at (3.468,4.736) {};
\node[transn] at (2.778,1.969) {};
\node[transn] at (1.116,3.651) {};
\node[transn] at (2.269,0.684) {};
\node[transn] at (3.142,3.870) {};
\node[transn] at (3.830,5.445) {};
\node[transn] at (1.839,3.707) {};
\node[termn] at (4.879,2.767) {};
\node[transn] at (3.232,5.440) {};
\node[hubn] at (4.859,2.490) {};
\node[transn] at (5.305,1.369) {};
\node[hubn] at (1.545,3.466) {};
\node[hubn] at (4.766,3.132) {};
\node[termn] at (2.028,0.000) {};
\node[termn] at (5.275,4.631) {};
\node[transn] at (2.412,4.338) {};
\node[transn] at (1.267,4.858) {};
\node[transn] at (2.169,5.884) {};
\node[transn] at (1.766,1.302) {};
\node[hubn] at (3.981,3.597) {};
\node[hubn] at (4.556,3.567) {};
\node[hubn] at (1.656,2.643) {};
\node[hubn] at (2.772,3.610) {};
\node[hubn] at (3.322,1.266) {};
\node[hubn] at (4.453,2.624) {};
\node[transn] at (2.043,1.945) {};
\node[hubn] at (3.451,5.106) {};
\node[termn] at (0.000,2.544) {};
\node[hubn] at (2.898,2.808) {};
\node[hubn] at (1.616,4.195) {};
\node[termn] at (3.151,6.270) {};
\node[transn] at (5.410,3.661) {};
\node[hubn] at (3.110,2.982) {};
\node[transn] at (0.955,2.277) {};
\node[hubn] at (4.399,3.906) {};
\node[hubn] at (3.311,3.156) {};
\node[transn] at (3.602,2.502) {};
\node[hubn] at (2.117,2.715) {};
\node[termn] at (3.484,0.504) {};
\node[hubn] at (4.101,2.320) {};
\node[hubn] at (3.329,2.774) {};
\node[hubn] at (2.471,4.141) {};
\node[hubn] at (4.459,4.277) {};
\node[hubn] at (1.489,2.048) {};
\node[transn] at (1.432,5.601) {};
\node[hubn] at (3.186,2.068) {};
\node[hubn] at (3.505,2.284) {};
\node[termn] at (4.829,1.794) {};
\node[hubn] at (2.099,3.186) {};
\node[termn] at (6.100,3.809) {};
\node[transn] at (4.548,1.688) {};
\node[transn] at (4.130,1.388) {};
\node[hubn] at (2.845,1.583) {};
\node[hubn] at (1.261,1.655) {};
\node[hubn] at (3.801,3.789) {};
\node[hubn] at (4.074,2.950) {};
\node[hubn] at (2.741,2.381) {};
\node[hubn] at (4.076,4.290) {};
\node[termn] at (4.919,0.125) {};
\node[hubn] at (3.976,2.663) {};
\node[hubn] at (2.518,3.336) {};
\node[hubn] at (2.424,2.677) {};
\node[transn] at (1.931,4.023) {};
\node[termn] at (1.907,4.487) {};
\node[hubn] at (2.356,2.031) {};
\node[hubn] at (3.591,2.958) {};
\node[hubn] at (0.622,1.822) {};
\node[hubn] at (2.923,4.527) {};
\node[transn] at (1.724,3.403) {};
\node[hubn] at (4.737,3.407) {};
\node[hubn] at (1.012,3.200) {};
\node[hubn] at (2.388,3.233) {};
\node[hubn] at (2.482,3.732) {};
\node[transn] at (1.736,1.828) {};
\node[hubn] at (3.011,3.408) {};
\node[hubn] at (1.359,2.762) {};
\node[transn] at (4.205,5.070) {};
\node[hubn] at (2.905,5.552) {};
\node[hubn] at (1.660,3.117) {};
\node[hubn] at (3.859,2.139) {};
\node[hubn] at (0.746,2.539) {};
\node[hubn] at (3.822,4.411) {};
\node[hubn] at (2.536,1.462) {};
\node[transn] at (3.052,4.381) {};
\node[transn] at (3.179,2.445) {};
\node[hubn] at (3.631,3.675) {};
\node[termn] at (2.776,0.717) {};
\node[hubn] at (4.699,4.159) {};
\node[hubn] at (3.129,4.677) {};
\node[srcn] at (3.822,4.411) {};
\node[srcn] at (3.142,3.870) {};
\node[srcn] at (1.267,4.858) {};
\node[srcn] at (2.039,1.195) {};
\node[srcn] at (2.482,3.732) {};
\node[srcn] at (1.931,4.023) {};
\node[tgtn] at (3.602,2.502) {};
\node[tgtn] at (3.451,5.106) {};
\node[tgtn] at (2.269,0.684) {};
\node[tgtn] at (1.545,3.466) {};
\node[tgtn] at (3.591,2.958) {};
\node[tgtn] at (4.829,1.794) {};
\node[anchor=north,font=\footnotesize] at (3.050,-0.25) {(b) ER ($n=93$, $g=48$)};
\end{tikzpicture}
\end{minipage}

\vspace{0.6em}
{\footnotesize
\tikz\node[hubn]{}; hub ($K(v)>0$)\quad
\tikz\node[transn]{}; transit ($K(v)=0$)\quad
\tikz\node[termn]{}; terminal ($K(v)<0$)\quad
\tikz\node[srcn]{}; source\quad
\tikz\node[tgtn]{}; target\\[0.2em]
coloured lines: shortest-path routes of the optimal assignment}
\caption{The two network families at $n\approx10^2$, with six particles routed
along shortest paths of the optimal assignment. (a) In the honeycomb every link
joins geometric neighbours; boundary nodes are transit vertices, interior nodes
are hubs of degree~$3$, and routes are long and evenly spread. (b) The
Erd\H{o}s--Rényi network (giant component of $G(n,p)$ with mean degree~$3$,
drawn with a Kamada--Kawai layout, which carries no intrinsic geometry) links
near and far nodes alike. Its degrees are heterogeneous, with terminals (red)
and hubs of degree up to~$8$, and routes are short and pass through hubs. The
genus densities are comparable ($g/n=0.38$ and $0.52$); the diameters ($19$ and
$9$) are not, and for these six particles the optimal mean route length is
$3.50$ against $3.00$ hops. The gap widens with $n$ (\cref{tab:simstruct,tab:simeff}).}
\label{fig:netexamples}
\end{figure}

\emph{Transport.} In each trial we place $k$ particles on distinct random
source nodes and $k$ target nodes ($k=24$ for $n\approx10^2$, $k=50$
otherwise), compute the hop-distance cost matrix $C$ by breadth-first search,
and solve:
\begin{enumerate}[label=(\alph*)]
\item the exact transport problem with uniform masses, whose optimal value
$W_{\rm opt}$ is the mean number of hops per particle;
\item the $F_q$-regularized problem for $q\in\{\tfrac12,1,2\}$ at relative
temperature $\tau=\gamma/\operatorname{med}(C)\in\{0.03,0.1,0.3\}$ (cost
normalized by its median);
\item the JSAI rounding~\cite{JSAI2026}, a one-to-one assignment maximizing
$\sum_iP_{i\sigma(i)}$;
\item routing of the optimal assignment along shortest paths, recording the
load on every node.
\end{enumerate}
Each configuration is repeated $20$ times. The Jacobian order $\kappa(G)$ is
computed exactly by sparse $LU$ factorization of the reduced Laplacian, except
for the $10^5$-node ER network, where stochastic Lanczos quadrature is used
(standard error $72$ on $\log\kappa\approx7.8\times10^4$; on all smaller
networks it agrees with the exact value to within $0.7\%$).

\subsection{Structural invariants: close in genus density and tree entropy, far apart metrically}\label{ssec:struct}

\begin{table}[!htbp]
\centering\small
\setlength{\tabcolsep}{4pt}
\begin{tabular}{l rrrr rrrr}
\toprule
 & \multicolumn{4}{c}{Honeycomb} & \multicolumn{4}{c}{ER ($c=3$, giant component)} \\
\cmidrule(lr){2-5}\cmidrule(lr){6-9}
$n$ & 96 & 1010 & 9938 & 99902 & 93 & 991 & 9999 & 99920 \\
$\lvert E\rvert$ & 131 & 1471 & 14767 & 149407 & 140 & 1602 & 15714 & 158765 \\
$g=\lvert E\rvert-n+1$ & 36 & 462 & 4830 & 49506 & 48 & 612 & 5716 & 58846 \\
\midrule
$g/n$                     & 0.375 & 0.457 & 0.486 & 0.496 & 0.516 & 0.618 & 0.572 & 0.589 \\
$\log\kappa(G)/n$         & 0.631 & 0.750 & 0.789 & 0.802 & 0.673 & 0.787 & 0.760 & 0.778 \\
$\diam(G)$                & 19 & 65 & 209 & 669 & 9 & 14 & 18 & 23 \\
mean distance             & 7.43 & 24.2 & 74.5 & 240.1 & 4.25 & 6.21 & 8.52 & 10.5 \\
terminals ($K<0$)         & 0 & 0 & 0 & 0 & 0.140 & 0.159 & 0.153 & 0.148 \\
transit ($K=0$)           & 0.271 & 0.087 & 0.028 & 0.009 & 0.280 & 0.223 & 0.246 & 0.240 \\
hubs ($K>0$)              & 0.729 & 0.913 & 0.972 & 0.991 & 0.581 & 0.618 & 0.602 & 0.612 \\
maximal degree            & 3 & 3 & 3 & 3 & 8 & 10 & 12 & 12 \\
\bottomrule
\end{tabular}
\caption{Structural invariants. The last three rows give the fractions of
terminal, transit and hub vertices in the sense of \cref{rem:curv}.}
\label{tab:simstruct}
\end{table}

\Cref{tab:simstruct} shows that the two families are close as far as the
divisor theory of \cref{sec:divisors} can see. The genus densities converge to
$g/n\approx0.50$ (honeycomb, where interior vertices have degree $3$) and
$g/n\approx0.59$ (ER, mean degree $\approx3.18$ on the giant component). By
\cref{prop:jac}, each degree carries $\kappa(G)=e^{hn}$ linear equivalence
classes, with spanning-tree entropy $h=\log\kappa/n$ close to $0.80$ for the
honeycomb (the infinite-lattice value is $0.8077$~\cite{Wu1977}, approached
from below because of the boundary) and $0.78$ for ER. The Jacobians are
therefore of almost the same exponential size. With respect to the invariants
compared here (genus density and spanning-tree entropy), the two networks are
\emph{topological twins}; we do not claim that they agree in every topological
invariant.

With respect to diameter, distance distribution and degree distribution they are \emph{metric opposites}. The honeycomb diameter grows as
$n^{0.507}$, a planar law, whereas the ER diameter grows by about $2$ per
factor $e$ in $n$, a small-world law. The canonical divisor also distinguishes
them. In the honeycomb almost every vertex is a hub of curvature $-\tfrac12$
and the total curvature $\chi=1-g$ is spread uniformly. In the ER network
about $15\%$ of the vertices are terminals ($K=-1$), and the hubs reach
$K=10$. The Gauss--Bonnet identity of \cref{rem:curv} holds in both, but the
curvature is homogeneous in one network and highly heterogeneous in the other.

\subsection{Transport efficiency and scaling}\label{ssec:eff}

\begin{table}[!htbp]
\centering\small
\setlength{\tabcolsep}{4pt}
\begin{tabular}{l rrrr rrrr}
\toprule
 & \multicolumn{4}{c}{Honeycomb} & \multicolumn{4}{c}{ER} \\
\cmidrule(lr){2-5}\cmidrule(lr){6-9}
$n$ & $10^2$ & $10^3$ & $10^4$ & $10^5$ & $10^2$ & $10^3$ & $10^4$ & $10^5$ \\
\midrule
$W_{\rm opt}$ (hops)          & 2.58 & 5.71 & 20.2 & 58.9 & 2.02 & 3.22 & 5.30 & 7.31 \\
\quad std.\ dev.              & 0.63 & 0.64 & 3.77 & 6.49 & 0.20 & 0.18 & 0.23 & 0.17 \\
mean cost $\overline C$       & 7.44 & 24.0 & 75.9 & 243.2 & 4.28 & 6.14 & 8.51 & 10.5 \\
gain $\overline C/W_{\rm opt}$ & 3.03 & 4.26 & 3.89 & 4.18 & 2.14 & 1.91 & 1.61 & 1.44 \\
cost dispersion (CV)          & 0.47 & 0.48 & 0.48 & 0.48 & 0.36 & 0.25 & 0.19 & 0.15 \\
equidistant targets           & 0.11 & 0.05 & 0.03 & 0.02 & 0.24 & 0.24 & 0.23 & 0.23 \\
hub attraction                & 1.02 & 1.01 & 1.00 & 1.00 & 1.15 & 1.23 & 1.32 & 1.38 \\
maximal node load             & 2.75 & 2.30 & 2.25 & 2.00 & 3.00 & 2.05 & 1.70 & 1.05 \\
\bottomrule
\end{tabular}
\caption{Optimal transport statistics (means over $20$ trials). ``Gain'' is
the ratio of the cost of a random assignment to the optimal cost. ``Equidistant
targets'' is the probability that two targets lie at the same distance from a
source. ``Hub attraction'' is the load-weighted mean degree on optimal routes
divided by the mean degree.}
\label{tab:simeff}
\end{table}

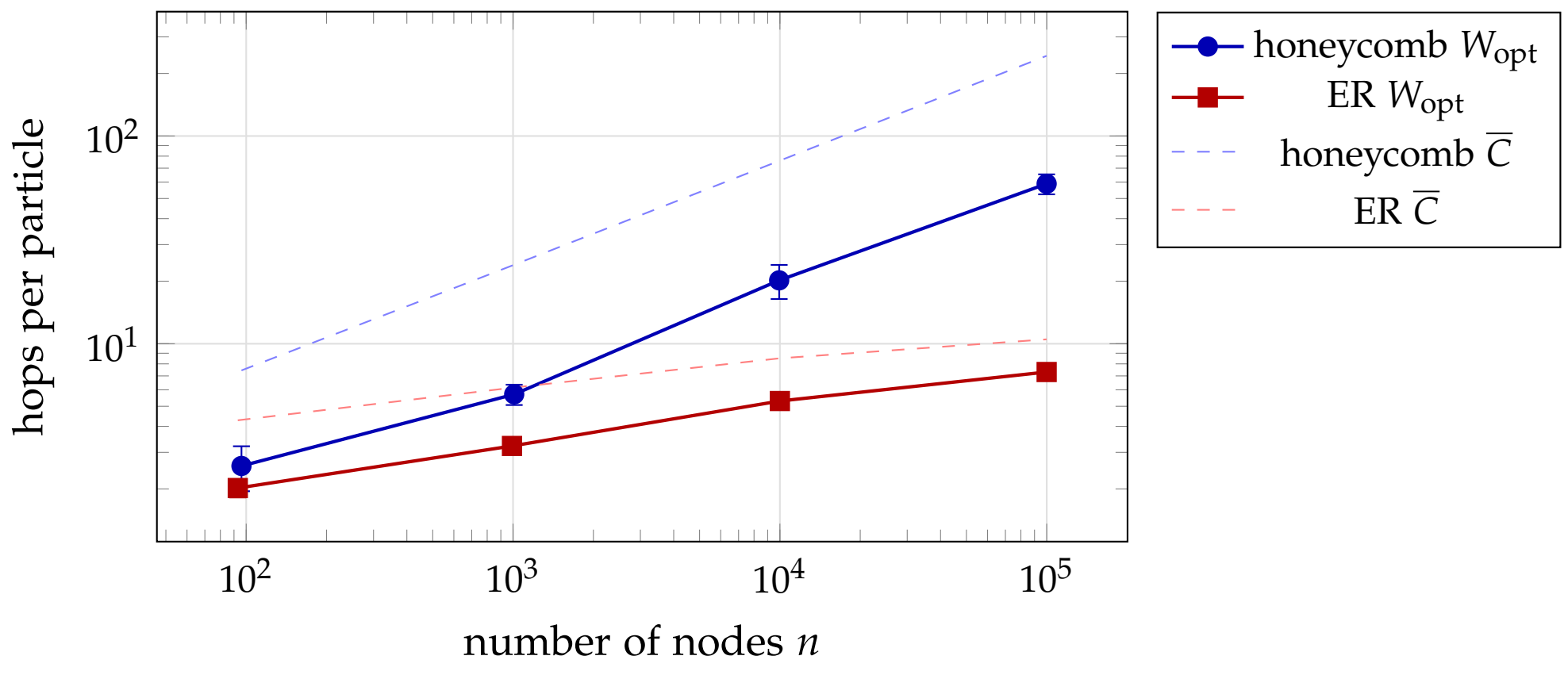
\begin{figure}[!htbp]
\centering
\begin{tikzpicture}
\begin{axis}[width=0.62\textwidth,height=6.0cm,xmode=log,ymode=log,
  xlabel={number of nodes $n$},ylabel={hops per particle},
  legend style={font=\footnotesize,at={(1.03,1.0)},anchor=north west},
  tick label style={font=\footnotesize},label style={font=\small},
  grid=major,grid style={black!12}]
\addplot[blue!70!black,thick,mark=*,error bars/.cd,y dir=both,y explicit]
  coordinates {(96,2.58)+-(0,0.63) (1010,5.71)+-(0,0.64) (9938,20.19)+-(0,3.77) (99902,58.88)+-(0,6.49)};
\addplot[red!70!black,thick,mark=square*,error bars/.cd,y dir=both,y explicit]
  coordinates {(93,2.02)+-(0,0.20) (991,3.22)+-(0,0.18) (9999,5.30)+-(0,0.23) (99920,7.31)+-(0,0.17)};
\addplot[blue!50,dashed] coordinates {(96,7.44) (1010,24.0) (9938,75.9) (99902,243.2)};
\addplot[red!50,dashed] coordinates {(93,4.28) (991,6.14) (9999,8.51) (99920,10.5)};
\legend{{honeycomb $W_{\rm opt}$},{ER $W_{\rm opt}$},{honeycomb $\overline C$},{ER $\overline C$}}
\end{axis}
\end{tikzpicture}
\caption{Optimal hops per particle (solid, $\pm1$ s.d.) and mean pairwise cost
(dashed). Least-squares fits over $n\approx10^3$--$10^5$: honeycomb
$W_{\rm opt}\propto n^{0.508\pm0.010}$; ER $W_{\rm opt}\approx(0.885\pm0.014)\log n+\mathrm{const}$.}
\label{fig:simscale}
\end{figure}

\Cref{tab:simeff} and \cref{fig:simscale} show the expected dichotomy. We
fitted the $60$ trials at $n\approx10^3,10^4,10^5$ by ordinary least squares.
These sizes all use $k=50$; the size $n\approx10^2$, which appears in the tables,
is excluded from the fits because it uses $k=24$ particles, and the cost depends
on $k$ (\cref{tab:robust}). Since only one network per family and size enters these fits,
the standard errors below measure variation over particle placements only. They
must not be read as uncertainty over the network ensemble; variation between
network realizations is assessed separately in \cref{tab:robust}. On the honeycomb,
$\log W_{\rm opt}=a+b\log n$ gives $b=0.508\pm0.010$ ($R^2=0.978$, against
$0.903$ for a fit linear in $\log n$). On the ER network a fit linear in
$\log n$ gives slope $0.885\pm0.014$ ($R^2=0.986$, against $0.969$ for a power
law, whose exponent is $0.178\pm0.004$). These are \emph{observations} over three
sizes, consistent with the diameter laws $n^{1/2}$ and $\log n$ of the two
families; they are not scaling theorems.

\emph{Robustness check.} This is a separate experiment, with the following
design.
\begin{itemize}
\item \emph{Networks.} Honeycomb: the same (unique) patch of each size as in
the main experiment. ER: five new realizations per size, seeds $1000$--$1004$,
generated as in \cref{app:computation}.
\item \emph{Placements.} For each network and each $k\in\{25,50,100\}$, $10$
placements of $k$ sources and $k$ targets, drawn as $2k$ distinct vertices
uniformly at random (generator seed $2026$). Only exact transport is computed.
\item \emph{Honeycomb exponent.} The OLS slope of $\log\overline W_{\rm opt}$ on
$\log n$, where $\overline W_{\rm opt}$ is the mean over the $10$ placements: three
points and one degree of freedom, so the quoted standard error is crude.
\item \emph{ER slope.} The OLS slope of the $15$ realization means
($5$ realizations $\times3$ sizes) on $\log n$.
\item \emph{Variance split (ER only).} For each $(n,k)$, a one-way
random-effects analysis of variance with the $5$ realizations as groups and the
$10$ placements as replicates. With between- and within-group mean squares
$\mathrm{MSB}$ and $\mathrm{MSW}$, the network variance is
$s^2_{\rm net}=\max\bigl(0,(\mathrm{MSB}-\mathrm{MSW})/10\bigr)$, and the network
share is $s^2_{\rm net}/(s^2_{\rm net}+\mathrm{MSW})$.
\end{itemize}
The results are in \cref{tab:robust}; the script \texttt{robust\_fit.py}
re-derives every entry from the stored data. The qualitative picture is
unchanged. The honeycomb exponent lies between $0.49$ and $0.52$ for all three
values of $k$. The ER logarithmic slope lies between $0.86$ and $0.91$,
decreasing slightly with $k$. The standard deviation of the five realization
means is at most $0.16$ hops, about $3\%$ of the mean; it contains placement
noise as well. The network component alone has $s_{\rm net}\le0.13$ hops and
accounts for $0$--$22\%$ of the variance over the nine $(n,k)$ cells; the rest
is placement variance. The honeycomb has no network component. At fixed
$n=10^5$ the honeycomb cost scales roughly as $k^{-0.46}$, close to the
$k^{-1/2}$, up to logarithmic factors, expected for random bipartite matchings
in the plane. The slopes therefore depend on the particle number, and we
report them as properties of this experimental design, not as universal
constants of the two families. The main-fit standard errors measure variation
over placements on the selected network only. The robustness study estimates
network-to-network variation for ER realizations. It cannot supply an ensemble
variance for the honeycomb, whose patch of a given size is unique. With three
sizes, several laws can fit nearly
as well (\cref{ssec:eff} compares a power law and a logarithmic law). The fitted
laws are observations within $10^3\le n\le10^5$, consistent with the stated
scalings; we do not claim that the families obey them asymptotically.

\begin{table}[!htbp]
\centering\small
\setlength{\tabcolsep}{3.4pt}
\begin{tabular}{c ccc c ccc c}
\toprule
 & \multicolumn{4}{c}{Honeycomb} & \multicolumn{4}{c}{ER (mean $\pm$ s.d.\ over 5 realizations)} \\
\cmidrule(lr){2-5}\cmidrule(lr){6-9}
$k$ & $10^3$ & $10^4$ & $10^5$ & exponent & $10^3$ & $10^4$ & $10^5$ & slope in $\log n$ \\
\midrule
25  & 8.58 & 29.3 & 81.1 & $0.489\pm0.028$ & $3.76\pm0.12$ & $5.87\pm0.04$ & $7.96\pm0.16$ & $0.911\pm0.015$ \\
50  & 5.92 & 18.6 & 64.7 & $0.520\pm0.011$ & $3.18\pm0.09$ & $5.17\pm0.09$ & $7.26\pm0.07$ & $0.885\pm0.012$ \\
100 & 4.47 & 14.6 & 42.6 & $0.491\pm0.016$ & $2.69\pm0.08$ & $4.62\pm0.07$ & $6.64\pm0.07$ & $0.857\pm0.010$ \\
\bottomrule
\end{tabular}
\caption{Robustness of the scaling observations: $W_{\rm opt}$ in hops at
$n\approx10^3,10^4,10^5$ (exact transport, $10$ placements per network). The ER
fits use all $15$ realization means.}
\label{tab:robust}
\end{table} At $n=10^5$ a particle travels $59$ hops on the honeycomb
and $7.3$ on the ER network. Optimization pays off on the honeycomb, where
the optimal assignment is about four times cheaper than a random one at every
scale. On the ER network the advantage shrinks from $2.1$ to $1.44$,
because distances concentrate: the coefficient of variation of the cost falls
from $0.36$ to $0.15$, and almost a quarter of all pairs of targets are
equidistant from a given source.

\emph{Interpretation.} By \cref{thm:chipmove}, $k\,W_{\rm opt}$ is exactly the
minimal number of single-chip moves that carry the source configuration to the
target configuration through effective configurations: about $2900$ moves on
the $10^5$-node honeycomb and $370$ on the ER network. Every upper bound
proved in \cref{sec:defect,sec:crossover} for a Wasserstein cost carries the
factor $\diam(G)$ (\cref{lem:compare}, \cref{thm:defect}(a), \cref{thm:AJ},
and the upper edge of the window in \cref{thm:crossover}(ii)). The factor
$\diam(G)\,g/n$ is $332$ on the large honeycomb and only $13.5$ on the large
ER network. Since the topological factor $g/n$ is nearly the same in both
families, the difference in transport cost cannot be attributed to these two
invariants. It is consistent with the metric differences (diameter, distance
and degree distributions) that the upper bounds expose, although the
experiment does not isolate a single cause.

\subsection{How particles travel: routes, hubs and curvature}\label{ssec:routes}

Optimal routes on the honeycomb spread evenly: the load-weighted degree equals
the mean degree (hub attraction $1.00$), and paths rarely meet (maximal node
load about $2$). On the ER network shortest paths are drawn towards
high-degree vertices: the hub attraction rises from $1.15$ to $1.38$ with $n$,
so a typical route node has $38\%$ more links than an average node.
Terminals carry no through traffic, since a leaf can only be the endpoint of a
shortest path. This is the transport content of the curvature reading of
\cref{rem:curv}. Positive $K(v)$ marks branching vertices that collect and
redistribute routes, while negative $K(v)$ marks dead ends; a curvature
distribution as heterogeneous as the ER one produces hub-centred transport,
and a uniform one produces diffuse transport. With $k=50$ particles the
networks are far from congestion, so the loads stay small in absolute terms;
the concentration on hubs is nonetheless systematic and grows with $n$.

\subsection{Regularization: three regimes and the role of \texorpdfstring{$q$}{q}}\label{ssec:reg}

\begin{table}[!htbp]
\centering\small
\setlength{\tabcolsep}{3.6pt}
\begin{tabular}{cc cccc cccc}
\toprule
 & & \multicolumn{4}{c}{Honeycomb, $n=99902$} & \multicolumn{4}{c}{ER, $n=99920$} \\
\cmidrule(lr){3-6}\cmidrule(lr){7-10}
$q$ & $\tau$ & excess & spread & zeros & rounded & excess & spread & zeros & rounded \\
\midrule
$\tfrac12$ & 0.03 & 3.64 & 47.7 & 0 & 1.011 & 1.43 & 49.9 & 0 & 1.000 \\
           & 0.10 & 4.03 & 49.8 & 0 & 1.001 & 1.44 & 50.0 & 0 & 1.000 \\
           & 0.30 & 4.13 & 50.0 & 0 & 1.000 & 1.44 & 50.0 & 0 & 1.000 \\
\addlinespace
$1$        & 0.03 & 1.05 & 3.5  & --- & 1.014 & 1.02 & 3.1  & --- & 1.002 \\
           & 0.10 & 1.27 & 8.2  & --- & 1.034 & 1.19 & 20.2 & --- & 1.002 \\
           & 0.30 & 2.07 & 24.0 & --- & 1.031 & 1.37 & 45.8 & --- & 1.000 \\
\addlinespace
$2$        & 0.03 & 1.000 & 1.23 & 0.94 & 1.000 & 1.000 & 1.37 & 0.96 & 1.005 \\
           & 0.10 & 1.000 & 1.26 & 0.96 & 1.000 & 1.000 & 1.37 & 0.96 & 1.005 \\
           & 0.30 & 1.001 & 1.42 & 0.96 & 1.000 & 1.000 & 1.37 & 0.96 & 1.005 \\
\bottomrule
\end{tabular}
\caption{Regularized plans at $n\approx10^5$ ($k=50$, means over $20$
trials). ``Excess'' is the expected cost divided by $W_{\rm opt}$; ``spread''
is $e^{H(P)}/k$, which equals $1$ for a permutation and $k=50$ for the product
coupling; ``zeros'' is the fraction of exact zeros (for $q=1$ all entries are
positive by \cref{thm:support}); ``rounded'' is the cost of the JSAI rounding
divided by $W_{\rm opt}$. The values at $n\approx10^3$ and $10^4$ are similar.}
\label{tab:simreg}
\end{table}

\Cref{tab:simreg} exhibits three regimes, and each is a theorem of the paper
read at scale.

\emph{(1) $q=\tfrac12$: information dominates.} The plans are essentially the
product coupling (spread $\approx50$), and the excess cost equals the gain of
\cref{tab:simeff} ($4.13$ against $4.18$ on the honeycomb, $1.44$ against
$1.44$ on ER): the plan has forgotten the geometry. By
\cref{thm:gauge,thm:crossover}, the informational correction is
$(\gamma/q)\ln_q(N)$ with $N=k^2=2500$ cells, and $\ln_{1/2}(2500)=98$,
$\ln_1(2500)=7.8$ and $\ln_2(2500)=0.9996$. At $\tau=0.03$ the correction is
$5.9$ for $q=\tfrac12$, against $0.23$ for $q=1$ and $0.015$ for $q=2$, in
units of the median cost. The entropic term therefore exceeds the whole cost
scale only when $q=\tfrac12$. This is the statement that $q$ is an
inverse-temperature multiplier, made quantitative through both $\gamma/q$ and
$\ln_q(N)$.

\emph{Fixed effective temperature.} To separate the factor $1/q$ from the shape
of $S_q$, we repeated the computation at fixed $\varepsilon=\gamma/q$, i.e.\ with
$\gamma=q\varepsilon$ (experiment~B; $10$ placements per network, generator seed
$31$; \cref{tab:fixeps}). The ordering of the three regimes persists:
\begin{itemize}
\item at $q=\tfrac12$ the plans remain close to the product coupling (spread
$41$--$50$);
\item at $q=2$ they remain essentially optimal and sparse;
\item $q=1$ is unchanged by construction.
\end{itemize}
The $q$-dependence that survives at fixed $\varepsilon$ is therefore carried by
the shape of $S_q$, quantified by $\ln_q(N)$, not by the rescaling $\gamma/q$
alone. These are observations for the networks, particle configurations and
parameter ranges tested here; we do not state them as a transition theorem in
$q$. One of the $540$ solves in this experiment returned a solver warning
(``optimal\_inaccurate''); the averages are insensitive to it.

\begin{table}[!htbp]
\centering\small
\setlength{\tabcolsep}{3.6pt}
\begin{tabular}{cc ccc ccc}
\toprule
 & & \multicolumn{3}{c}{Honeycomb, $n=99902$} & \multicolumn{3}{c}{ER, $n=99920$} \\
\cmidrule(lr){3-5}\cmidrule(lr){6-8}
$q$ & $\varepsilon$ & excess & spread & zeros & excess & spread & zeros \\
\midrule
$\tfrac12$ & 0.03 & 2.99 & 41.5 & 0 & 1.41 & 49.3 & 0 \\
           & 0.10 & 3.73 & 49.2 & 0 & 1.43 & 50.0 & 0 \\
           & 0.30 & 3.93 & 49.9 & 0 & 1.44 & 50.0 & 0 \\
\addlinespace
$2$        & 0.03 & 1.000 & 1.31 & 0.96 & 1.000 & 1.36 & 0.96 \\
           & 0.10 & 1.000 & 1.32 & 0.96 & 1.000 & 1.36 & 0.96 \\
           & 0.30 & 1.001 & 1.53 & 0.96 & 1.000 & 1.40 & 0.96 \\
\bottomrule
\end{tabular}
\caption{Experiment~B at $n\approx10^5$: regularized plans at fixed effective
temperature $\varepsilon=\gamma/q$ (relative to the median cost; means over $10$
placements). The values at $n\approx10^3$ and $10^4$ are similar (supplementary
file \texttt{fixeps.json}).}
\label{tab:fixeps}
\end{table}

\emph{(2) $q=2$: geometry dominates, and plans are sparse.} The excess cost is
$1.000$ and $94$--$96\%$ of the entries are exact zeros, whereas for $q\le1$ no
entry vanishes. The second fact is guaranteed by \cref{thm:support}(i). The
first is an observation: for $q>1$ the theorem permits zeros but does not
guarantee them, since whether they occur depends on the costs, the marginals
and $\gamma$. Here they occur in most entries.

\emph{(3) $q=1$: a temperature-driven crossover whose scale is set by cost
dispersion.} At the same relative temperature the ER plan spreads much
faster than the honeycomb plan (spread $45.8$ against $24.0$ at $\tau=0.3$),
yet its excess cost stays smaller ($1.37$ against $2.07$). Both facts follow
from \cref{thm:crossover}. In the normalized high-temperature limit the
distance cost drops out directly; what remains is determined by the marginals
and, for the gap $\Gamma$, by the linear systems involved. For the plans of this
experiment the marginals are uniform on the $k$ particles, so the limiting plan
is the uniform product coupling on both networks. Graph structure would re-enter
through the marginals or the linear systems, neither of which varies here.
The low-temperature limit is the metric cost. The
temperature at which entropy takes over is therefore set by the spread of
costs on the scale of the median, which is the dispersion row of
\cref{tab:simeff}. Concentrated ER distances (CV $0.15$, $23\%$ ties) leave
little for the cost term to discriminate, so the crossover occurs at a lower
relative temperature, but it is also cheap. Dispersed honeycomb distances (CV
$0.48$) resist entropy for longer and make it expensive once it wins.

\emph{Rounding hides all of this.} After the JSAI rounding, every plan at $n\approx10^5$ is
within $3.4\%$ of the optimum on the honeycomb and within $0.5\%$ on ER
(at most $4.0\%$ and $0.9\%$ over all sizes),
even at $q=\tfrac12$ where the plan itself is a product coupling. Differences
between regularizers that are measured \emph{after} rounding to a one-to-one
assignment, as in the total distances of~\cite{JSAI2026}, are therefore
second-order. Combined with \cref{thm:gauge}, which shows that $F_q$ and the
power-law regularizer coincide up to $\gamma\mapsto\gamma/q$, this is a further
reason to read those tables with care.

\subsection{Where the Riemann--Roch theorem acts: particle density}\label{ssec:density}

All the transport experiments above use $k\le50$ particles on networks with
$g\approx n/2$. In divisor language the particle configuration is a divisor of
degree $d=k<g$ (indeed $d\ll g$ for $n\ge10^3$), which lies deep in the \emph{special} range. There
\cref{cor:RR}(i) gives only $r(D)\ge d-g<0$. The upper bound
$\min(1,g/d)$ of \cref{thm:defect} is informative only for $d>g$, every class
is reachable by an effective configuration only for $d\ge g$, and
configuration independence requires $d\ge2g-1$. (The lower bound $1/d$ and the
existence of a positive gap hold for every $d\ge1$, but they say only that the
Jacobian is non-trivial.) Translated into densities of particles per node:
\begin{gather*}
  \text{steerability:}\quad \frac dn\ge\frac gn\approx0.50\ \text{(honeycomb)},\ 0.59\ \text{(ER)};\\
  \text{configuration independence:}\quad \frac dn\ge\frac{2g-1}n\approx0.99\ \text{(honeycomb)},\ 1.18\ \text{(ER)}.
\end{gather*}
Riemann--Roch phenomena therefore require of the order of one particle per
node. In the sparse regime of practical particle transport (the
experiments of~\cite{JSAI2026} use $12$--$15$ particles on $60$ nodes, a
density of $0.20$--$0.25$), the Riemann--Roch bound $r(D)\ge d-g$ is vacuous, so the theory guarantees no
freedom to steer particle configurations by chip-firing. It is consistent with
this that the large-scale observations are described by metric quantities
($\diam(G)$, cost dispersion, curvature heterogeneity) rather than by divisor
theory. It is also why the topological gap of \cref{sec:crossover}
must be probed at high density, as we do next.

\subsection{An exact test at matched genus}\label{ssec:matched}

To see the theorems act directly, we compare the honeycomb patch with
$n=16$ nodes and $m=19$ edges ($g=4$) with six connected random graphs
$G(16,19)$, which have the same $n$, $m$ and $g$. For $D=dv_1$ we compute
exactly the rank, the linear system, $\delta_{\TV}$, $d\,\delta_{W_1}$ (by
\cref{thm:chipmove}) and, for the worst target $E$, the $q=1$ crossover
temperature $\gamma^\ast$ of \cref{thm:crossover}. We use $d=4=g$ (special
range) and $d=7=2g-1$ (non-special range).

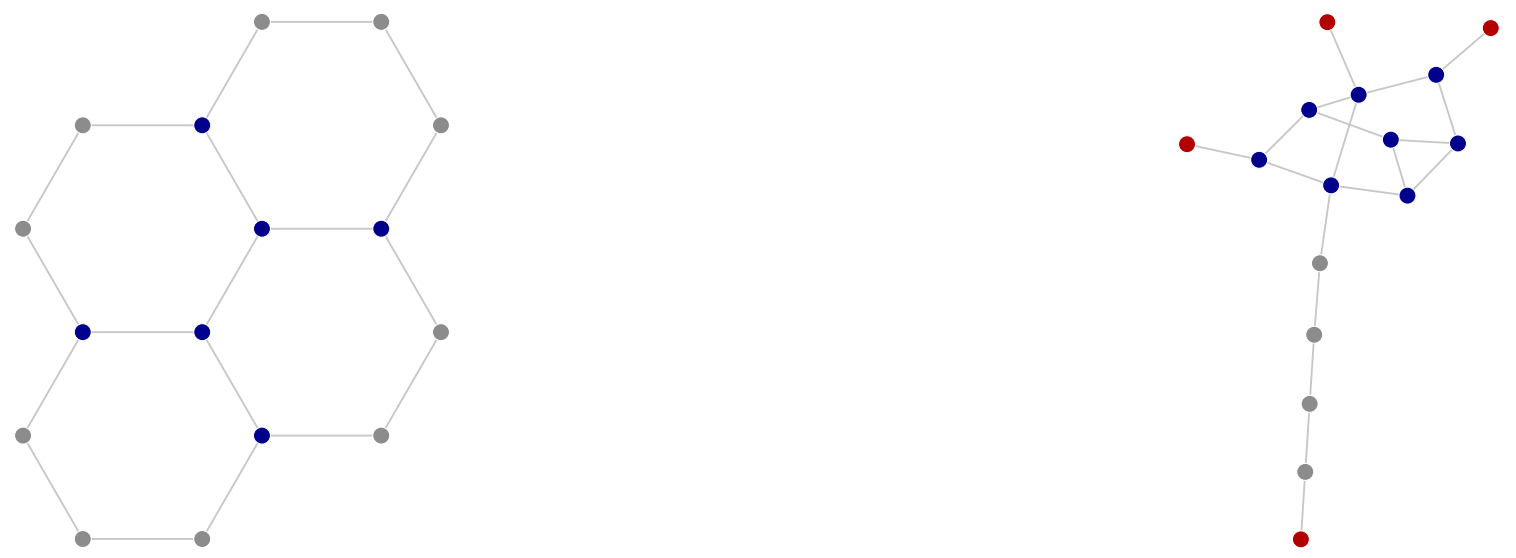
\begin{figure}[!htbp]
\centering
\begin{minipage}[b]{0.42\textwidth}\centering
\begin{tikzpicture}[x=1cm,y=1cm]
\draw[black!22,line width=0.35pt] (0.395,0.000)--(0.000,0.684);
\draw[black!22,line width=0.35pt] (0.395,0.000)--(1.185,0.000);
\draw[black!22,line width=0.35pt] (0.000,0.684)--(0.395,1.368);
\draw[black!22,line width=0.35pt] (0.395,1.368)--(0.000,2.052);
\draw[black!22,line width=0.35pt] (0.395,1.368)--(1.185,1.368);
\draw[black!22,line width=0.35pt] (0.000,2.052)--(0.395,2.736);
\draw[black!22,line width=0.35pt] (0.395,2.736)--(1.185,2.736);
\draw[black!22,line width=0.35pt] (1.185,0.000)--(1.580,0.684);
\draw[black!22,line width=0.35pt] (1.580,0.684)--(1.185,1.368);
\draw[black!22,line width=0.35pt] (1.580,0.684)--(2.369,0.684);
\draw[black!22,line width=0.35pt] (1.185,1.368)--(1.580,2.052);
\draw[black!22,line width=0.35pt] (1.580,2.052)--(1.185,2.736);
\draw[black!22,line width=0.35pt] (1.580,2.052)--(2.369,2.052);
\draw[black!22,line width=0.35pt] (1.185,2.736)--(1.580,3.420);
\draw[black!22,line width=0.35pt] (1.580,3.420)--(2.369,3.420);
\draw[black!22,line width=0.35pt] (2.369,0.684)--(2.764,1.368);
\draw[black!22,line width=0.35pt] (2.764,1.368)--(2.369,2.052);
\draw[black!22,line width=0.35pt] (2.369,2.052)--(2.764,2.736);
\draw[black!22,line width=0.35pt] (2.764,2.736)--(2.369,3.420);
\node[transn] at (0.395,0.000) {};
\node[transn] at (0.000,0.684) {};
\node[hubn] at (0.395,1.368) {};
\node[transn] at (0.000,2.052) {};
\node[transn] at (0.395,2.736) {};
\node[transn] at (1.185,0.000) {};
\node[hubn] at (1.580,0.684) {};
\node[hubn] at (1.185,1.368) {};
\node[hubn] at (1.580,2.052) {};
\node[hubn] at (1.185,2.736) {};
\node[transn] at (1.580,3.420) {};
\node[transn] at (2.369,0.684) {};
\node[transn] at (2.764,1.368) {};
\node[hubn] at (2.369,2.052) {};
\node[transn] at (2.764,2.736) {};
\node[transn] at (2.369,3.420) {};
\node[anchor=north,font=\footnotesize] at (1.382,-0.25) {(c) honeycomb patch, $g=4$};
\end{tikzpicture}
\end{minipage}\qquad
\begin{minipage}[b]{0.42\textwidth}\centering
\begin{tikzpicture}[x=1cm,y=1cm]
\draw[black!22,line width=0.35pt] (1.648,3.072)--(1.135,2.940);
\draw[black!22,line width=0.35pt] (1.648,3.072)--(1.792,2.618);
\draw[black!22,line width=0.35pt] (1.648,3.072)--(2.009,3.381);
\draw[black!22,line width=0.35pt] (1.792,2.618)--(1.348,2.643);
\draw[black!22,line width=0.35pt] (1.792,2.618)--(1.458,2.273);
\draw[black!22,line width=0.35pt] (1.135,2.940)--(0.928,3.420);
\draw[black!22,line width=0.35pt] (1.135,2.940)--(0.953,2.341);
\draw[black!22,line width=0.35pt] (1.135,2.940)--(0.808,2.840);
\draw[black!22,line width=0.35pt] (0.808,2.840)--(1.348,2.643);
\draw[black!22,line width=0.35pt] (0.808,2.840)--(0.477,2.510);
\draw[black!22,line width=0.35pt] (0.782,0.446)--(0.811,0.896);
\draw[black!22,line width=0.35pt] (0.782,0.446)--(0.753,0.000);
\draw[black!22,line width=0.35pt] (0.477,2.510)--(0.000,2.613);
\draw[black!22,line width=0.35pt] (0.477,2.510)--(0.953,2.341);
\draw[black!22,line width=0.35pt] (1.348,2.643)--(1.458,2.273);
\draw[black!22,line width=0.35pt] (0.841,1.353)--(0.811,0.896);
\draw[black!22,line width=0.35pt] (0.841,1.353)--(0.879,1.826);
\draw[black!22,line width=0.35pt] (0.879,1.826)--(0.953,2.341);
\draw[black!22,line width=0.35pt] (1.458,2.273)--(0.953,2.341);
\node[hubn] at (1.648,3.072) {};
\node[hubn] at (1.792,2.618) {};
\node[hubn] at (1.135,2.940) {};
\node[termn] at (0.928,3.420) {};
\node[hubn] at (0.808,2.840) {};
\node[transn] at (0.782,0.446) {};
\node[hubn] at (0.477,2.510) {};
\node[termn] at (0.000,2.613) {};
\node[termn] at (2.009,3.381) {};
\node[hubn] at (1.348,2.643) {};
\node[transn] at (0.841,1.353) {};
\node[transn] at (0.879,1.826) {};
\node[hubn] at (1.458,2.273) {};
\node[termn] at (0.753,0.000) {};
\node[transn] at (0.811,0.896) {};
\node[hubn] at (0.953,2.341) {};
\node[anchor=north,font=\footnotesize] at (1.005,-0.25) {(d) ER 2, $g=4$};
\end{tikzpicture}
\end{minipage}
\caption{Matched genus ($n=16$, $m=19$, $g=4$). (c) The honeycomb patch: no
leaves, every vertex a hub or transit vertex, $\kappa=1092$. (d) The random graph
``ER~2'' of \cref{tab:matched}: four terminals (red) and $\kappa=154$. Its
diameter is $8$ against $7$, and it has the largest $W_1$-defect ($7\delta_{W_1}=27$) and
the latest crossover ($\gamma^\ast=2.20$) of the family, while
$\delta_{\TV}=g/d=4/7$ is the same for both.}
\label{fig:matchedpair}
\end{figure}

\begin{table}[!htbp]
\centering\small
\setlength{\tabcolsep}{4.2pt}
\begin{tabular}{l ccccc cc ccc}
\toprule
 & & & & & & \multicolumn{2}{c}{$d=4$ (special)} & \multicolumn{3}{c}{$d=7$ (non-special)} \\
\cmidrule(lr){7-8}\cmidrule(lr){9-11}
Graph & $\kappa$ & $\diam$ & $\rho$ & leaves & max deg & $r(4v_1)$ & $\delta_{\TV}$ &
$\delta_{\TV}$ & $7\delta_{W_1}$ & $\gamma^\ast$ \\
\midrule
honeycomb & 1092 & 7 & 6 & 0 & 3 & 1 & 0.75 & 4/7 & 16 & 1.67 \\
ER 1   & 216  & 6 & 5 & 4 & 5 & 1 & 0.75 & 4/7 & 11 & 1.15 \\
ER 2   & 154  & 8 & 3 & 4 & 4 & 1 & 0.75 & 4/7 & 27 & 2.20 \\
ER 3   & 229  & 7 & 6 & 4 & 5 & 0 & 1.00 & 4/7 & 11 & 1.22 \\
ER 4   & 88   & 6 & 4 & 6 & 6 & 0 & 1.00 & 4/7 & 18 & 1.54 \\
ER 5   & 440  & 7 & 5 & 3 & 4 & 0 & 1.00 & 4/7 & 16 & 1.31 \\
ER 6   & 273  & 6 & 5 & 4 & 4 & 0 & 1.00 & 4/7 & 12 & 1.20 \\
\bottomrule
\end{tabular}
\caption{Exact divisor and transport invariants at matched genus $g=4$
($n=16$, $m=19$). In every case $\rho\le7\delta_{W_1}\le\diam\cdot g$
(\cref{thm:AJ}).}
\label{tab:matched}
\end{table}

\Cref{tab:matched} (with two of the graphs drawn in \cref{fig:matchedpair}) separates cleanly what is topological from what is
metric.

\emph{Special versus non-special.} At $d=4=g$ the rank of $4v_1$ is $1$ on
three graphs and $0$ on four, and $\delta_{\TV}$ follows it ($0.75$ or $1$):
in the special range the configuration and the graph matter. At $d=7=2g-1$
every graph has $r(7v_1)=3$ by \cref{cor:RR}(ii), and every graph has
$\delta_{\TV}=4/7=g/d$ exactly. On these seven graphs the upper bound of
\cref{thm:defect}(a) is attained in every case, and the total-variation defect
depends only on the genus, however different the networks. This is an
observation for these graphs; we do not claim it for all non-special classes
(the class $K+v_1$ on $G''$ is a counterexample; see \cref{sec:conclusion}).

\emph{The metric defect is not topological.} The value $7\delta_{W_1}$ ranges
from $11$ to $27$, always between $\rho(G)$ and $\diam(G)\,g$ as
\cref{thm:AJ} requires. Its largest value occurs on the graph of largest
diameter (ER 2, $\diam=8$), where a few constrained chips must travel
far. The Jacobian orders differ by a factor of $12$ (from $88$ to $1092$; the
honeycomb has by far the largest Jacobian, having no leaves) without ordering
the defects.

\emph{The crossover tracks the metric defect.} Across the seven graphs the
crossover temperature $\gamma^\ast$ correlates with $7\delta_{W_1}$ with
Pearson coefficient $0.955$ (Spearman $0.87$). This is the content of
\cref{thm:crossover}(i)--(ii): the low-temperature value of the gap is
$N_E/d$, the window $\gamma_c\propto N_E/d$ scales with it, and so the
temperature at which information overrides topology is governed by how far the
chips constrained by topology must travel. Topology decides \emph{whether} a
gap exists (\cref{thm:phase}, $g\ge1$); geometry decides \emph{how long} it
survives.

\subsection{Synthesis}\label{ssec:synthesis}

\begin{table}[!htbp]
\centering\small
\renewcommand{\arraystretch}{1.18}
\begin{tabular}{>{\raggedright\arraybackslash}p{3.5cm}>{\raggedright\arraybackslash}p{2.1cm}>{\raggedright\arraybackslash}p{8.1cm}}
\toprule
Result & Evidence & Observation \\
\midrule
\Cref{thm:defect,thm:serre,thm:vertexrank} & exact test &
$1784$ classes (\cref{ssec:pdnum}); $\delta_{\TV}=g/d$ in all seven
matched-genus graphs at $d=2g-1$ (\cref{tab:matched}). \\
\Cref{thm:AJ} & exact test &
$\rho\le d\,\delta_{W_1}\le\diam\cdot g$ in \cref{tab:AJ,tab:matched}. \\
\Cref{thm:crossover,thm:phase,thm:TI} & exact test &
Gap window, decomposition and crossover on the small test graphs and the
matched-genus graphs; crossover correlates with the $W_1$-defect
($r=0.955$). \\
\Cref{thm:minimax,cor:threelevels} & exact test & \Cref{tab:levels}. \\
\Cref{thm:gauge} & large networks &
Ordering $q=\tfrac12\ll1\ll2$ at fixed $\gamma$ and at fixed $\gamma/q$, carried by
$\ln_q(N)$ ($98$, $7.8$, $1.0$ at $N=2500$). \\
\Cref{thm:support,prop:legendre} & large networks &
No zeros for $q\le1$, as proved; for $q=2$, where zeros are permitted but not
guaranteed, $94$--$96\%$ of entries vanish at $n=10^5$. \\
\Cref{thm:face} & large networks &
All $40$ instances degenerate ($\dim O=25$--$56$); $q=\tfrac12$ selects the whole
face, $q=2$ a boundary point in $26$ of $40$. \\
\Cref{prop:jac,prop:invisible}, \cref{lem:compare} & consistency only &
Close genus density and tree entropy, transport costs differing by a factor
$8$ at $n=10^5$; metric bounds scale with $\diam(G)$ ($n^{1/2}$ against
$\log n$). \\
\Cref{rem:curv} & consistency only &
Diffuse routes on the uniform honeycomb, hub-centred routes on ER (attraction
$1.38$). \\
\Cref{thm:chipmove} & identity &
$k\,W_{\rm opt}$ is the number of chip moves ($\approx2900$ against $\approx370$ at
$n=10^5$). \\
\bottomrule
\end{tabular}
\caption{Results and the kind of evidence provided for each. ``Exact test'':
direct test by exact computation over linear systems on small graphs. ``Large
networks'': direct test of a statement about regularized plans. ``Consistency
only'': the large-network data are compatible with the result but do not test
it, since no minimization over linear systems is involved and $d\ll g$.}
\label{tab:synthesis}
\end{table}

The large-scale comparison is thus consistent with the central thesis of the
paper, within the limits stated in \cref{tab:synthesis}. Honeycomb and Erd\H{o}s--Rényi
networks of equal size are close in genus density and tree entropy, and entirely
distinguishable to transport. The information geometry of $F_q$ acts on
both in the same way, through $\gamma/q$ and $\ln_q(N)$. Topology becomes
visible in transport only at particle densities comparable to $g/n$, and its
persistence against entropy is then measured by the metric defect.

\section{Concluding remarks and open problems}\label{sec:conclusion}

The information geometry generated by $F_q$ is the Tsallis geometry at
temperature $\gamma/q$, with the correction term acting as a gauge. Since it is
defined without reference to the firing lattice, it carries no information on
linear equivalence. Topology reappears arithmetically, in
transport costs minimized over linear systems, and it competes with entropy
through a gap whose low- and high-temperature limits are explicit and whose
existence, for divisors of every degree $d\ge1$, is equivalent to $g\ge1$. We close
with questions.

\begin{enumerate}[label=(\arabic*)]
\item Can $d\,\delta_{W_1}(D)$ be characterized by distance invariants of
$\Jac(G)$, for instance a covering radius of the classes of $\Div^d_+$ in the
Cayley graph of \cref{thm:AJ}, with effectivity accounted for? Is it eventually
periodic in $d$, with period dividing
the exponent of $\Jac(G)$? Is its eventual maximum a graph invariant between
$\rho(G)$ and $\diam(G)g$?
\item Which graphs satisfy $\rho(G)=\diam(G)g$? Cycles do. More generally,
characterize the classes with $\delta_{\TV}(D)=(d-r(D))/d$. The upper bound
itself is proved; its sharpness is a separate question, and the finite
enumerations reported here are evidence, not a classification theorem. The criterion
$r(D)=\min_vm_v(D)$ of \cref{thm:vertexrank} is sufficient and covers $1768$ of
the $1769$ equality cases found among $1784$ classes; the classes $4v_1$ on $K_4$
and $3v$ on $C_3$ (\cref{prop:smallgenus}) show that it is not necessary. Equality
holds for every class in genus at most one; is the same true in genus two, as
all $198$ genus-two classes computed suggest? Equality fails, for example, for $K+v_1$ on
$G''$ and for several classes of degree $2g-2$ and $2g-1$ on $K_4$ and $K_{3,3}$.
\item Bound the topological term $\mathsf T_{q,\gamma}(D,E)$ at \emph{finite}
temperature in terms of $r(D)$, $r(K-D)$ and the word metric of $\Jac(G)$. At
present only its limits are known (\cref{thm:TI}), together with the upper bound
inherited from \cref{thm:serre}.
\item Is $\mathsf I_{q,\gamma}(E)=0$ on an interval $(0,\gamma_0(E))$ for every target
$E$, as observed, and can $\gamma_0$ be expressed through the Legendre dual? Is
$q_{\rm gap}(\gamma)$ always increasing, and when is the integrality gap of
\cref{thm:minimax} the whole of $\mathsf T$, as on $C_7$?
\item Find a closed form for the crossover temperature $\gamma^\ast(q)$. The
value $16.0$ at $q=2$ suggests that the quadratic case is solvable.
\item In unbalanced transport the correction term of $F_q$ is a chemical
potential. Does it shift $\gamma^\ast$, giving it a role beyond gauge?
\item Can the torsion $\widehat{\Jac(G)}$ among dual potentials be read off from
entropic potentials $(\alpha,\beta)$, so that Sinkhorn-type iterates detect
linear equivalence classes?
\item Exact computations on seven graphs of genus four suggest that the
crossover temperature at matched genus is an increasing function of the $W_1$-defect. Can this be proved, and
does a dense-particle regime ($d/n\gtrsim g/n$) on large honeycomb and Erd\H{o}s--Rényi
networks show the same law?
\item Do these results extend to metric graphs and tropical
curves~\cite{GathmannKerber2008,MikhalkinZharkov2008}, where Riemann--Roch
theorems hold and transport is continuous?
\end{enumerate}

\appendix
\section{Worked examples}\label{app:exercises}

\begin{exercise}
On $C_7$ let $D=2v_1+v_3+v_5$. Then $\deg D=4$, $g=1$, $\MobIdx(D)=4$ and
$K-D=-2v_1-v_3-v_5$, so $r(K-D)=-1$ and $r(D)=3$. By \cref{thm:defect}(c),
$\delta_{\TV}(D)=\tfrac14$, and by \cref{thm:AJ}, $\delta_{W_1}(D)=\tfrac34$.
\end{exercise}

\begin{exercise}
For $D=5v_1$ on $T_7$, $C_7$ and $G''$, we have $d>2g-2$, so $r=5-g$ and
$\MobIdx=6-g$, giving $(\MobIdx,r)=(6,5),(5,4),(3,2)$. The defect bounds $g/d$
are $0,\tfrac15,\tfrac35$, and all three are attained. By \cref{thm:phase}, a
positive-gap regime exists on $C_7$ and $G''$ but not on $T_7$.
\end{exercise}

\begin{exercise}
On the four-cycle $C_4$ let $D=2v_1$. Then $K=0$, so $\mu_K$ is undefined and
\eqref{eq:naive} has no content. On the other hand $\MobIdx(D)=2$, $r(D)=1$,
$\delta_{\TV}(D)=\tfrac12$, and, since $\Jac(C_4)\cong\Z/4\Z$ has Cayley diameter
$2=\diam(C_4)$, \cref{thm:AJ} gives $\delta_{W_1}(D)=1$.
\end{exercise}

\section{Computational method}\label{app:computation}

\medskip\noindent\emph{Classes.}\;Fix $q=v_1$ and the reduced Laplacian $L_q$,
with $\kappa=\det L_q$. A degree-zero divisor is principal if and only if its
restriction $x$ to $V\setminus\{q\}$ lies in $L_q\Z^{n-1}$, and since
$L_q^{-1}=\adj(L_q)/\kappa$, the vector $\adj(L_q)x\bmod\kappa$ is a complete
class invariant. It is computed in exact integer arithmetic.

\medskip\noindent\emph{Ranks.}\;$r(D)\ge k$ is tested by checking that
$D-E$ lies in an effective class for every $E\in\Div^k_+$. This check is
independent of the Riemann--Roch theorem, and \eqref{eq:RR} was verified in
every case.

\medskip\noindent\emph{Defects and Jacobians.}\;$\TV$ is computed directly.
$W_1$ and $W_2^2$ come from transportation linear programs (HiGHS) and,
equivalently, from breadth-first search in the chip-move graph
(\cref{thm:chipmove}); the two methods agree. $\rho(G)$ is obtained by
breadth-first search on the Cayley graph and $N_w$ by Dijkstra's algorithm.

\medskip\noindent\emph{Regularized transport.}\;The problems $\Wq$ are solved
with the conic solver Clarabel through CVXPY, and crossover temperatures are
located by bisection to relative precision $10^{-8}$. On $C_7$ the vanishing
high-temperature slope makes bisection unreliable beyond $\gamma\approx10^2$;
we report only that no sign change occurs for $\gamma\le5$.

\medskip\noindent\emph{Reproducibility.}\;All computations were run with
Python~3.13, NumPy, SciPy~1.18.1 (HiGHS linear programming), NetworkX~3.6.1 and
CVXPY~1.9.3 with the Clarabel interior-point solver, on a machine with $2$
virtual CPUs and $7$\,GB of memory. The scripts and the resulting data (JSON)
form the supplementary archive. The settings are as follows.
\begin{itemize}
\item \emph{Networks.} The honeycomb network is
\texttt{hexagonal\_lattice\_graph}$(m,k)$ with $m=\operatorname{round}(\sqrt{n/2})-1$
and $k=\operatorname{round}(n/(2(m+1)))-1$. The ER network is the giant
component of \texttt{fast\_gnp\_random\_graph}$(n_0,3/(n_0-1))$ with
$n_0=\operatorname{round}(n/0.9405)$, seeded by the first integer drawn from a
NumPy generator with seed $n$. In the main experiment
(\cref{tab:simstruct,tab:simeff,tab:simreg,tab:fixeps}) one network is
generated per family and size, so the trial variation there is over particle
placements, not over networks; the honeycomb patch of a given size is unique in
any case. Network-to-network variation is assessed only in the robustness study
of \cref{tab:robust}: five further ER networks per size (seeds $1000$--$1004$),
$k\in\{25,50,100\}$, and $10$ placements per network.
\item \emph{Structural invariants.} Diameters are double-sweep estimates (four
sweeps), which are lower bounds. Mean distances use $60$ random sources
(generator seed $1$). $\log\kappa$ is computed exactly by sparse $LU$ except for
the $10^5$-node ER network (stochastic Lanczos quadrature, $30$ probes, $80$
steps).
\item \emph{Transport trials.} Generator seed $7$; $20$ trials per family and
size; $k=24$ particles at $n\approx10^2$ and $k=50$ otherwise. The regularized
problems are posed on the $k\times k$ cost matrix between particle positions,
so they have $k^2\le2500$ variables at every $n$. The size $n$ enters only
through breadth-first search. The dominant memory is the pair of $k\times n$
distance and predecessor arrays (about $80$\,MB at $n=10^5$). The complete run
of \cref{tab:simeff,tab:simreg} took $164$\,s, of which $32$\,s and $45$\,s
were the $10^5$-node honeycomb and ER cases.
\item \emph{Solvers.} Clarabel with default tolerances ($10^{-8}$ on duality gap
and feasibility). For $q\notin\{1,2\}$ in \cref{sec:pd}, exact power cones are
used, and problems are restricted to the supports of the marginals. In
\cref{sec:pd} every solve returned status \emph{optimal}; in \cref{sec:sim} the
marginals are uniform and strictly positive, the case in which no inaccuracy
was observed. Linear programs use HiGHS with default tolerances, except the
relative-interior computation of \cref{tab:face} (feasibility tolerance
$10^{-10}$).
\item \emph{Divisor classes.} Classes are enumerated by grouping all of
$\Div^d_+(G)$ according to the exact class invariant described above. Ranks are
computed by the brute-force test above, $m_v(D)$ as the maximum of $D'(v)$ over
the enumerated linear system, and $\delta_{\TV}(D)$ from the integer overlaps
$\max_{D'}\sum_v\min(D'(v),E(v))$; all comparisons are exact. The file \texttt{classes\_all.csv} has one row per class and the columns
\texttt{graph}, \texttt{n}, \texttt{g}, \texttt{d}; \texttt{rep} (a representative
$D$, vertex by vertex); \texttt{size\_linear\_system} ($\lvert\lin D\rvert$);
\texttt{r}, \texttt{r\_K\_minus\_D}; \texttt{m\_v} (all $m_v(D)$) and
\texttt{min\_m\_v}; \texttt{delta\_TV}; \texttt{lower} ($1-\min_vm_v/d$) and
\texttt{upper} ($(d-r(D))/d$); and the flags \texttt{upper\_attained},
\texttt{lower\_attained} and \texttt{criterion} ($r(D)=\min_vm_v$). The script
\texttt{verify\_classes.py} is an independent re-implementation. It builds the
graphs with NetworkX, keys classes by the exact rational inverse of the reduced
Laplacian, and computes ranks by its own brute-force routine. It recomputes
every column of all $1784$ rows, finds no mismatch, and checks that for each
graph and degree the classes partition $\Div^d_+(G)$. Running \texttt{classdump.py} from the archive regenerates the file
row for row, identically, in about $90$\,s. \texttt{reproduce\_counts.py} re-derives,
and asserts, every count quoted in the paper. The archive is self-contained,
its scripts use paths relative to their own folder, and a \texttt{README}
maps each script to the tables and statements it reproduces. All scripts were
re-run from a fresh copy of the archive before submission.
\item \emph{Robustness runs} (\cref{tab:robust}). Exact transport only; ER
realizations with seeds $1000$--$1004$, placements from a generator with seed
$2026$. The estimators are defined in \cref{ssec:eff} and implemented in
\texttt{robust\_fit.py}.
\item \emph{Small graphs.} Matched-genus random graphs are the first six
connected outputs of \texttt{gnm\_random\_graph}$(16,19)$ with seeds drawn from a
generator with seed $3$. The optimal-face instances of \cref{tab:face} use seed
$11$. All divisor-theoretic quantities are computed in exact integer arithmetic.
\end{itemize}

\Cref{tab:repro} collects, for the large-network experiments, where each item
needed for reproduction is specified.

\begin{table}[!htbp]
\centering\small
\renewcommand{\arraystretch}{1.12}
\begin{tabular}{>{\raggedright\arraybackslash}p{3.6cm}>{\raggedright\arraybackslash}p{10.2cm}}
\toprule
Item & Specification \\
\midrule
Network generation, seeds & Honeycomb: \texttt{hexagonal\_lattice\_graph}, unique per size. ER: giant component of $G(n_0,3/(n_0-1))$; main experiment seed derived from $n$, robustness seeds $1000$--$1004$. \\
Vertices, edges, genus & \Cref{tab:simstruct}, rows $n$, $\lvert E\rvert$, $g$. \\
Independent networks & Main experiment: $1$ per family and size. Robustness: honeycomb $1$, ER $5$ per size. \\
Particle placements & $2k$ distinct vertices drawn uniformly without replacement; the first $k$ are sources, the rest targets. Main: $20$ per size (seed $7$), $k=24$ at $n\approx10^2$, else $k=50$. Robustness: $10$ per network and $k\in\{25,50,100\}$ (seed $2026$). \\
Costs & Hop distances by breadth-first search; for regularized plans, divided by their median. \\
Regularization & $q\in\{\tfrac12,1,2\}$; $\tau=\gamma/\operatorname{med}(C)\in\{0.03,0.1,0.3\}$ (\cref{tab:simreg}); fixed $\varepsilon=\gamma/q$ in \cref{tab:fixeps}. \\
Solvers, stopping & Exact transport: \texttt{linear\_sum\_assignment} (exact). Regularized: Clarabel, tolerance $10^{-8}$ on duality gap and feasibility; one warning in $540$ solves, reported in \cref{ssec:reg}. Zeros counted below $10^{-6}/k^2$. \\
Regression & Honeycomb: $\log W_{\rm opt}=a+b\log n$; ER: $W_{\rm opt}=a+b\log n$; OLS over $n\approx10^3,10^4,10^5$ (\cref{ssec:eff}). Standard errors are OLS standard errors. \\
Variance split & One-way random-effects ANOVA, \cref{ssec:eff}. \\
\bottomrule
\end{tabular}
\caption{Reproducibility checklist for \cref{sec:sim}.}
\label{tab:repro}
\end{table}

For the small-graph computations, the archive also contains scripts that
re-solve the regularized problems without CVXPY: Sinkhorn iteration for $q=1$,
and the smooth dual for $q=2$ and for the right-hand side of \cref{thm:minimax}.
They reproduce the gap curves and crossover temperatures of \cref{tab:phase}
and the entries of \cref{tab:levels}. The \texttt{README} of the archive lists,
for each script, the tables and statements it reproduces.

\end{document}